\documentclass[authoryear,preprint,review,12pt]{elsarticle}

\usepackage{amssymb}
\usepackage{amsmath}
\usepackage{hyperref}
\usepackage{booktabs}
\usepackage{adjustbox} % package to adjust table width
\usepackage{multirow}
\usepackage{tabularx}
\usepackage{longtable}
\usepackage{booktabs}
\usepackage{ragged2e} % for better text alignment
\usepackage{xltabular}
\usepackage{pifont}
\usepackage{tablefootnote}
\usepackage{threeparttable}
\usepackage{pdflscape}
\journal{Chemical Engineering Science}

\begin{document}

\begin{frontmatter}



\title{A comprehensive approach to incorporating intermolecular dispersion into the openCOSMO-RS model. Part 2: Atomic polarizabilities}


\author[inst1,inst11]{Daria Grigorash}
\affiliation[inst1]{organization={Department of Chemistry, Technical University of Denmark},%Department and Organization
            city={Kgs. Lyngby},
            postcode={2800}, 
            country={Denmark}}
            
\affiliation[inst11]{organization={Center for Energy Resources Engineering, Technical University of Denmark},%Department and Organization
            city={Kgs. Lyngby},
            postcode={2800}, 
            country={Denmark}}
            
\author[inst2]{Simon Müller}
\affiliation[inst2]{organization={Institute of Thermal Separation Processes, Hamburg University of Technology},%Department and Organization
            city={Hamburg},
            postcode={21073}, 
            country={Germany}}
            
\author[inst6]{Esther Heid}
\affiliation[inst6]{organization={Institute of Material Chemistry, TU Wien},%Department and Organization
            city={Vienna},
            postcode={1060},
            country={Austria}}
            
\author[inst5]{Frank Neese}
\author[inst7]{Dimitrios Liakos}
\affiliation[inst5]{organization={FAccTs GmbH},
            %addressline={}, 
            city={Cologne},
            postcode={50677},
            country={Germany}}
\affiliation[inst7]{organization={Max-Planck-Institut für Kohlenforschung},
            %addressline={}, 
            city={Muelheim an der Ruhr},
            postcode={45470},
            country={Germany}}
\author[inst5]{Christoph Riplinger}
\author[inst5]{Miquel Garcia-Ratés}

\author[inst3]{Patrice Paricaud}
\affiliation[inst3]{organization={UCP, ENSTA Paris, Institut Polytechnique de Paris},%Department and Organization
            city={Palaiseau},
            postcode={91762}, 
            country={France}}
            
\author[inst1,inst11]{Erling H. Stenby}
\author[inst2]{Irina Smirnova}
\author[inst1,inst11]{Wei Yan\corref{corresponding author}}
\cortext[corresponding author]{Corresponding author at: Department of Chemistry, Technical University of Denmark, 2800 Kgs. Lyngby, Denmark.\ead{weya@kemi.dtu.dk}}
\begin{abstract}
openCOSMO-RS is an open-source predictive thermodynamic model that can be applied to a broad range of systems in various chemical and biochemical engineering domains. This study focuses on improving openCOSMO-RS by introducing a new dispersion term based on atomic polarizabilities. We evaluate different methods for processing polarizability data, including scaling and combining it to compute segment-segment dispersion interaction energies, with a focus on halocarbon systems. The results demonstrate that the modified model outperforms our previous method developed in the first part of this work \citep{grigorash2024comprehensiveapproachincorporatingintermolecular} , while at the same time requiring fewer adjustable parameters. The approach was applied to a broad dataset of over 50,000 data points, consistently increasing the accuracy across a variety of data types. These findings suggest that atomic polarizability is a valuable descriptor for refining dispersion interactions in predictive thermodynamic models.
\end{abstract}



\begin{highlights}
\item A new dispersion term based on atomic polarizabilities was implemented in openCOSMO-RS, improving its accuracy.
\item Various methodologies for handling polarizabilities were evaluated on halocarbon systems.  
\item The updated model outperforms previous versions with fewer parameters and can be applied to diverse mixtures beyond halocarbons. \end{highlights}

\end{frontmatter}


\section{Introduction}
\label{sec:sample1}

%%       \citet{<label>} ==> Jones et al. (2015)
%%       \citep{<label>} ==> (Jones et al., 2015)

Predictive thermodynamic models that integrate molecular-level quantum chemical (QC) information have become essential tools in both academic research and industrial applications. A prominent approach in this domain is the COSMO-RS (COnductor-like Screening Model for Realistic Solvents) model, developed by combining QC calculations of screening surface charge densities with statistical thermodynamics \citep{Klamt1995Conductor-likeMolecules, Klamt1998RefinementCOSMO-RS}. It offers a predictive framework for estimating thermodynamic properties directly from molecular descriptors.

Since the initial development of COSMO-RS, several implementations have been proposed including  COSMO-SAC \citep{Lin2002AModel}, COSMO-RS(ol) \citep{Grensemann2005PerformanceMethods} and openCOSMO-RS \citep{Gerlach2022AnDescriptors}. These models have been successfully applied to predicting thermodynamic properties such as vapor-liquid equilibria (VLE) \citep{Klamt2007PredictionCOSMOtherm, Mambo-Lomba2021PredictionsApproaches, HSIEH201090, Hsieh2014ConsideringBehavior, Grensemann2005PerformanceMethods}, infinite dilution activity coefficients (IDACs)  \citep{Fingerhut2017ComprehensiveEquilibria, Gerlach2022AnDescriptors}, solvation free energies \citep{MULLER2025114250, Saidi2020PredictionsApproaches, Saidi2020PredictionsApproachesb, PAES2024124641}, partitioning coefficients \citep{Gerlach2022AnDescriptors, Klamt1998RefinementCOSMO-RS} and phase behavior across diverse chemical families \citep{Jiriste2022PredictingCOSMO-RS, Klajmon2022PurelyCOSMO-RS, Peng2022}, including both polar and nonpolar \citep{Klamt2002PredictionCOSMO-RS, Klamt2003PredictionCOSMO-RS, Eckert2003PredictionCOSMO-RS}, and electrolyte systems \citep{Gerlach2018,Muller2019EvaluationIons, kroger_prediction_2020, Muller2020CalculationSystems, GonzalezdeCastilla2021OnCoefficients, ritter_influence_2016, arrad_thermodynamic_2024, gonzalez_de_castilla_analogy_2022}. 

Central to these thermodynamic models is the accurate representation of intermolecular interactions, including hydrogen bonding, electrostatic and dispersion forces. Dispersion forces, in particular, arise from interactions between instantaneous dipoles generated by fluctuations in polarizable electron clouds \citep{Prausnitz1999MolecularEquilibria}. These forces are intrinsically linked to the polarizability of atoms or molecules, a property quantifying how easily an electron cloud can be distorted under an electric field. Theories link polarizability to van der Waals interactions, which are proportional to the product of the polarizabilities of the interacting species and scale inversely with the sixth power of their separation distance, as described by London's potential \citep{London1937TheForces}. Polarizability may serve as a link between quantum mechanical electron behavior and macroscopic thermodynamic properties. Therefore, the dispersion contribution in COSMO-RS could be enhanced by incorporating polarizabilities.

In the first part of this work \citep{grigorash2024comprehensiveapproachincorporatingintermolecular}, we developed a dispersion contribution for the openCOSMO-RS model based on a small set of general adjustable atomic parameters, resulting in significant improvements in modeling the phase equilibrium of halocarbons. Building on that foundation, the focus of this work is the development of a generalized openCOSMO-RS model that incorporates a dispersion term based on a novel descriptor—atomic polarizability—which, to the best of our knowledge, has not been previously implemented in the COSMO-RS framework. Atomic polarizability tensors were calculated using the latest release of ORCA 6.0  \citep{Neese2012TheSystem, Neese2020ThePackage}, as detailed in Section \ref{sec:COSMO_calculations}, and subsequently projected onto molecular cavities following the procedure outlined in Section \ref{sec:Polarizability_projections methods}. The approach for incorporating polarizabilities into the COSMO-RS equations is discussed in Section \ref{sec:COSMORS}, followed by the description of the regression procedure used to fine-tune the modified model to experimental data (Section \ref{sec:regression_methods}). As in the first part of this study, we initially focused on halocarbons \ref{sec:halocarbons}, as these systems are particularly sensitive to dispersive interactions and frequently require a special consideration of the dispersion term for an adequate modeling \citep{Klamt2007PredictionCOSMOtherm}. In Section \ref{sec:Intermolecular interactions in fluorocarbon mixtures}, we discuss these challenging interactions in depth, focusing primarily on hydrocarbon - fluorocarbon mixtures. To determine the optimal approach for incorporating polarizabilities into the dispersion term using halocarbon data, we evaluate various methodologies in Sections \ref{sec:Isotropic polarizabilities} and \ref{sec:Polarizability projections}. Additionally, we investigate the impact of the F-atom radius used in C-PCM (Conductor-like Polarizable Continuum Model) calculations (Section \ref{sec:F-atom radius}) and different polarizability combining strategies in Sections \ref{sec:adjusted power} and \ref{sec:ionization potentials}. Finally, we extend the models that demonstrated the best performance for halocarbon data (Section \ref{sec:conclusions_halocarbons}) to a comprehensive dataset (Section \ref{sec:Parameterizations_all_data_discussion}) of nearly 50,000 data points, covering compounds of diverse chemical nature, which results in the improved general openCOSMO-RS model.
%%%% 

\section{Methods}
\label{sec:Methods}

\subsection{QC calculations}
\label{sec:COSMO_calculations}
For QC C-PCM calculations, we used the open-source workflow from the first part of this work and detailed in \cite{Gerlach2022AnDescriptors}, now with the latest ORCA 6.0 release. The new version includes methods for calculating molecular static polarizability tensors and decomposing them into atomic components. These properties are calculated by calling an additional block for electrical properties, along with the reference method, such as Hartree-Fock or DFT (Density Functional Theory), to solve for the electronic structure. 

An example input might contain the following block:
\begin{verbatim}
%elprop
    Polar 1
    Polaratom 1
end
\end{verbatim}
The polarizability is calculated analytically, with the index "1" in the input block specifying this. Since DFT is the basis for electrostatics in the C-PCM approach used here, it serves as the reference method. Polarizabilities were calculated in a perfect conductor to keep the calculations consistent with the rest of the workflow. The resulting values are higher than those obtained in vacuum, but behave qualitatively the same.

The polarizability tensor is given by: 
\begin{equation}
\alpha_{KL} = \sum_{\mu\nu} \frac{\partial P_{\mu\nu}}{\partial r_{K}} \langle \mu \vert r_{L} \vert \nu \rangle.
\end{equation}

Here, \(\alpha_{KL}\) is the polarizability tensor (\(K,L=x,y,z\)), \(\mu\), \(\nu\) are basis functions, \(P_{\mu\nu}\) is the density matrix. The electric field \(E\) interacts with the dipole moment of the molecule as \(-Ed\), where the dipole operator is given by:
\begin{equation}
d = \sum_{A} Z_{A} R_{A} - \sum_{i} r_{i}.    
\end{equation}
Here, \(A\) denotes nuclei at positions \(R_A\) with charge \(Z_A\) and \(i\) denotes electrons with position operator \(r_i\). The derivative of the density matrix is the response density with respect to an electric field. In real space it reads: 
\begin{equation}
\rho^{(K)}(\mathbf{r}) = \sum_{\mu\nu} \frac{\partial P_{\mu\nu}}{\partial r_{K}} \mu(\mathbf{r}) \nu(\mathbf{r}).
\end{equation}
In order to arrive at a local polarizability for a given atom \(A\), we decompose the real-space response density by multiplying it with a basic function \(f^{(A)}(\mathbf{r})\) to get an atomic response density \(\rho^{(A;K)}(\mathbf{r})=f^{(A)}(\mathbf{r})\rho^{(K)}(\mathbf{r})\). The basic function must be chosen such that \(\sum_{A} f^{A}=1\). A particularly simple choice and the one we have adopted in this work is: 
\begin{equation}
f^{(A)}(\mathbf{r}) = \frac{\tilde{\rho}^{(A)} (r)}{\sum_{B}\tilde{\rho}^{(B)} (r)}, 
\end{equation}
where, \(\tilde{\rho}^{(A)} (r)\) is the spherically symmetric, tabulated density of a neutral atom \(A\). Thus, in real-space the atomic polarizability tensor is defined as: 
\begin{equation}
\alpha^{(A)}_{KL} = \int \rho^{(A;K)}(\mathbf{r})r_Ldr,
\end{equation}
which fulfills the requirement that \(\sum_{A}\alpha^{(A)}_{KL}=\alpha_{KL}\). The necessary integration is carried out numerically using the same grid as used in the preceding DFT calculation. 

The "Polaratom" command in the input block decomposes the molecular polarizability tensor into atomic components. 
For an initial assignment of polarizability values to each segment, we considered isotropic polarizabilities, \(\alpha_{\textrm{iso}}\), calculated from the diagonal elements (\(\alpha_{xx}, \alpha_{yy}, \alpha_{zz}\)) of an atomic polarizability tensor:
\begin{equation}
\alpha_{\textrm{iso}} = \frac{\alpha_{xx} + \alpha_{yy} + \alpha_{zz}}{3}. \label{eq:isotropic_polarizability}
\end{equation}
Additionally, atomic polarizability tensors were projected onto segments. Using these as a new segment descriptor is inspired by the London attractive potential between two spherically symmetric systems \citep{London1937TheForces}:
\begin{equation}
 \phi^{\textrm{London}}_{ij} = -\frac{3}{2} \frac{\alpha_{i} \alpha_{j}}{r_{ij}^6} \cdot \frac{h \nu_i h \nu_j}{h(\nu_i + \nu_j)}, \label{eq:London}
\end{equation}
where \(\alpha_{i}\) and \(\alpha_{j}\) are their polarizabilities, and \(r_{ij}\) is the distance between them. Equation \ref{eq:London} further demonstrates the effect of zero-point vibrations with frequencies \(\nu_i\) and \(\nu_j\), which can be approximated by ionization potentials \citep{Hudson1960IntermolecularRules}. 

\subsection{Projection of Atomic Polarizability Tensors onto Molecular Cavities}
\label{sec:Polarizability_projections methods}
\begin{figure}
    \centering
    \includegraphics[width=0.5\linewidth]{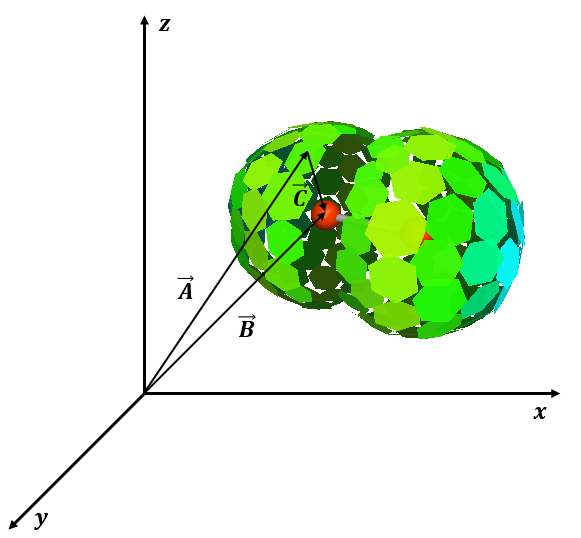}
    \caption{The molecular cavity of a diatomic molecule (Br\(_{2}\)) with its segments and vectors connecting the center of the coordinate system and a segment center (\(\vec{A}\)), an atom center (\(\vec{B}\)), and the center of the segment to the atom center (\(\vec{C}\)). The molecular cavity for the image is created using COSMOtherm v.24.}
    \label{fig:polarizability_projections}
\end{figure}

Figure \ref{fig:polarizability_projections} illustrates the projection of the polarizability tensor onto the molecular cavity. Vector \(\vec{A}\) connects the coordinate system center to the center of segment \textit{i}, and vector \(\vec{B}\) to the atom center. Vector \(\vec{C}\), connecting these two centers, is calculated by subtracting the coordinates of \(\vec{A}\) from those of \(\vec{B}\). We first take the absolute value of the coordinates of \(\vec{C}\) and then normalize \(\vec{C}\) by the sum of its coordinates to equalize all vectors from other segments relative to their location with respect to the atom center. For each atom, ORCA electrical properties calculations provide a 3-dimensional polarizability tensor in the form of a \(3 \times 3\) matrix. The projection of the atomic polarizability tensor \(\boldsymbol{\alpha}\) onto the atom’s segments is then performed via matrix multiplication of this tensor with vector \(\vec{C}\):
\begin{equation}
\vec{D} = \vec{C} \cdot \boldsymbol{\alpha}.
\label{eq:projection_equation}
\end{equation}

To derive a scalar descriptor, we compute the Euclidean norm of the resulting vector \(\vec{D}\), ensuring a physically meaningful positive value. This norm is then used as a local descriptor of the segment’s polarizability, hereafter referred to as the polarizability projection \(\alpha_{i}^{I}\) of segment \(i\) on atom \(I\). 

\begin{figure}
    \centering
    \includegraphics[width=0.5\linewidth]{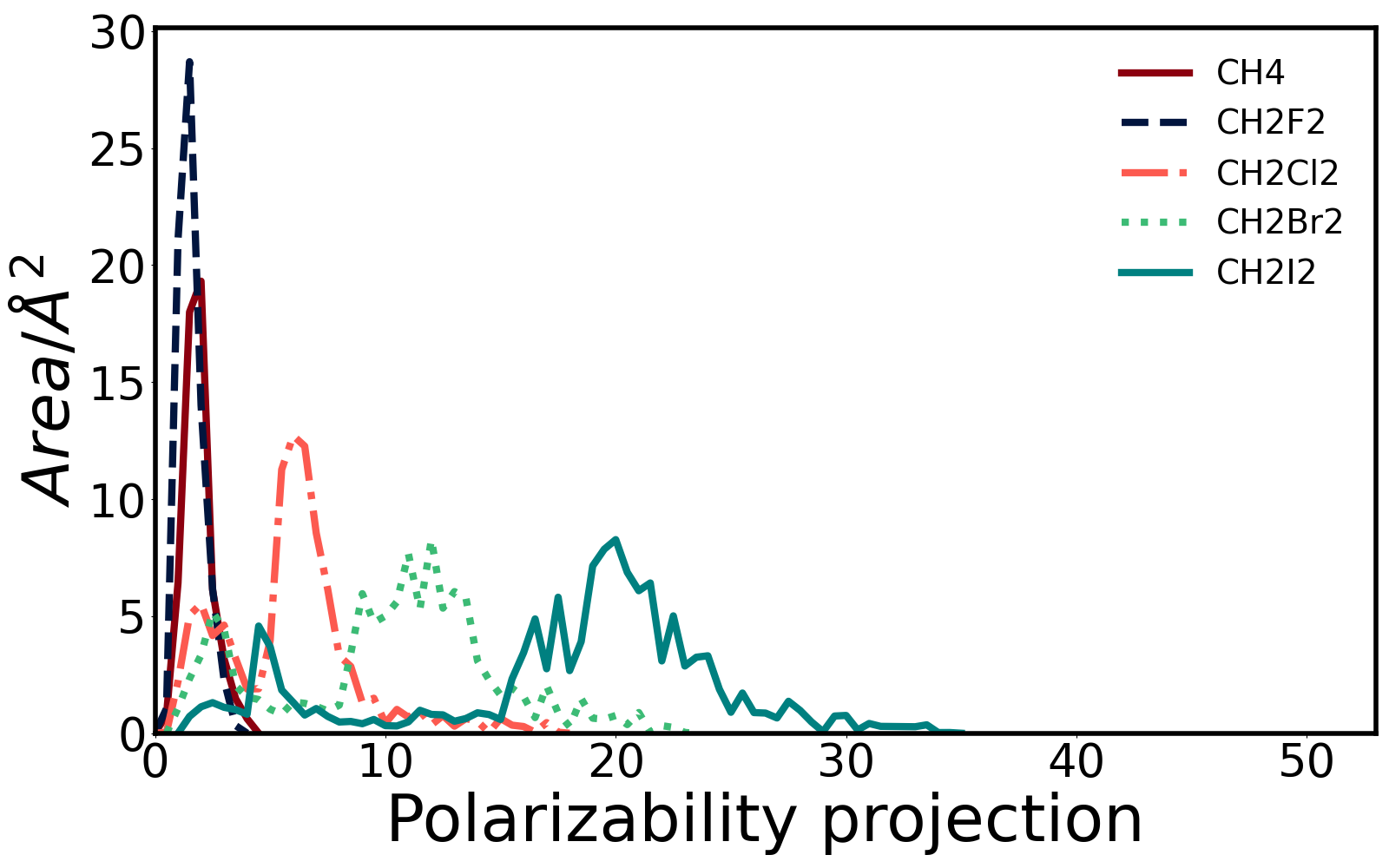}
    \caption{\(\alpha\)-profiles of methane and dihaloalkanes.}
    \label{fig:alpha_profiles_dihalo}
\end{figure}

This procedure results in a large number of possible polarizability projection values within a single molecule, which significantly increases computational time due to the expanded variety of segment types. To keep computational time manageable, we performed the clustering of \(\alpha_{i}^{I}\) with further details in Section \ref{sec:Polarizability projections}. In a similar manner as the screening surface charge densities are clustered to generate the \(\sigma\)-profiles in COSMO-RS, we can analogously plot \(\alpha\)-profiles as illustrated in Figure \ref{fig:alpha_profiles_dihalo}. It can be observed that as the atomic number of the halogen increases, the \(\alpha\)-profiles of the corresponding haloalkanes shift toward higher polarizability projection values. This trend aligns with the expected physical properties of halogens and haloalkanes.

To incorporate more molecular information to the new descriptor, we evaluated five scaling factors \((w_{1} - w_{5})\) for each projection \(\alpha_{i}^{I}\). 

\begin{itemize}
    
    \item[\(w_{1}\)]: Normalization by the number of grids (or segments) C-PCM calculations assigned to atom \(I\), expressed as 
    \begin{equation}
    w_{1} = \frac{1}{n_{\textrm{grid}}^{I}}.
    \label{eq:w1}
    \end{equation}

    \item[\(w_{2}\)]: The average number of grids across all atoms is used in Eq.~\ref{eq:w1}.

    \item[\(w_{3}\)]: The ratio of the segment area \(a_{i}\) to the total area of atom \(I\):
    \begin{equation}
    w_{3} = \frac{a_{i}}{A_{I}}.
     \label{eq:w3}
    \end{equation}

    \item[\(w_{4}\)]: The average atomic area across the molecule is used in the denominator of Eq ~\ref{eq:w3}.

   \item[\(w_{5}\)]: Based on the COSMO-RS model’s \(\sigma\)-averaging method, this scaling substitutes \(\sigma\) with \(\alpha_{i}^{I}\):
    \begin{equation}
    \alpha_{i}^{I*} = \frac{\sum_{j}^{N} \alpha_{j}^{I} \frac{r_{j}^{2} r_{\textrm{av}}^{2}}{r_{j}^{2} + r_{\textrm{av}}^{2}} \exp\left(\frac{-d_{ij}^{2}}{r_{j}^{2} + r_{\textrm{av}}^{2}}\right)}{\sum_{j}^{N} \frac{r_{j}^{2} r_{\textrm{av}}^{2}}{r_{j}^{2} + r_{\textrm{av}}^{2}} \exp\left(\frac{-d_{ij}^{2}}{r_{j}^{2} + r_{\textrm{av}}^{2}}\right)}.
    \label{eq:w5}
    \end{equation}

\end{itemize}
In the first four cases, the scaled polarizability is obtained by simply multiplying \(\alpha_{i}^{I}\) by one of the scaling factors, \(w_{i}\). Using the scaling factor \(w_{1}\), we can distinguish between segments located on different atoms while sharing the same \(\alpha_{i}^{I}\). The inverse proportionality to distance, as indicated in Eq.~\ref{eq:London}, is captured through the scaling  factor \(w_{3}\), which, along with \(w_{4}\), also accounts for the area of an individual segment. Similar to the original \(\sigma\)-averaging procedure, Eq.~\ref{eq:w5} smears out segments and their corresponding \(\alpha_{i}^{I}\).  

\subsection{COSMO-RS}
\label{sec:COSMORS}
For a detailed explanation of the model, we refer the reader to foundational works by Klamt \citep{Klamt1998RefinementCOSMO-RS, Klamt2000COSMO-RS:Liquids} and to our previous studies \citep{Gerlach2022AnDescriptors, grigorash2024comprehensiveapproachincorporatingintermolecular}. Here, we highlight only the aspects of the model that were modified or differ from the original implementation.

In COSMO-RS, thermodynamic properties are calculated based on the interaction energies between molecular segments \( E_{ij} \), with three contributing components:
\begin{equation}
 E_{ij} = E_{\textrm{mf}}(\sigma_i,\sigma_j) + E_{\textrm{hb}}(\sigma_i,\sigma_j) - E_{\textrm{vdW}}.
\end{equation}
These contributions include the repulsive misfit energy \(E_{\textrm{mf}}(\sigma_i,\sigma_j)\), the hydrogen-bonding energy \(E_{\textrm{hb}}(\sigma_i,\sigma_j)\), and the dispersive energy \(E_{\textrm{vdW}}\).
The electrostatic misfit free energy is typically calculated as:
\begin{equation}
E_{\textrm{mf}}(\sigma_i,\sigma_j) = 0.5 a_{\textrm{eff}} \alpha_{\textrm{mf}} \left[ (\sigma_i + \sigma_j)^2 + f_{\textrm{corr}}(\sigma_i + \sigma_j)(\sigma_i^\perp + \sigma_j^\perp) \right],
\end{equation}
which incorporates both the screening charge density \(\sigma\) and the correlation screening charge density \(\sigma^\perp\) to account for the influence of the surrounding segments. Here, \(f_{\textrm{corr}}\) represents a correlation correction factor adjusted to dielectric energy data \citep{Klamt1998RefinementCOSMO-RS}. The general COSMO-RS parameters \(\alpha_{\textrm{mf}}\) and \(a_{\textrm{eff}}\) are the misfit prefactor and the effective contact area of a segment, respectively.

In this study, to enhance predictions for fluorinated molecules, the misfit free energy was corrected by an additional repulsive energy term \(E_{\textrm{corr}}^{\textrm{F}}\) to all segment-segment interactions involving an F-atom (F - segments). This modification leads to the following adjusted equation:
\begin{equation}
E_{\textrm{mf}}(\sigma_i,\sigma_j) = 0.5 a_{\textrm{eff}} \alpha_{\textrm{mf}} \left[ (\sigma_i + \sigma_j)^2 + f_{\textrm{corr}}(\sigma_i + \sigma_j)(\sigma_i^\perp + \sigma_j^\perp) \right] + E_{\textrm{corr}}^{\textrm{F}}.
\label{eq:misfit_correction_factor}
\end{equation}

The dispersive energy contribution, following the approach described in the first part of this work, was incorporated as an additional term to the segment-segment interaction energy. The dispersive interactions between segments \(i\) and \(j\) are calculated based on a new descriptor, which is either the atomic isotropic polarizability of the corresponding atom or the polarizability projection:
\begin{equation}
 E_{\textrm{vdW}} = a_{\textrm{eff}}m_{\textrm{vdW}} f(\alpha_i, \alpha_j). \label{eq:Evdw}
\end{equation}
For the polarizability combining function \(f(\alpha_i, \alpha_j)\), we evaluated several forms, including direct multiplication, the square root, an adjusted power of the root, and a logarithmic function defined as \(\ln{\left(1 + \frac{\alpha_i \alpha_j}{\textrm{a.u.}^6}\right)}\). In the logarithmic function, the constant "1" is added to \(\alpha_i \alpha_j\) to ensure that the logarithm yields a zero value when one of the polarizability projections is zero. Additionally, \(\alpha_i \alpha_j\) is divided by \(\textrm{a.u.}^6\) to maintain the dimensionless nature of the logarithmic argument, however, in the forthcoming discussion, we will omit it for simplicity.
In certain cases, the dispersive interactions were differentiated between F - segments and all other segments. For this purpose, two distinct scaling factors were applied: \(m_{\textrm{vdW}}^{\textrm{F}}\) for interactions involving an F-atom and \(m_{\textrm{vdW}}\) otherwise.

For the combinatorial contribution, we considered the Flory-Huggins (FH) term \citep{Flory1942ThemodynamicsSolutions}, following the assessment results of the first part of this study. Furthermore, it has been shown in \citet{KROOSHOF2024114146} that the FH term is the only physically consistent, purely combinatorial model among other commonly used combinatorial terms.

\subsection{Regression Procedure}
\label{sec:regression_methods}
To optimize the parameters of openCOSMO-RS, we considered various types of phase equilibrium data: infinite dilution activity coefficients (IDAC), mole fraction based partition coefficients of component \(i\) between water and an organic liquid (\(K_i^{\textrm{org/w}}\)), VLE, LLE, and solvation free energy (\(\Delta G_{\textrm{solv}}\)).

For VLE, LLE, IDAC and \(K_i^{\textrm{org/w}}\), we used the same objective function as in Part 1 of this work.

The solvation free energy, \(\Delta G_{\textrm{solv}}\), was calculated using:
\begin{equation}
\Delta G_{\textrm{solv}} = E_{\textrm{diel}} + RT\ln{\gamma_i^{\infty}} - \sum_{\alpha} \tau_{\alpha}A_{\alpha} - \omega_{\textrm{ring}}n_{\textrm{ring}} - RT\ln{\frac{\nu_{\textrm{IG}}}{\nu_{\textrm{liquid}}}} - \eta, \label{eq:Gsolv}
\end{equation}
where \(E_{\textrm{diel}}\) is the dielectric energy involved in transferring the solute from the gas phase to an ideal conductor. The second term corresponds to the liquid phase chemical potential of the solute at infinite dilution, with the ideal conductor as the reference state, as predicted by openCOSMO-RS. The third term accounts for the energy associated with cavity formation, incorporating a van der Waals-like contribution calculated as the product of each atom's surface area \(A_{\alpha}\) and an atomic-number-dependent factor \(\tau_{\alpha}\). To correct for ring-containing molecules, the fourth term introduces a contribution proportional to the number of rings \(n_{\textrm{ring}}\), scaled by a general parameter \(\omega_{\textrm{ring}}\). The fifth term adjusts for the change in reference states from mole fractions to molar concentrations, with the molar volume of the ideal gas - \(\nu_{\textrm{IG}}\) and of liquid phase - \(\nu_{\textrm{liquid}}\). Lastly, \(\eta\) is an adjustable parameter. The molar volumes were calculated using a predictive QSPR model developed by \citet{MULLER2025114250}. This model is based on molecular descriptors, such as \(\sigma-\)moments derived from \(\sigma-\)profiles, number of atoms and molecular surface area.

The models' performance was assessed using the Average Absolute Deviation (AAD), calculated as:
\begin{equation}
\mathrm{AAD} = \frac{1}{N_p} \sum_{s=1}^{N_s} \overline{\left|\ln(X_i^{\textrm{calc}}) - \ln(X_i^{\textrm{exp}})\right|}, \label{eq:AAD_LLE_avg}
\end{equation}
where \(X_i\) represents \(\ln{\gamma_i^{\infty}}, \ln{\gamma_i^{\textrm{VLE}}}, K_i^{\textrm{org/w}}, K_i^{\textrm{LLE}},\) or \(\Delta G_{\textrm{solv}}\) and \(N_p\) is the number of respective data points. The overline indicates an average for binary mixture properties when two measurements are available (e.g., \(\ln{\gamma_1^{\textrm{VLE}}},~ \ln{\gamma_2^{\textrm{VLE}}}\)). The total \(\textrm{AAD}_\textrm{TOT}\), was computed as the sum of absolute deviations across all data types, normalized by the total number of data points.

\section{Results and Discussion}

\begin{figure}
    \centering
    \includegraphics[width=1\linewidth]{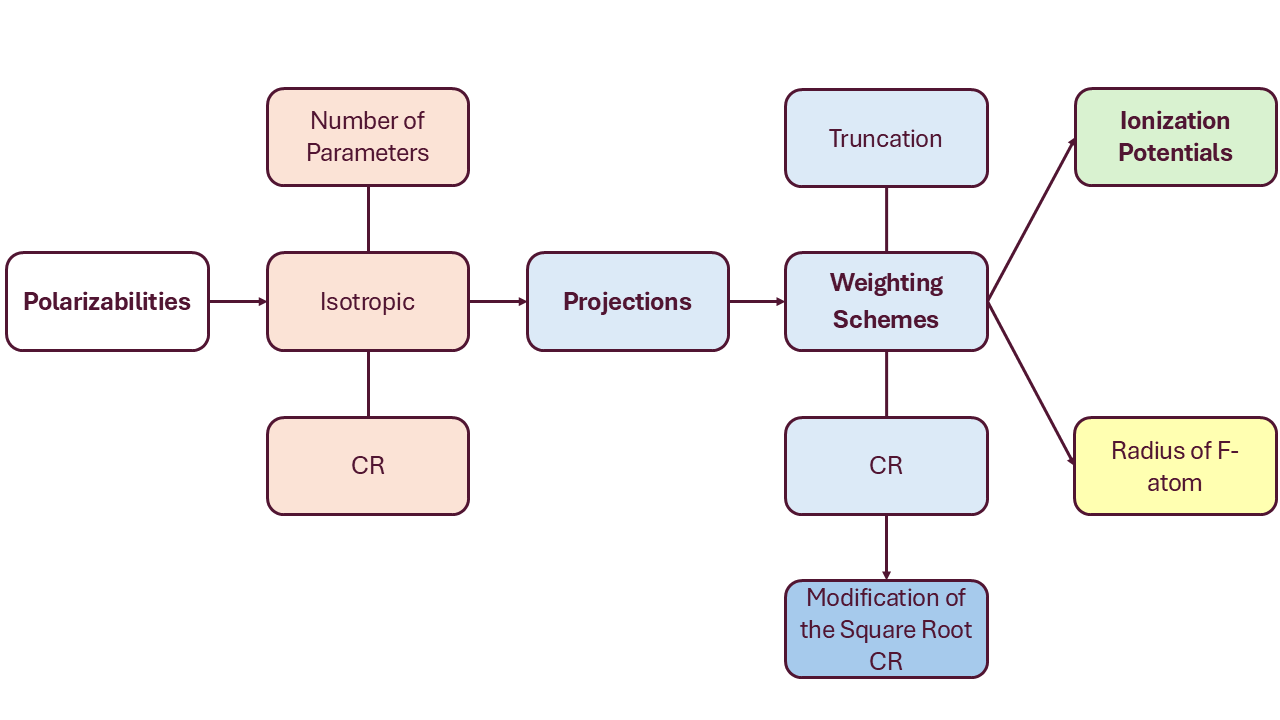}
    \caption{Schematic workflow for incorporating atomic polarizabilities into openCOSMO-RS using halocarbon data.}
    \label{fig:polarizability_workflow}
\end{figure}

In the following sections, we discuss the challenges associated with modeling intermolecular interactions in fluorocarbon mixtures and explore various approaches to address these issues. To guide the reader through the detailed development process, Figure~\ref{fig:polarizability_workflow} provides a schematic representation of the methods used to incorporate atomic polarizabilities into openCOSMO-RS applied to halocarbon data. 

Following this, the most promising approaches (highlighted in bold in Figure \ref{fig:polarizability_workflow}) were evaluated on the entire dataset, extending beyond halocarbons to ensure the general applicability of the model.

\label{sec:Results and discussion}

%%       \citet{<label>} ==> Jones et al. (2015)
%%       \citep{<label>} ==> (Jones et al., 2015)
\subsection{Halocarbons}
\label{sec:halocarbons}
\subsubsection{Intermolecular Interactions in Fluorocarbon Mixtures}
\label{sec:Intermolecular interactions in fluorocarbon mixtures}

The physical nature of intermolecular interactions between alkanes and fluoroalkanes is still not completely understood. Specifically, the thermodynamic behavior of these mixtures, such as their immiscibility, contradicts expectations based on their similar structure and polarity. Various explanations have been proposed \citep{Scott1958, ROWLINSON1982132, Morgado2017, Siebert1971, Song2003, MURRAY2021106382}, with a general consensus that the observed interactions are weaker than anticipated. For example, a molecular simulation study by  \citet{Song2003},  demonstrated that reducing intermolecular interactions involving fluorine by 25\% led to a significant improvement in the model's accuracy, indicating that these interactions are indeed weaker than expected. This finding aligns with an earlier review by \citet{Scott1958}, who attributed it to the highly repulsive nature of fluorine. Consequently, a standard square root combining rule (CR) alone is insufficient to adequately describe the intermolecular forces in fluorine-containing systems.

To enhance models such as COSMO-RS for fluorinated compounds, explicit consideration of dispersive interactions has been proposed, as these were not included in the original implementations. For instance, Paricaud et al. \citep{Mambo-Lomba2021PredictionsApproaches} improved the COSMO-SAC-dsp model by reducing the dispersion parameter for F-atom, which improved VLE modeling of refrigerant blends. In our prior work \citep{grigorash2024comprehensiveapproachincorporatingintermolecular}, we introduced atomic dispersion parameters, which substantially enhanced the accuracy of thermodynamic property predictions for fluorinated mixtures. A similar approach has been implemented in COSMOtherm, as noted in \citet{Klamt2007PredictionCOSMOtherm, SACHSENHAUSER201489}.

\begin{figure}
    \centering
    \includegraphics[width=0.45\linewidth]{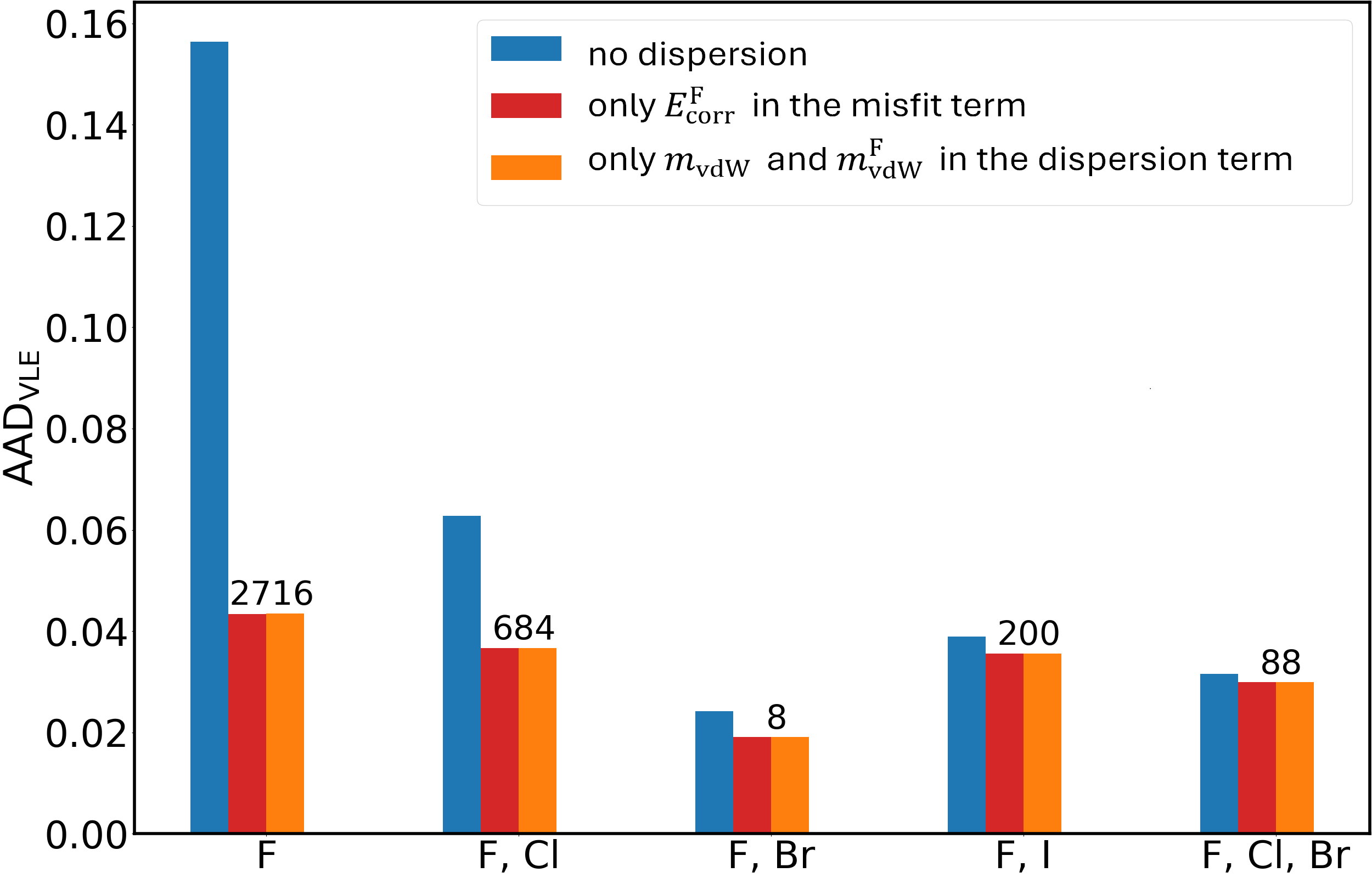}
    \includegraphics[width=0.45\linewidth]{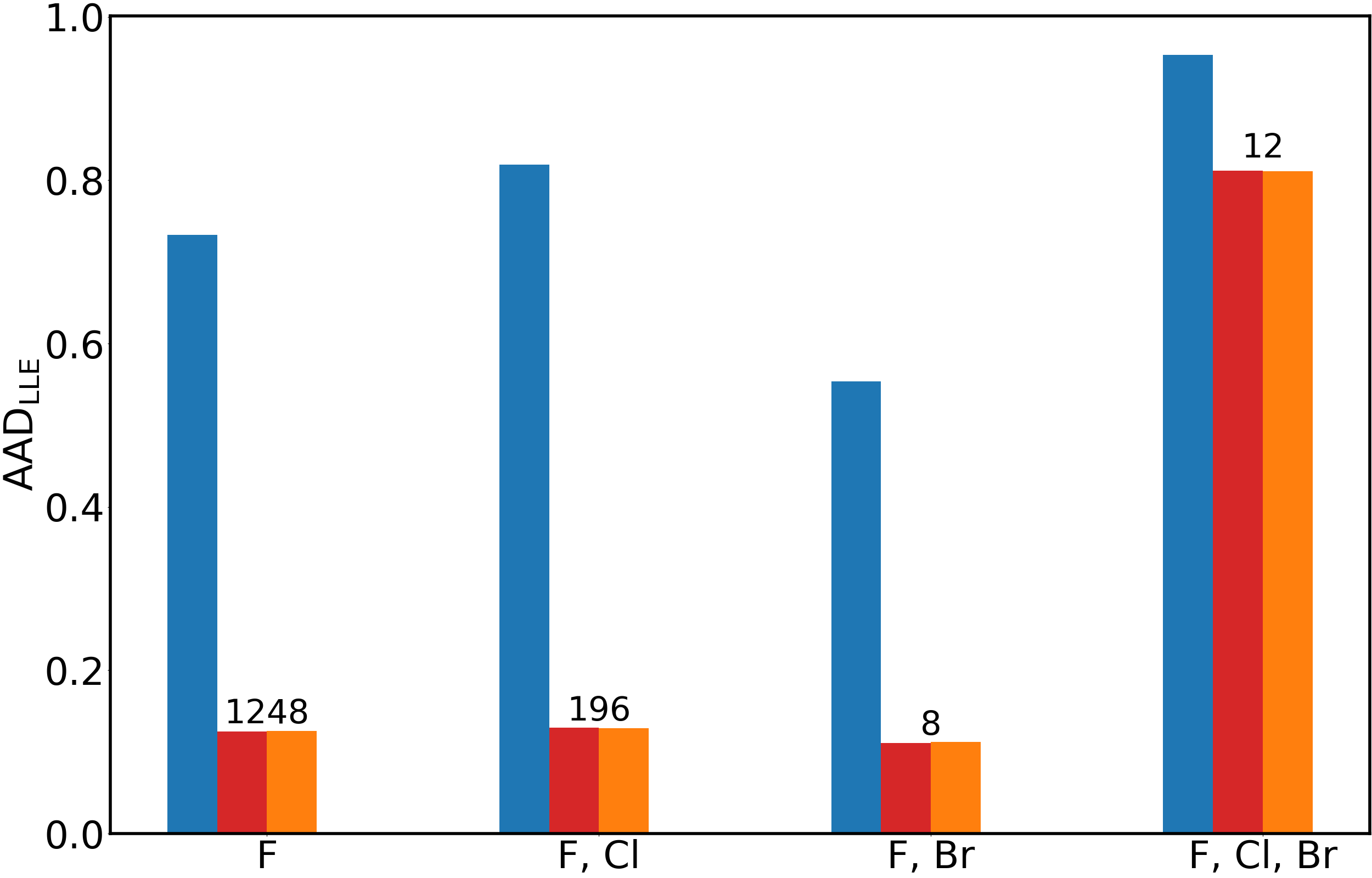}
    \includegraphics[width=0.45\linewidth]{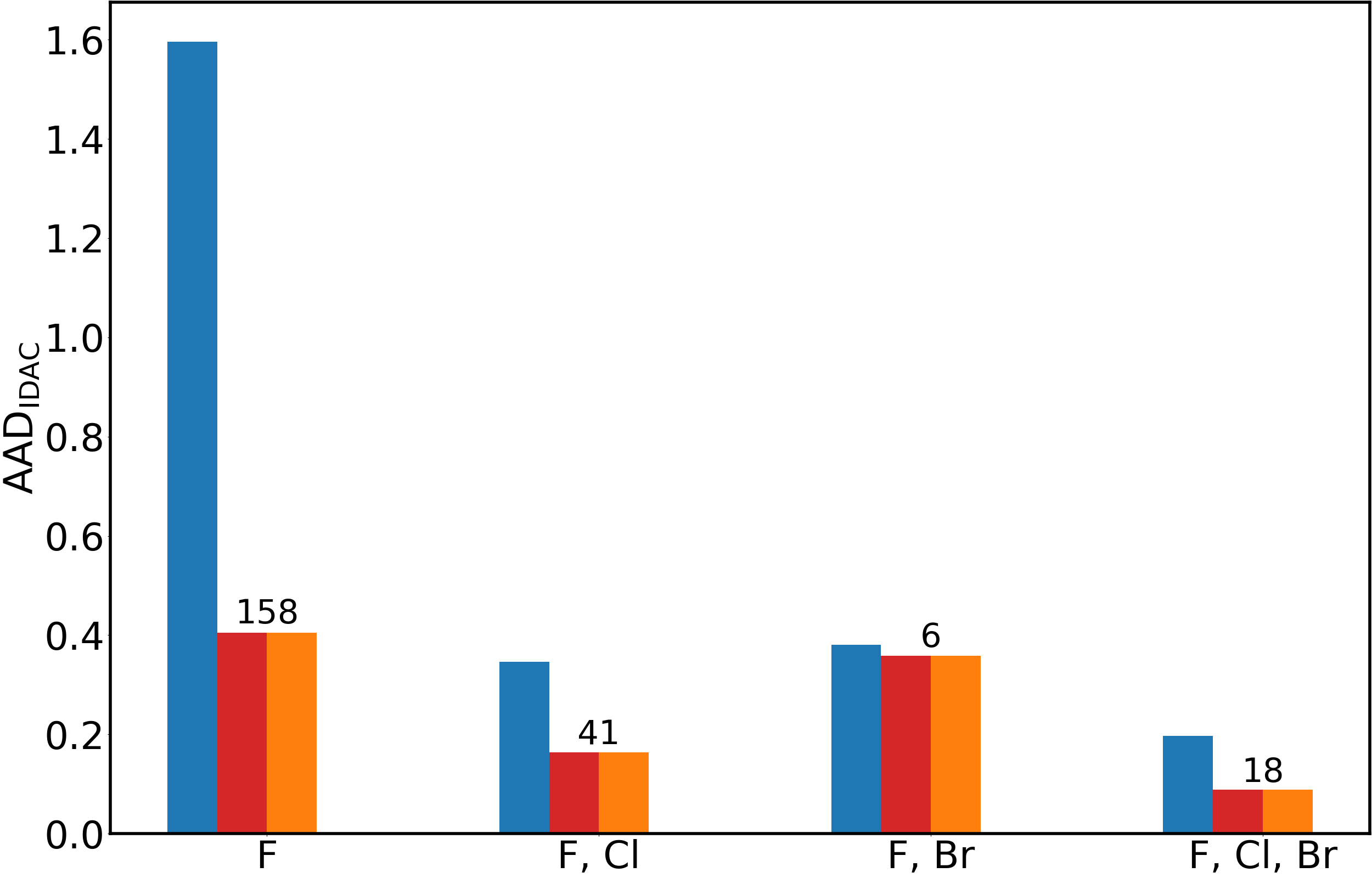}
    \caption{Bar plots comparing the AAD for VLE (left), IDAC (middle), and LLE (right) data across three models: the original model openCOSMO-RS 24a, and models incorporating corrections for F-atom interactions either through the misfit term or the dispersion term. The evaluated data is categorized based on the types of halogen atoms present in the molecules. Numbers of data points are indicated above columns for each category.}
    \label{fig:AAD_bar_f_corr_or_two_param_disp}
\end{figure}

To further examine interactions involving fluorinated species, we expanded our existing halocarbon dataset by incorporating 1861 additional IDAC data points from \citet{Gmehling2008ActivityCoefficients}. To account for the weaker dispersion interactions associated with fluorinated species within openCOSMO-RS, we implemented two modifications detailed in Section \ref{sec:COSMORS}. The first modification introduces repulsive energy (\(E_{\textrm{corr}}^{\textrm{F}}, [\textrm{J/mol}]\)) for segment-segment interactions involving F-atom in the misfit term, as shown in Eq. \ref{eq:misfit_correction_factor}. This parameter was optimized using experimental data for fluorinated compounds. Alternatively, the second modification adjusts the dispersion term (Eq. \ref{eq:Evdw}) by selectively reducing the strength of dispersive interactions involving F-atom relative to those with other atoms. This approach required regressing two dispersion parameters: \(m_{\textrm{vdW}}^{\textrm{ F}}\) for interactions involving F-segments and \(m_{\textrm{vdW}}\) for all other interactions not accounting for \(f(\alpha_i, \alpha_j)\) at this stage. As shown in Figure \ref{fig:AAD_bar_f_corr_or_two_param_disp} and detailed in the first section of Table \ref{tab:isotropic_polarizabilities}, both modifications led to similar improvements for data groups containing fluorinated compounds. However, as may be expected, for non-fluorinated groups, there was no noticeable enhancement compared to the original model. Since the dispersion parameters for the respective atoms remained unchanged, any residual contributions to the activity coefficient calculations were effectively canceled out.

\begin{table}[htbp]
    \centering
    \caption{Overview of parametrization approaches, showing AADs and parameter values. Atomic isotropic polarizabilities are integrated into the dispersion term.}
    \begin{adjustbox}{width=\textwidth} % Adjust table width to text width
    \begin{threeparttable}
    \begin{tabular}{lcccccccccc}
        \toprule
        \multicolumn{3}{c}{Parameterizations\tnote{1}} & \multicolumn{3}{c}{Parameters\tnote{2}} & \multicolumn{4}{c}{AAD} \\
        \cmidrule(lr){1-3}
        \cmidrule(lr){4-6}
        \cmidrule(lr){7-10}
        the misfit correction & the dispersion term & CR & \(E_{\textrm{corr}}^{\textrm{F}}\) &  \(m_{\textrm{vdW}}\) &  \(m_{\textrm{vdW}}^{\textrm{F}}\) & TOT & IDAC & LLE & VLE\\
        \midrule
        \multicolumn{3}{c}{no dispersion/the original model\tnote{3}} &  &  &  & 0.194 & 0.415 & 0.727 & 0.068 \\
        \ding{51}&  &
        & 327.76 &  &  & 0.091 & 0.320 & 0.142 & 0.035 \\
       & \ding{51} &
        &  & 388.494 & 319.457 & 0.091 & 0.320 & 0.142 & 0.035 \\ 
        \midrule
        \multirow{5}{*}{}
        & \ding{51} & \multirow{5}{*}{\(\sqrt{\alpha_{i}\alpha_{j}}\)} &  & 3.271 &  & 0.179 & 0.361 & 0.690 & 0.063 \\
        \ding{51}& \ding{51} &  & 327.76 & 2.729 &  & 0.079 & 0.266 & 0.133 & 0.032 \\
        \ding{51}& \ding{51} &  & 321.24 & 2.738  &  & 0.078 & 0.266 & 0.125 & 0.032 \\
        \ding{51}& \ding{51} &  & 291.92 & 2.709  & 1.813 & 0.078 & 0.265 & 0.126 & 0.032 \\
        & \ding{51} &  &   & 5.026 & 2.640E-06 & 0.132 & 0.348 & 0.389 & 0.048 \\        
        \midrule
        \multirow{5}{*}{}
        & \ding{51} & \multirow{5}{*}{\(\alpha_{i}\times\alpha_{j}\)} &  & 0.042 &  & 0.179 & 0.350 & 0.704 & 0.064 \\
        \ding{51}& \ding{51} &  & 327.76 & 0.037 &  & 0.078 & 0.259 & 0.132 & 0.032 \\
        \ding{51}& \ding{51} &  & 323.50 & 0.037  &  & 0.077 & 0.260 & 0.127 & 0.032 \\
        \ding{51}& \ding{51} &  & 313.78 & 0.037  & 2.0E-04 & 0.077 & 0.259 & 0.128 & 0.032 \\
        & \ding{51} &  &   & 0.050  & 1.066E-07 & 0.174 & 0.350  & 0.668 & 0.063 \\           
        \midrule
        \multirow{5}{*}{}
        & \ding{51} & \multirow{5}{*}{\(\ln{(1 + \alpha_i \alpha_j)}\)} &  & 779.954  &  & 0.179 & 0.388 & 0.692 & 0.059 \\
        \ding{51}& \ding{51} &  & 327.76  & 444.426  &  & 0.080 & 0.270  & 0.148 & 0.031 \\  
        \ding{51}& \ding{51} &  & 316.85 & 458.098  &  & 0.079 & 0.271 & 0.138 & 0.030 \\
        \ding{51}& \ding{51} &  & 181.75 & 459.129 & 451.599 & 0.077 & 0.269 & 0.129 & 0.030 \\
        & \ding{51} &  &   & 472.255  & 454.964 & 0.078 & 0.273  & 0.138 & 0.029 \\     
        \bottomrule
    \end{tabular}
    \begin{tablenotes}
        \item[1] Parametrization characteristics involve: 1) the misfit correction for F-segments using \(E_{\textrm{corr}}^{\textrm{F}}\) parameter (Eq. \ref{eq:misfit_correction_factor}); 2) the dispersion term (Eq. \ref{eq:Evdw}), distinguishing F interactions is indicated by non-zero value of \(m_{\textrm{vdW}}^{\textrm{F}}\); 3) the combining rule for polarizabilities used in a dispersion term.
         \item[2] The general parameters for openCOSMO-RS 24a with FH combinatorial term (\(a_{\textrm{eff}}=4.754~[\textrm{Å}^{2}] ,\:\alpha_{\textrm{mf}}=7287~[\textrm{kJ}/(\textrm{mol}\cdot\textrm{Å}^{2})/e^{2}],\:c_{\textrm{hb}}=43345~[\textrm{kJ}/(\textrm{mol}\cdot\textrm{Å}^{2})/e^{2}],\:\sigma_{\textrm{hb}}=0.008918~[e/\textrm{Å}^{2}],\:r_{\textrm{av}}=0.5~[\textrm{Å}],\:f_{\textrm{corr}}=2.4,\:c_{\textrm{hb}}^{T}=1.5\)) were regressed using the IDAC, partitioning coefficients and solvation free energies from \citet{MULLER2025114250}. These general parameters are used for all evaluations of halocarbons.
        \item[3] The original openCOSMO-RS 24a model.
    \end{tablenotes}
    \end{threeparttable}
    \end{adjustbox}
    \label{tab:isotropic_polarizabilities}
\end{table}

\subsubsection{Atomic Polarizabilities}
\label{sec:Isotropic polarizabilities}

\begin{figure}
    \centering
    \includegraphics[width=0.45\linewidth]{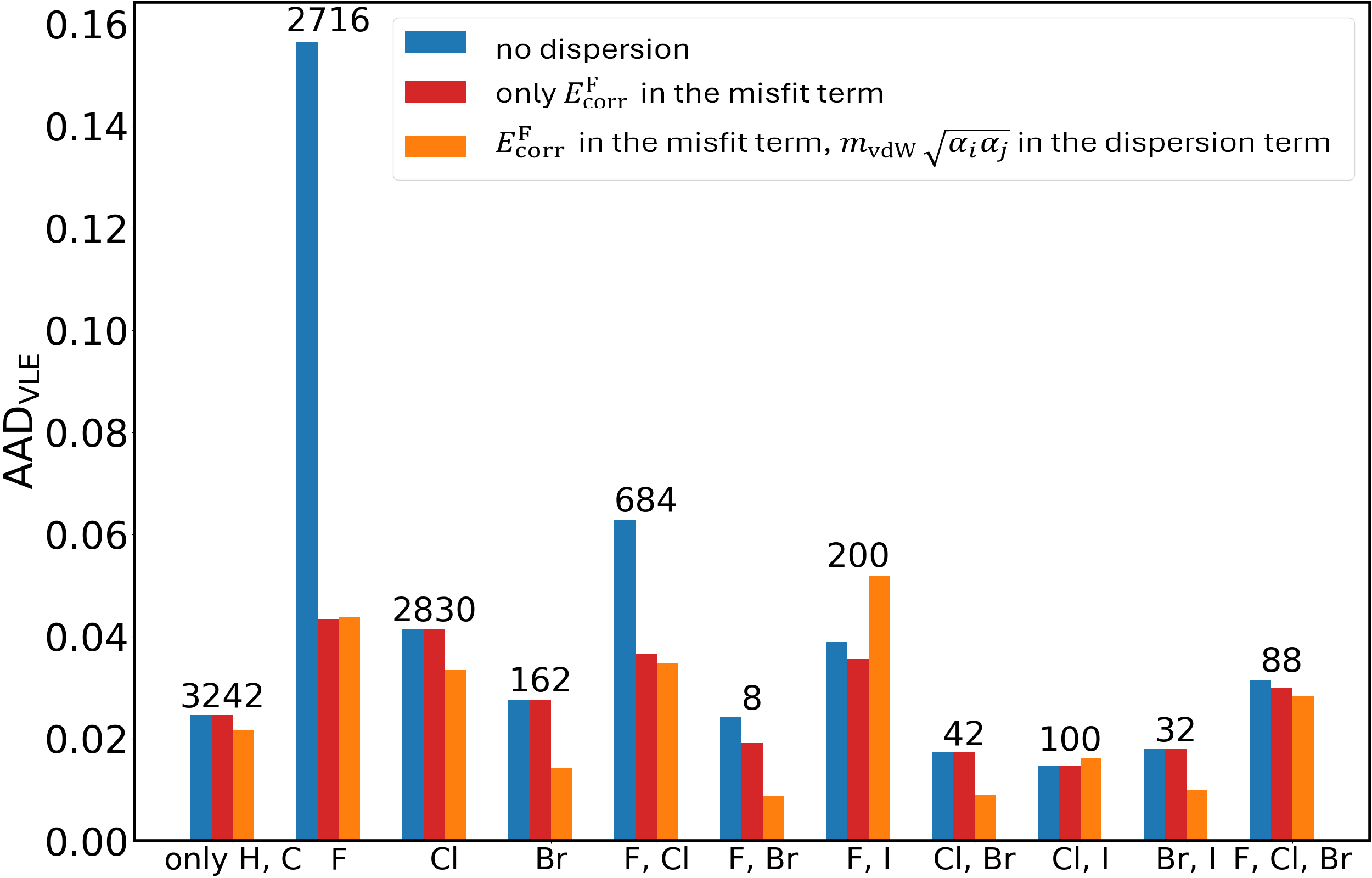}
    \includegraphics[width=0.45\linewidth]{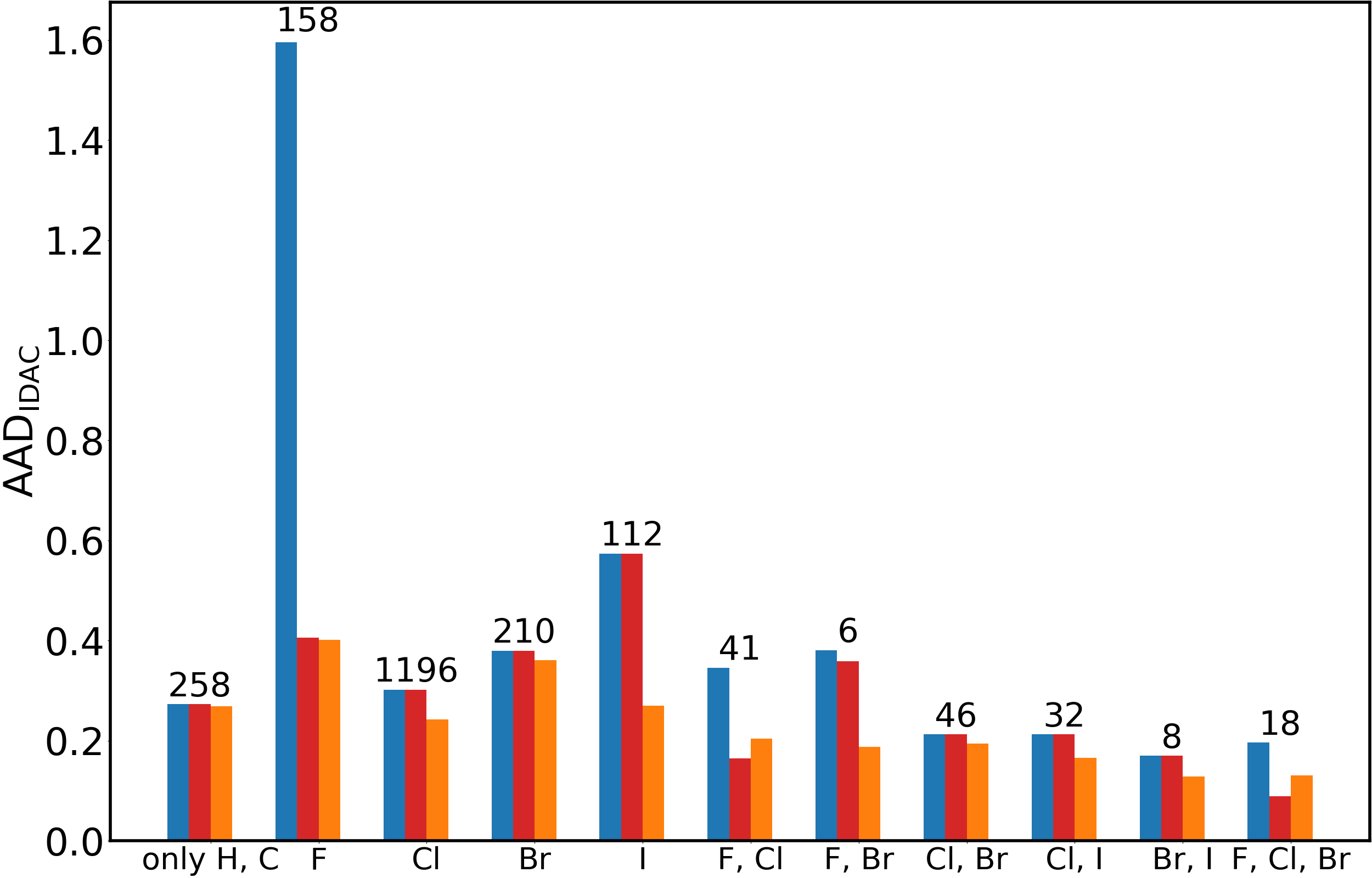}
    \includegraphics[width=0.45\linewidth]{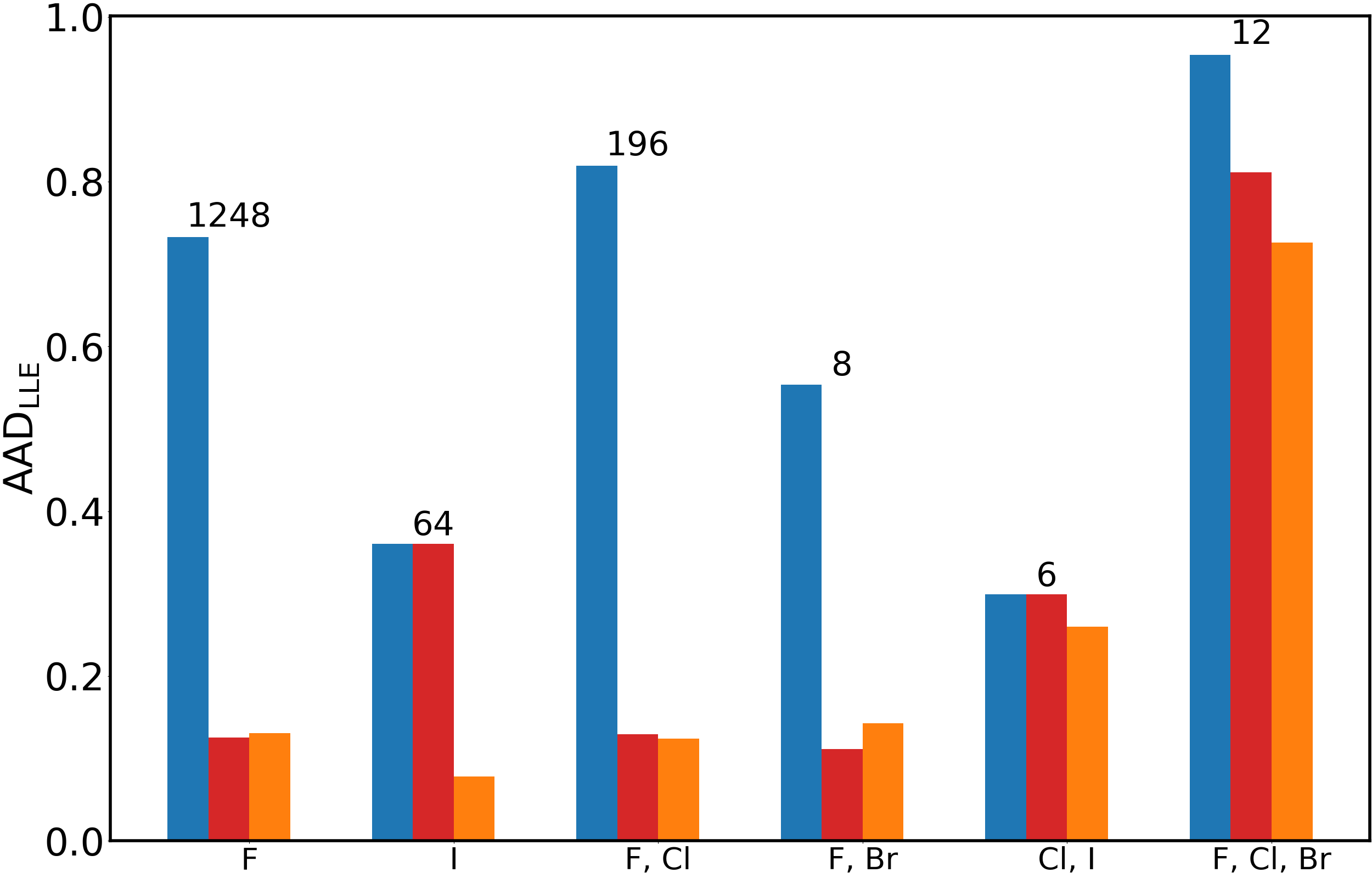}
    \caption{Bar plots comparing the AAD for VLE (left), IDAC (middle), and LLE (right) data across three models: the original model openCOSMO-RS 24a, the model incorporating corrections for F-atom interactions through the misfit term alone, and the model including both the misfit correction and a dispersion term with isotropic atomic polarizabilities. The evaluated data is categorized based on the types of halogen atoms present in the molecules. Numbers of data points are indicated above columns for each category.}
    \label{fig:AAD_bar_isotropic_sqrt}
\end{figure}

To improve the modeling of all halocarbons, including both fluorinated and non-fluorinated species, we incorporated atomic isotropic polarizabilities into the dispersion term. We then applied the three previously mentioned combining rules, each with a single adjustable scaling factor, \(m_{\textrm{vdW}}\), for all types of segment-segment interactions, irrespective of the atom to which they belong. It should be noted that the units of \(m_{\textrm{vdW}}\) depend on the chosen CR, as isotropic polarizabilities are expressed in atomic units cubed (\(\textrm{a.u.}^3\)). Hereafter, the units of \(m_{\textrm{vdW}}\) are as follows: for the multiplication CR, \([\textrm{J}/\textrm{Å}^2/\textrm{a.u.}^6]\); for the square-root CR, \([\textrm{J}/\textrm{Å}^2/\textrm{a.u.}^3]\); and for the logarithmic CR, \([\textrm{J}/\textrm{Å}^2]\). The results are summarized in Table \ref{tab:isotropic_polarizabilities}. It can be seen that without any additional correction for F-atom interactions, the overall improvement across all data categories remains modest compared to models incorporating the \(E_{\textrm{corr}}^{\textrm{F}}\) parameter, regardless of the combining rule applied. However, models that integrate both the \(E_{\textrm{corr}}^{\textrm{F}}\) correction in the misfit term and isotropic polarizabilities in the dispersion term show consistent improvements across all combining rules, with an example for the square root CR depicted in Figure \ref{fig:AAD_bar_isotropic_sqrt}. There is a consistent improvement of AADs in groups containing Cl-, Br-, or I- atoms. For data groups involving F-atoms, however, the results are more controversial, which may indicate a deviation from the square root CR. For instance, the AAD for the VLE data involving both F- and I-atoms increases with the inclusion of polarizabilities, while it decreases for the F- and Br-atom group across most data types. Other groups, such as those with F and Cl, Cl and I, or F, Cl, and Br, show improvement in two out of three data types but worsen in the third. Nevertheless, the addition of the dispersion term with isotropic polarizabilities improves the overall AADs, as well as those for specific data types.

Furthermore, in this approach, both \(E_{\textrm{corr}}^{\textrm{F}}\) and \(m_{\textrm{vdW}}\) can be regressed simultaneously, which leads to a slight improvement in the overall results, as shown in Table \ref{tab:isotropic_polarizabilities}. However, this also increases the regression time. Therefore, in the subsequent model evaluations, we fix the \(E_{\textrm{corr}}^{\textrm{F}}\) parameter, and for the best-performing models, all other parameters will be adjusted simultaneously later. Additionally, segment-segment interactions involving F-atoms can be distinguished in the dispersion term by using a distinct scaling factor \(m_{\textrm{vdW}}^{\textrm{F}}\). Subsequently, all three parameters—\(E_{\textrm{corr}}^{\textrm{F}}\), \(m_{\textrm{vdW}}\), and \(m_{\textrm{vdW}}^{\textrm{F}}\)—were adjusted simultaneously. Although this approach demonstrates superior AADs, the improvement over adjusting only two parameters is minimal. Therefore, the third parameter will not be used in further evaluations

\begin{figure}
    \centering
    \includegraphics[width=0.3\linewidth]{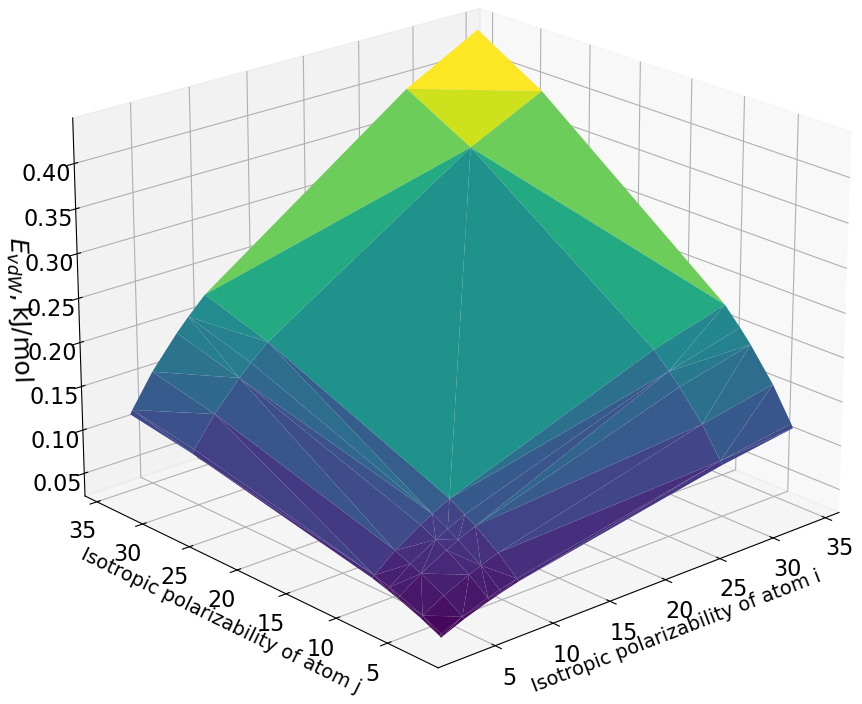}
    \includegraphics[width=0.3\linewidth]{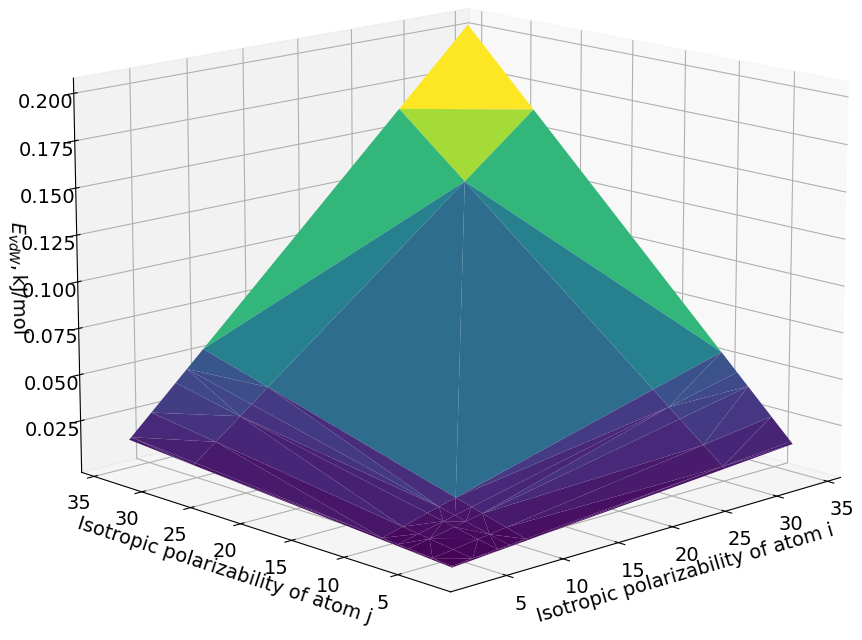}
    \includegraphics[width=0.3\linewidth]{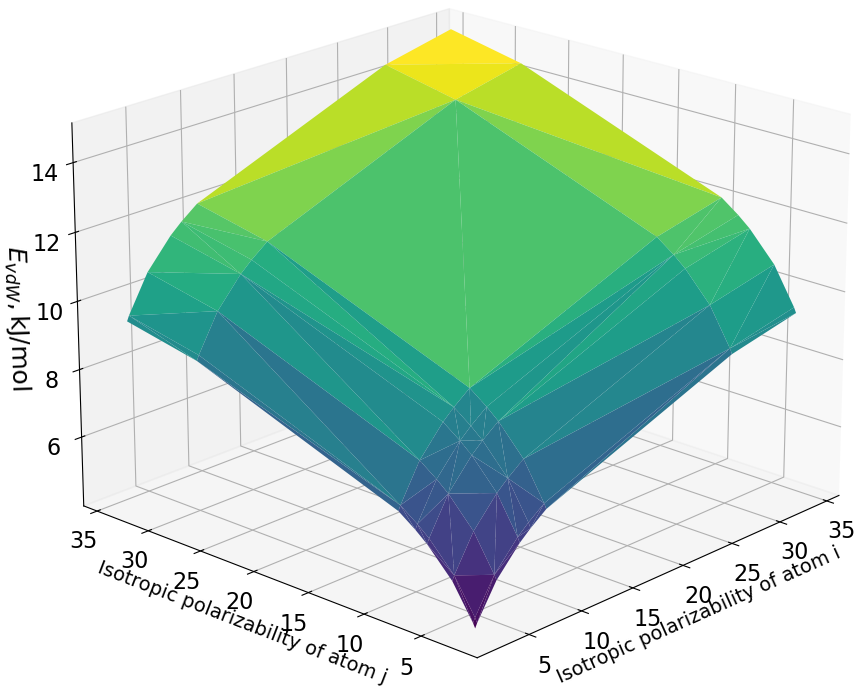}
    \caption{The 3D dispersion energy (\( E_{\textrm{vdW}} \)) surfaces as functions of atomic isotropic polarizabilities for the C\(_2\)HBrClF\(_3\) molecule, calculated using the square root (left), multiplication (middle), and logarithmic (right) combining rules.}
    \label{fig:Evdw_surfaces_diff_comb_rules}
\end{figure}

As we showed earlier, the additional repulsive repulsive energy in the misfit term can be replaced with two adjustable parameters in the dispersion term, leading to equivalent results. However, this is not feasible using either the square root or multiplication combining rules. The scaling factor for interactions involving the F-atom tends to zero, resulting in worse total AADs than those obtained with the misfit term correction. On the other hand, this approach can be applied with the logarithmic combining rule, which smoothens the difference between \( \alpha_i \) and \( \alpha_j \) to the greatest extent among the considered combining rules. To understand the effect of the CR for the polarizabilities on the dispersion energy contribution to the interaction matrix, we evaluated the 3D dispersion energy surfaces for the investigated molecules. For instance, Figure \ref{fig:Evdw_surfaces_diff_comb_rules} illustrates such a surface for the C\(_2\)HBrClF\(_3\) molecule, which includes nearly all halogens and thus covers a broad range of atomic polarizabilities. The surfaces were calculated for all possible pairs of \( \alpha_i \) and \( \alpha_j \), using the scaling factors \( m_{\textrm{vdW}} \), regressed with a fixed \( E_{\textrm{corr}}^{\textrm{F}} \) from Table \ref{tab:isotropic_polarizabilities}. Noticeably, the energy range across these surfaces varies substantially, with the multiplication and square root CR resulting in lower energy values compared to the logarithmic combining rule. Nevertheless, this variation is not critical, as the activity coefficient represents a residual property. Therefore, the difference in the absolute values of \( E_{\textrm{vdW}} \) will be negated when subtracting the reference state activity coefficient.
 
\subsubsection{Local Polarizability Projections}
\label{sec:Polarizability projections}

Another approach, described in Section \ref{sec:Polarizability_projections methods}, involves projecting the atomic polarizability tensors onto the molecular surface. This method assigns a local polarizability to each segment, as opposed to using a single isotropic atomic polarizability for individual atoms. In addition, we assessed different scaling strategies for the local polarizabilities of the segments that may account for the deviations from the chosen combining rule.

\begin{figure}
    \centering
    \includegraphics[width=0.45\linewidth]{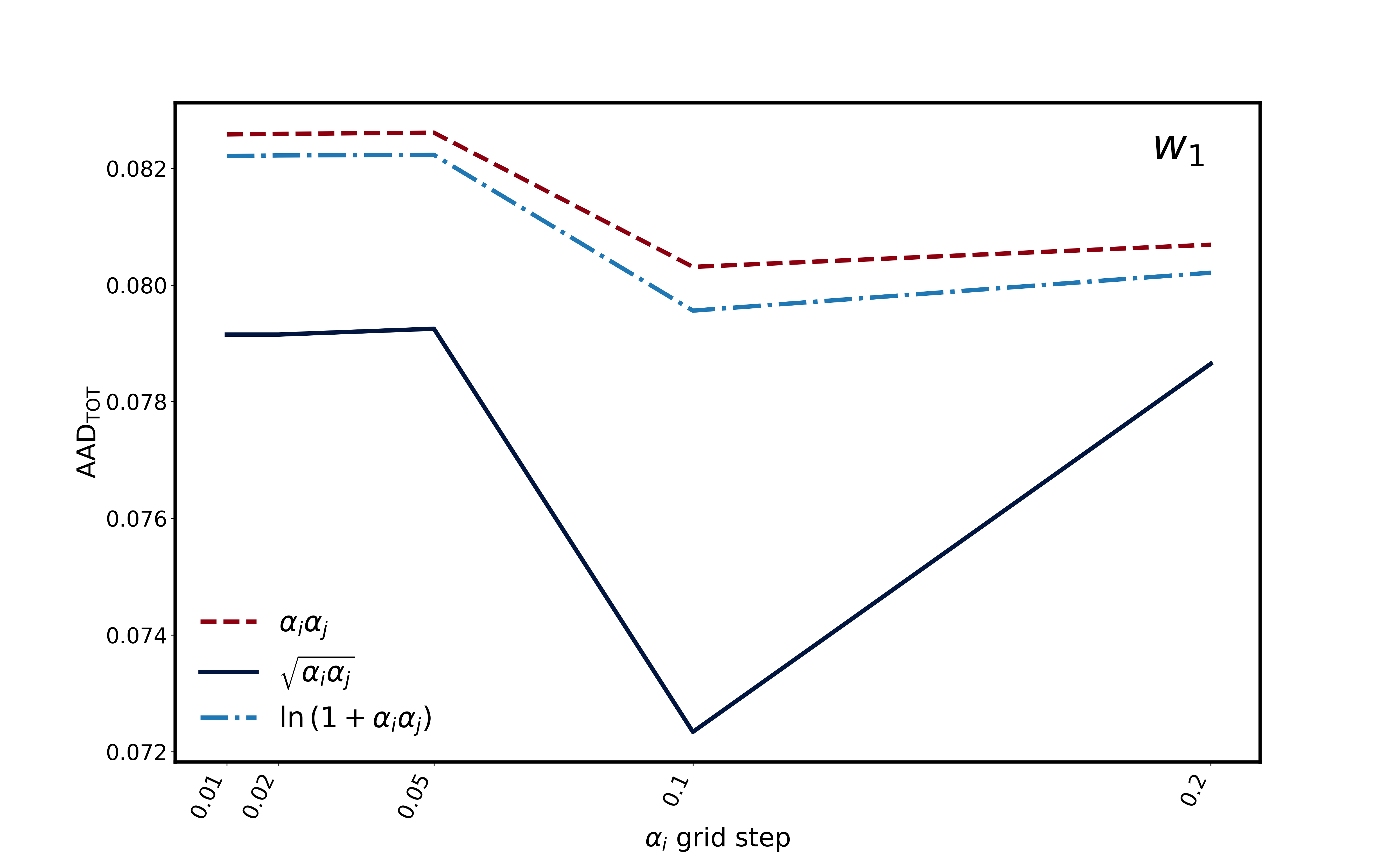}
    \includegraphics[width=0.45\linewidth]{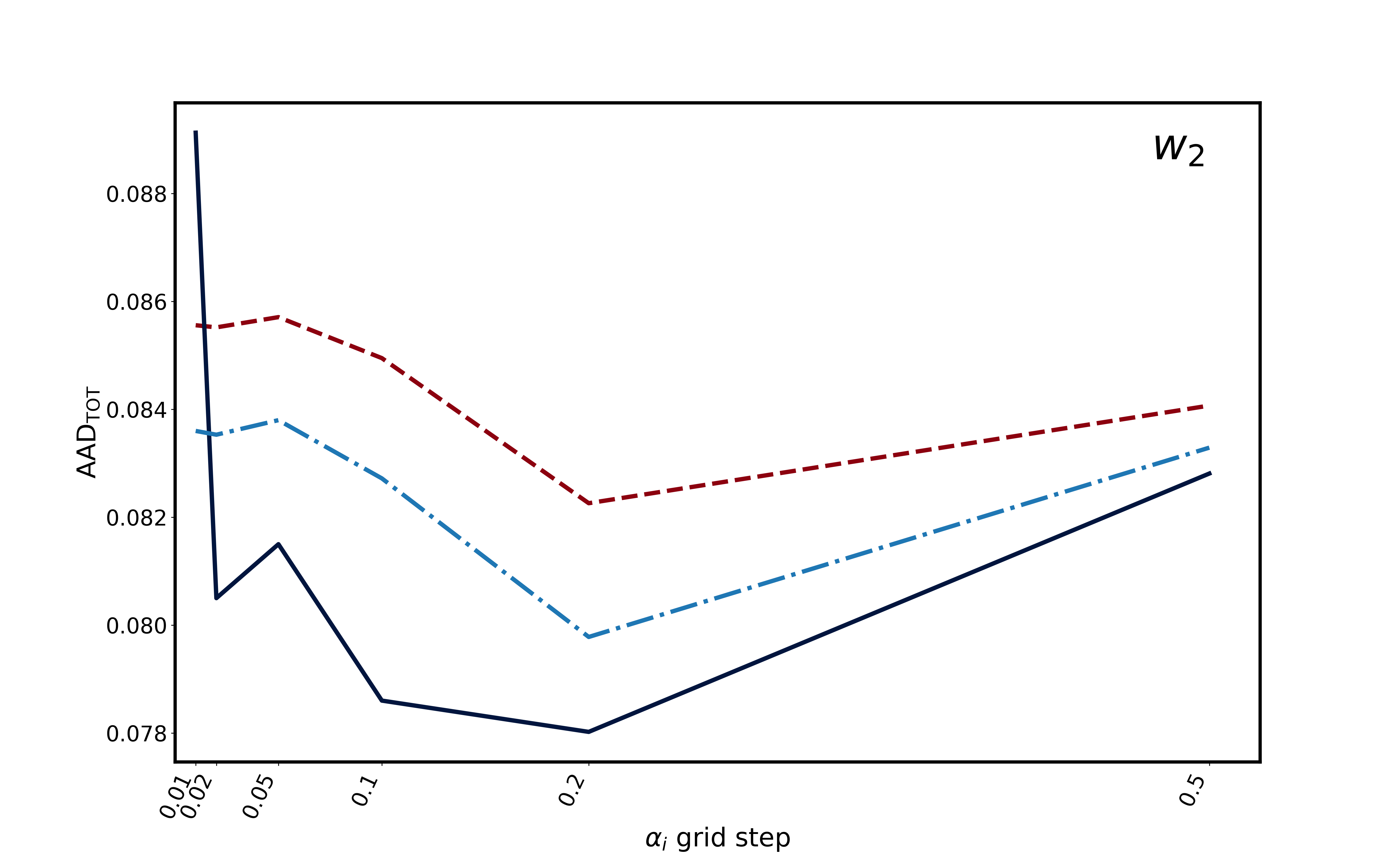}
    \includegraphics[width=0.45\linewidth]{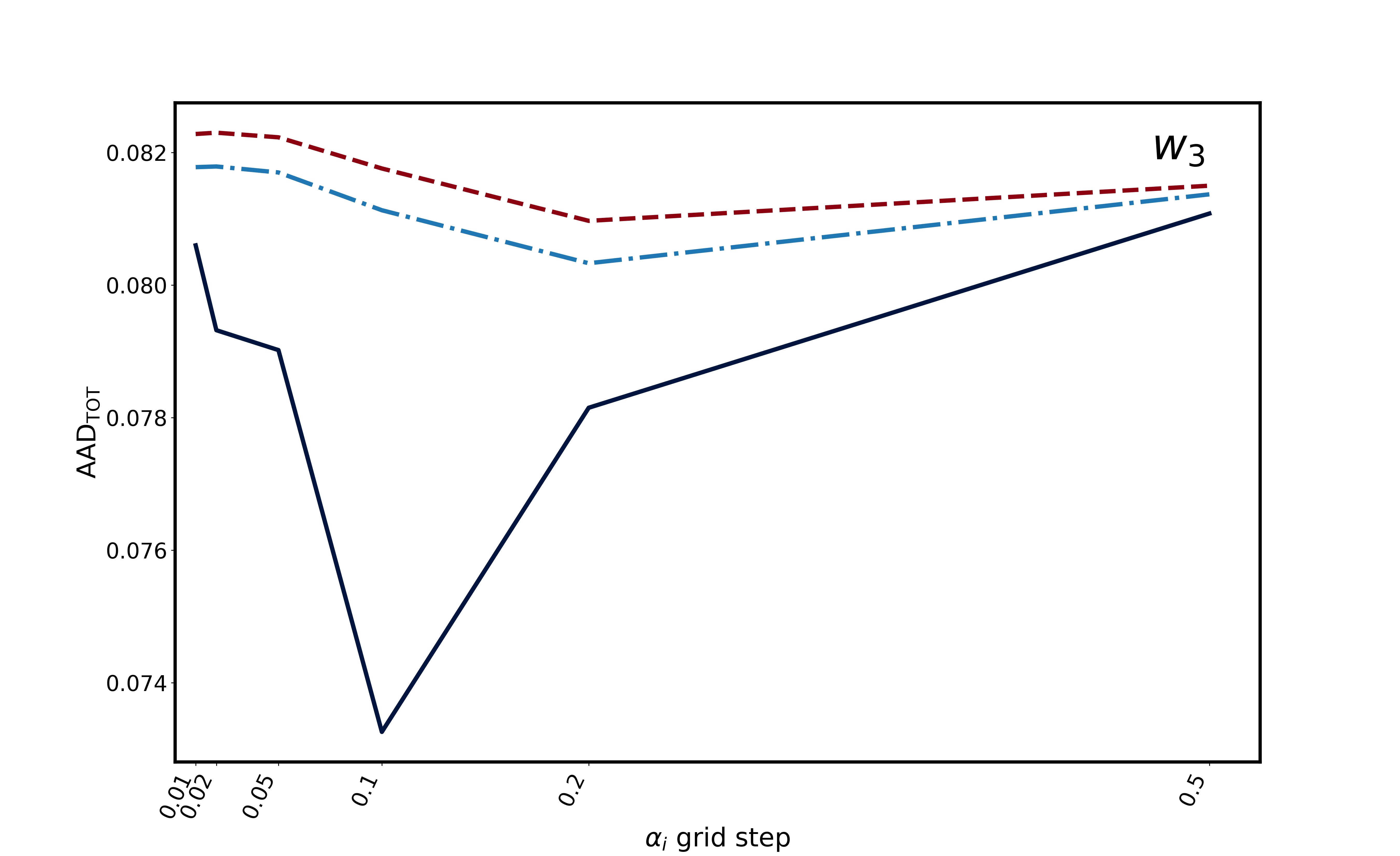}
    \includegraphics[width=0.45\linewidth]{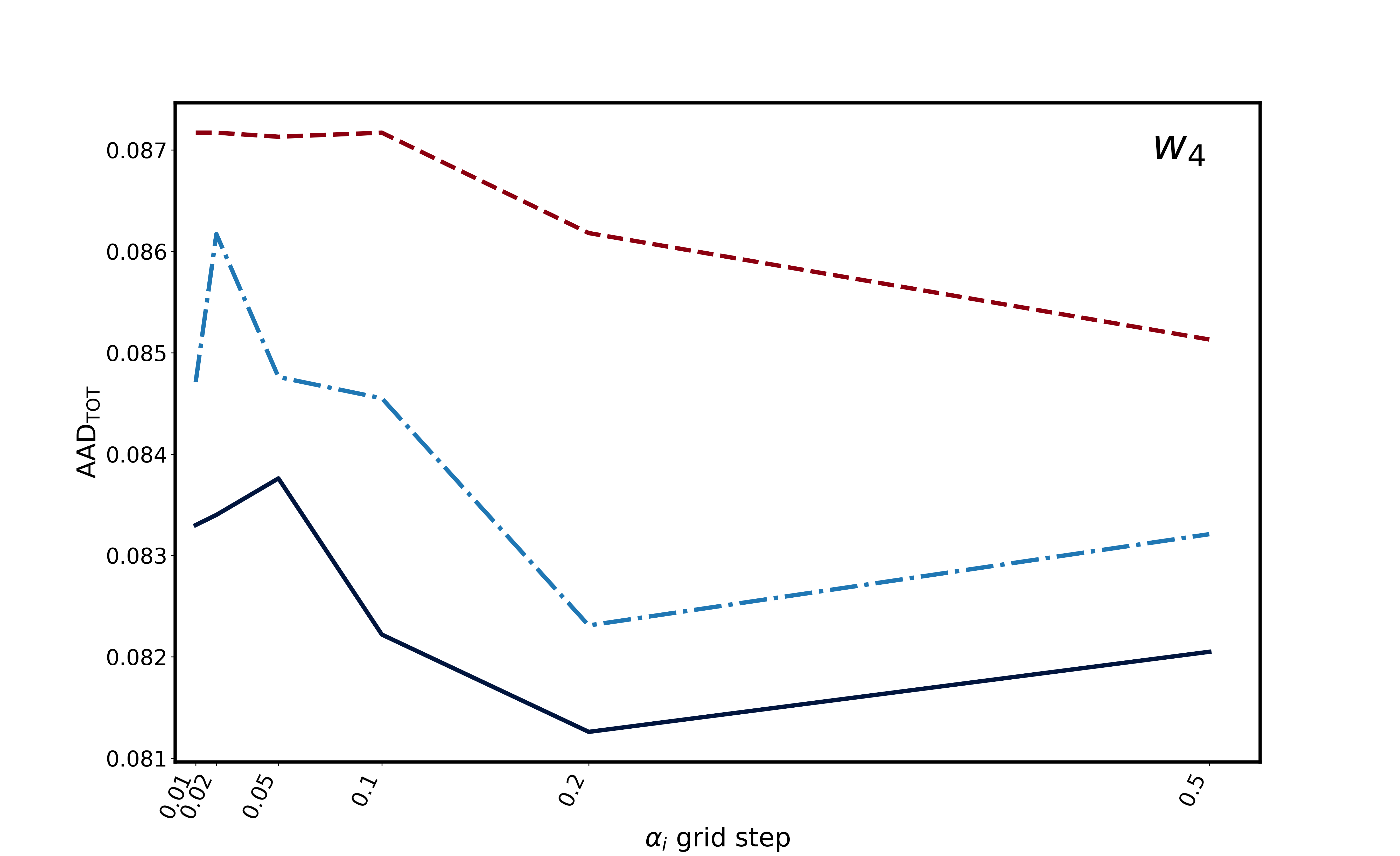}
    \includegraphics[width=0.45\linewidth]{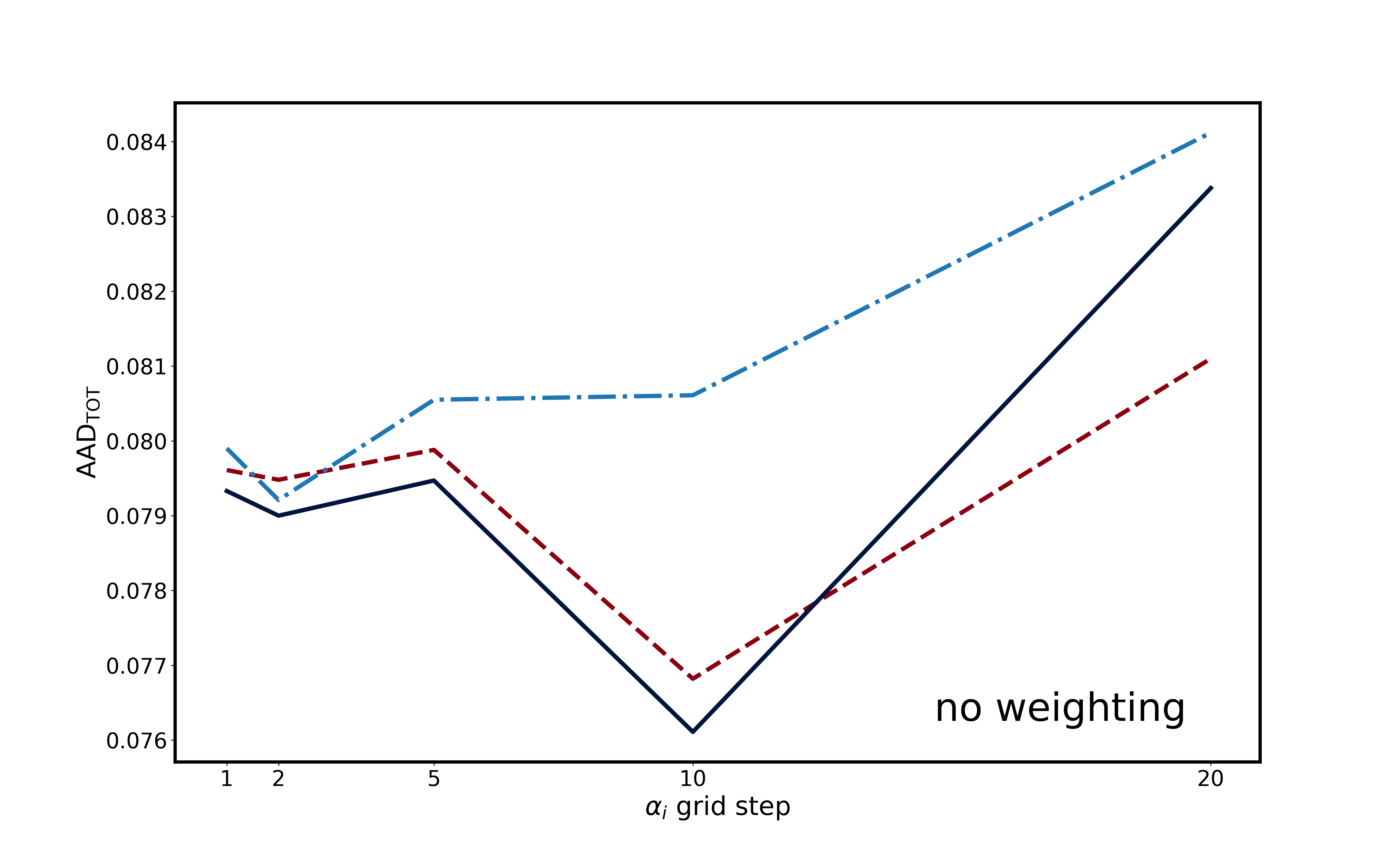}
    \includegraphics[width=0.45\linewidth]{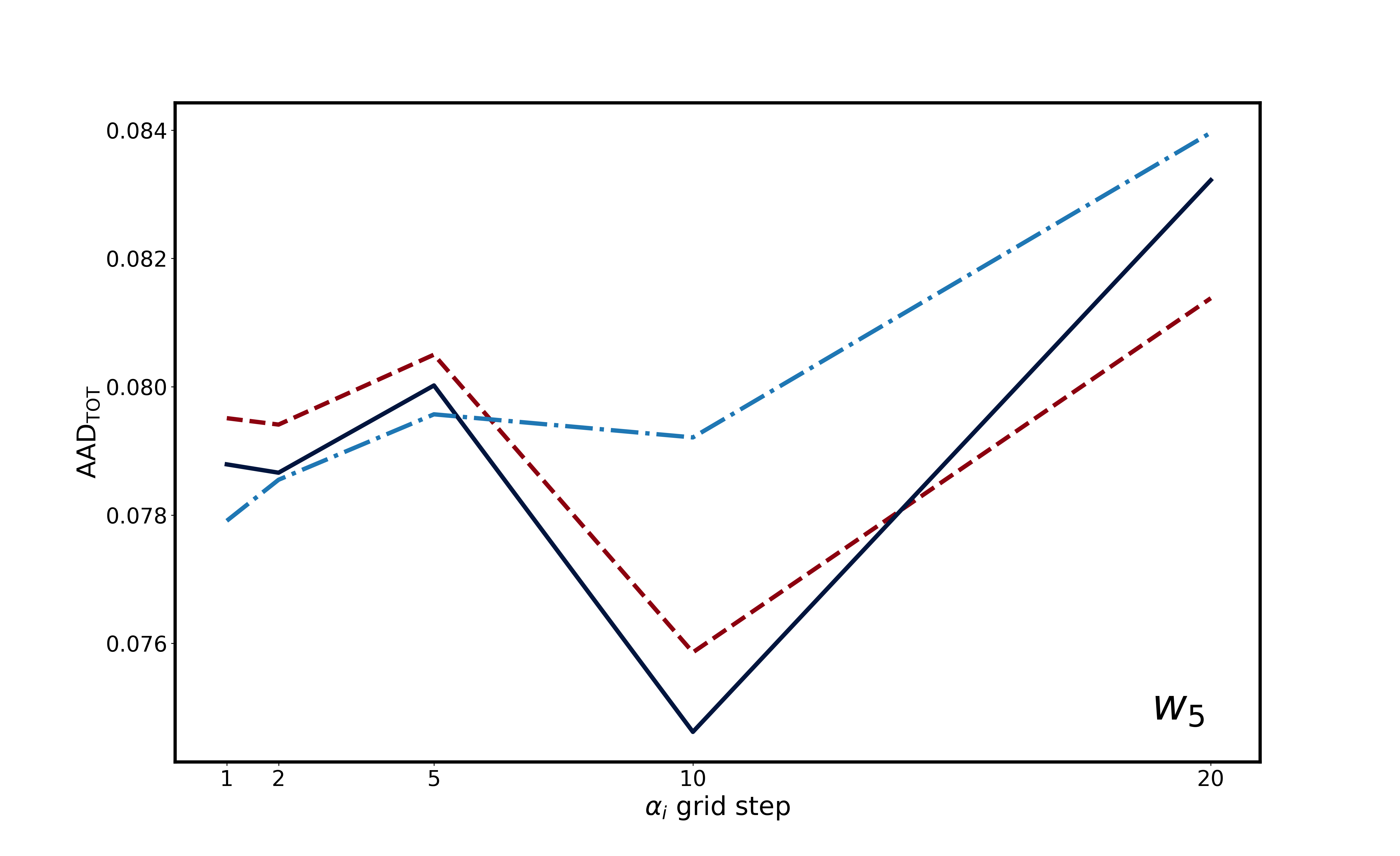}
    \caption{The total AADs calculated for models, involving different scaling factors (\(w_{1}-w_{5}\)) and CRs as a function of \(\alpha_{i}\) grid step sizes.}
    \label{fig:truncation_AAD}
\end{figure}

\begin{figure}
    \centering
    \includegraphics[width=0.45\linewidth]{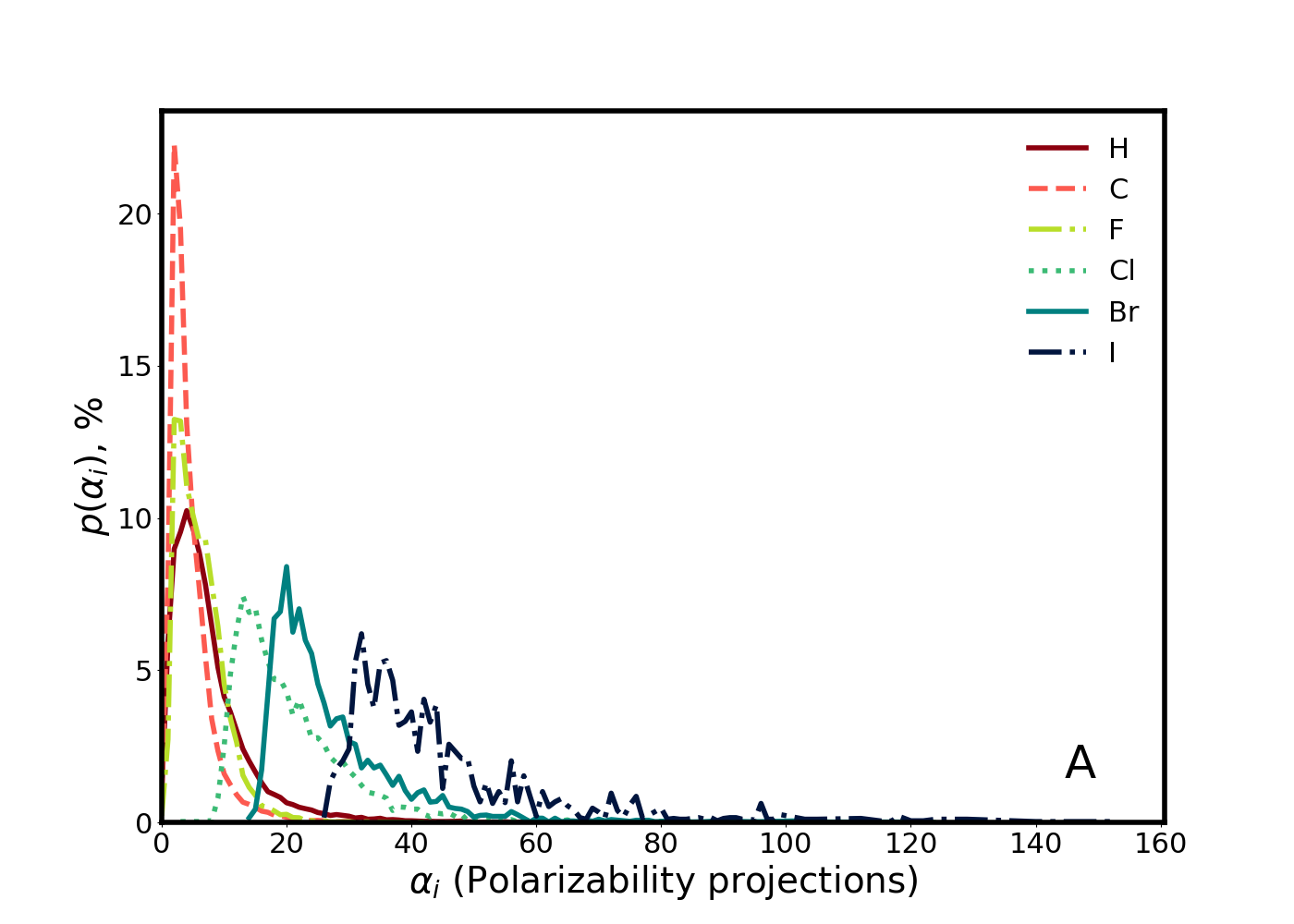}
    \includegraphics[width=0.45\linewidth]{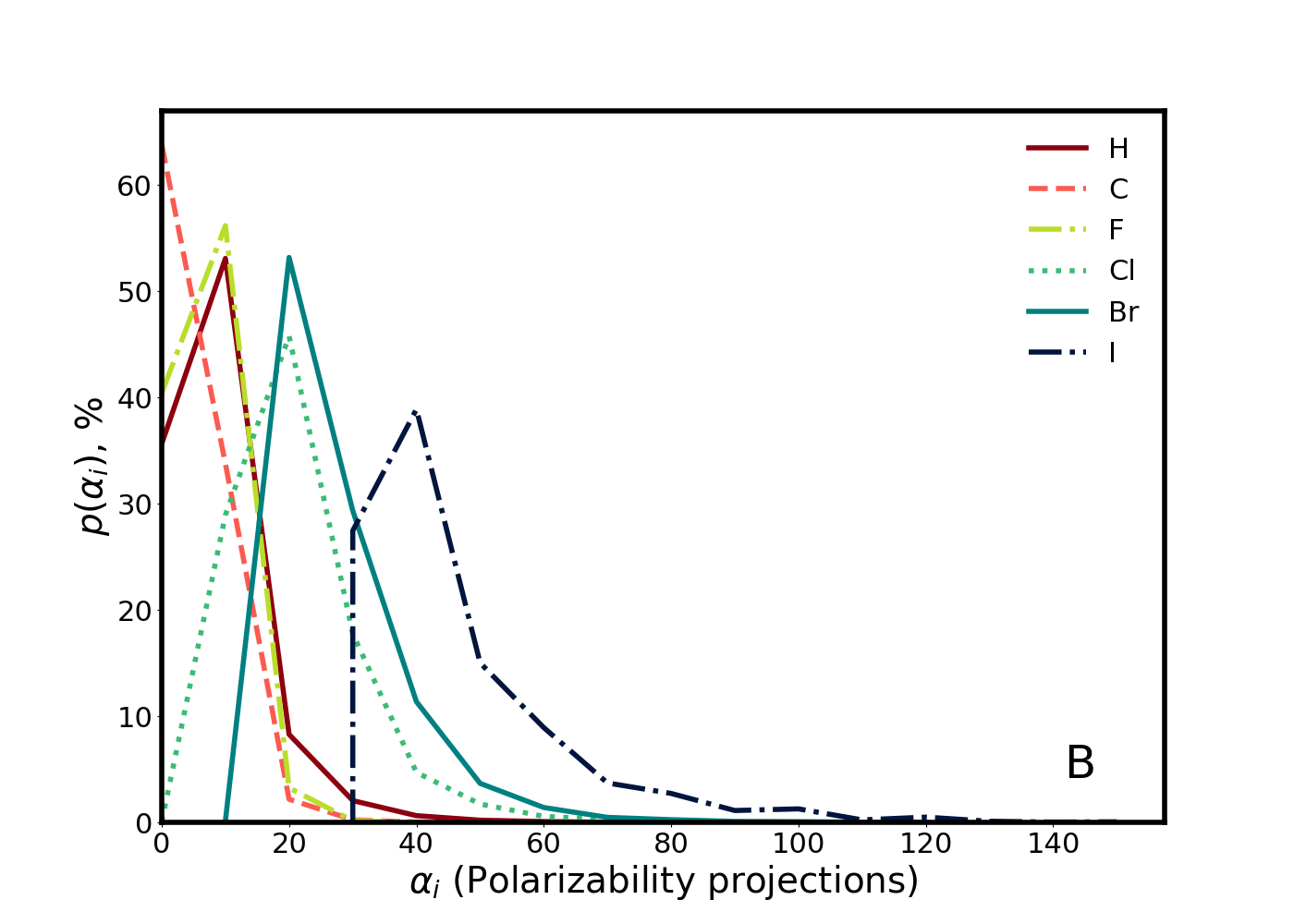}
    \caption{The distribution of unscaled \(\alpha_{i}\) (\(p(\alpha_{i})\)) across all segments assigned to a particular atom (H, C, F, Cl, Br or I) in halocarbon molecules within the collected data. The clustering was performed using \(\alpha_{i}\) grid step of 1 (A) and of 10 (B).}
    \label{fig:projection_distr_no_weighting_diff_grids}
\end{figure}

In analogy to clustering of the screening surface charge densities, we determined the optimal grid for polarizability projections by performing regressions with the fixed \(E_{\textrm{corr}}^{\textrm{F}} = 327.76\) and adjusting the scaling factors \(m_{\textrm{vdW}}\) for each \(\alpha_{i}\) grid. The magnitude of the \(\alpha_{i}\) values depends on the chosen scaling factor. To explore this, we considered two groups of grids. When \(\alpha_{i}\) is unscaled or scaled using Eq. \ref{eq:w5} (with \(\alpha_{i}\) not exceeding 160), we tested grid step sizes of 1, 2, 5, 10, and 20. For scaling factors 1-4, which reduce \(\alpha_{i}\) by approximately two orders of magnitude, we used step sizes of 0.01, 0.02, 0.05, 0.1, 0.2, and 0.5. This evaluation was performed for all CRs described earlier. Figure \ref{fig:truncation_AAD} illustrates the resulting total AADs across all data types for different grid steps. Surprisingly, the finest grids do not always provide the most accurate predictions. Instead, optimal grids are often those that are 10 to 20 times coarser. To examine the factors behind this trend, we plotted the probability \(p(\alpha_{i})\) of finding a segment with a specific unscaled \(\alpha_{i}\) on atoms (H, C, F, Cl, Br, or I) across all halocarbon molecules. In Figure \ref{fig:projection_distr_no_weighting_diff_grids}A, the segments were clustered using a grid step of 1, and in Figure \ref{fig:projection_distr_no_weighting_diff_grids}B, a grid step of 10 was used. Both plots show the expected polarizability trends for the atoms, however, with the coarser grid, nearly half of the segments associated with H-, C-, and F-atoms yield \(\alpha_{i}\) values of zero. This suggests that the current version of openCOSMO-RS might implicitly account for the dispersion term within other terms, as a result of adjusting general model parameters without an explicit dispersion term. This could lead to double-counting of dispersion interactions, particularly for the most abundant atoms in the halocarbon systems (H, C, and F). This issue can be further addressed by simultaneously tuning the misfit parameters and dispersion parameters across a larger data set. A coarser grid also substantially reduces computational time by decreasing the number of segment types per system. Consequently, in further evaluations, we will focus on the coarse grids that yielded the lowest overall AADs. Figure \ref{fig:truncation_AAD} indicates that using the square root CR along with \(E_{\textrm{corr}}^{\textrm{F}}\) in the misfit term might be the optimal modeling approach at this stage. In addition, the AADs across various scalings of \(\alpha_{i}\) do not differ drastically. However, scaling factors 2 and 4 lead to higher deviations compared to 1 and 3, regardless of the CR or grid step size. This shows that the scaling factor should likely depend on the quantity of a particular atom, rather than the average value across the entire molecule (Eqs. \ref{eq:w1} and \ref{eq:w3}). Therefore, scaling factors 2 and 4 will not be considered within the upcoming analysis.

\begin{table}[htbp]
    \centering
    \caption{Comparison of parametrization approaches incorporating local polarizability projections with different scalings (\(w_{i}\)) on coarse \(\alpha_{i}\) grids. The table includes AADs (Total, IDAC, LLE, VLE) and corresponding parameter values (\(m_{\textrm{vdW}}\)). The value of \(E_{\textrm{corr}}^{\textrm{F}}\) is set to 327.76.}
    \begin{adjustbox}{width=\textwidth} % Adjust table width to text width
    \begin{tabular}{lccccccc}
        \toprule
        \(w_{i}\) &  \(m_{\textrm{vdW}}\) & AAD\textsubscript{TOT} & AAD\textsubscript{IDAC} & AAD\textsubscript{LLE} & AAD\textsubscript{VLE}\\
        \midrule
        None &  3.139 & 0.076 & 0.257 & 0.139 & 0.029 \\
        1 (Eq. \ref{eq:w1}) & 811.338 & 0.072 & 0.239 & 0.124 & 0.030 \\
        3 (Eq. \ref{eq:w3}) & 716.554 & 0.073 & 0.250 & 0.127 & 0.029 \\
        5 (Eq. \ref{eq:w5}) &  3.090 & 0.075 & 0.242 & 0.145 & 0.029 \\
        \bottomrule
    \end{tabular}
    \end{adjustbox}
    \label{tab:different_weights_best_trunc}
\end{table}

The model parameters and AADs for the best-performing approaches, identified based on the total AADs shown in Figure \ref{fig:truncation_AAD}, are provided in Table \ref{tab:different_weights_best_trunc}. These models, which incorporate the square root CR and use coarse grids with step sizes of 0.1 for scaling factors 1 and 3, and 10 for the others, form the basis for further assessments. Compared to the approach with only \(E_{\textrm{corr}}^{\textrm{F}}\) in the misfit term, the total AADs of all models in Table \ref{tab:different_weights_best_trunc} are reduced by 16\% - 20\%, along with consistent improvements in AADs for specific data types. Furthermore, this approach provides better results than the one based on isotropic polarizabilities. This may result from considering local properties, using an effective scaling strategy, and selecting an appropriate grid step size.
%%To further elaborate on the difference between the scalings, we will present additional plots to demonstrate that the observed improvements are not simply due to more segments being zeroed out with better scalings. These plots will illustrate that the key factor for the improved performance is the appropriate distribution and choice of grid step size, rather than the number of zeroed segments.

\subsubsection{The Evaluation of the F-atom Radius Used in C-PCM Calculations}
\label{sec:F-atom radius}

\begin{figure}
    \centering
    \includegraphics[width=0.45\linewidth]{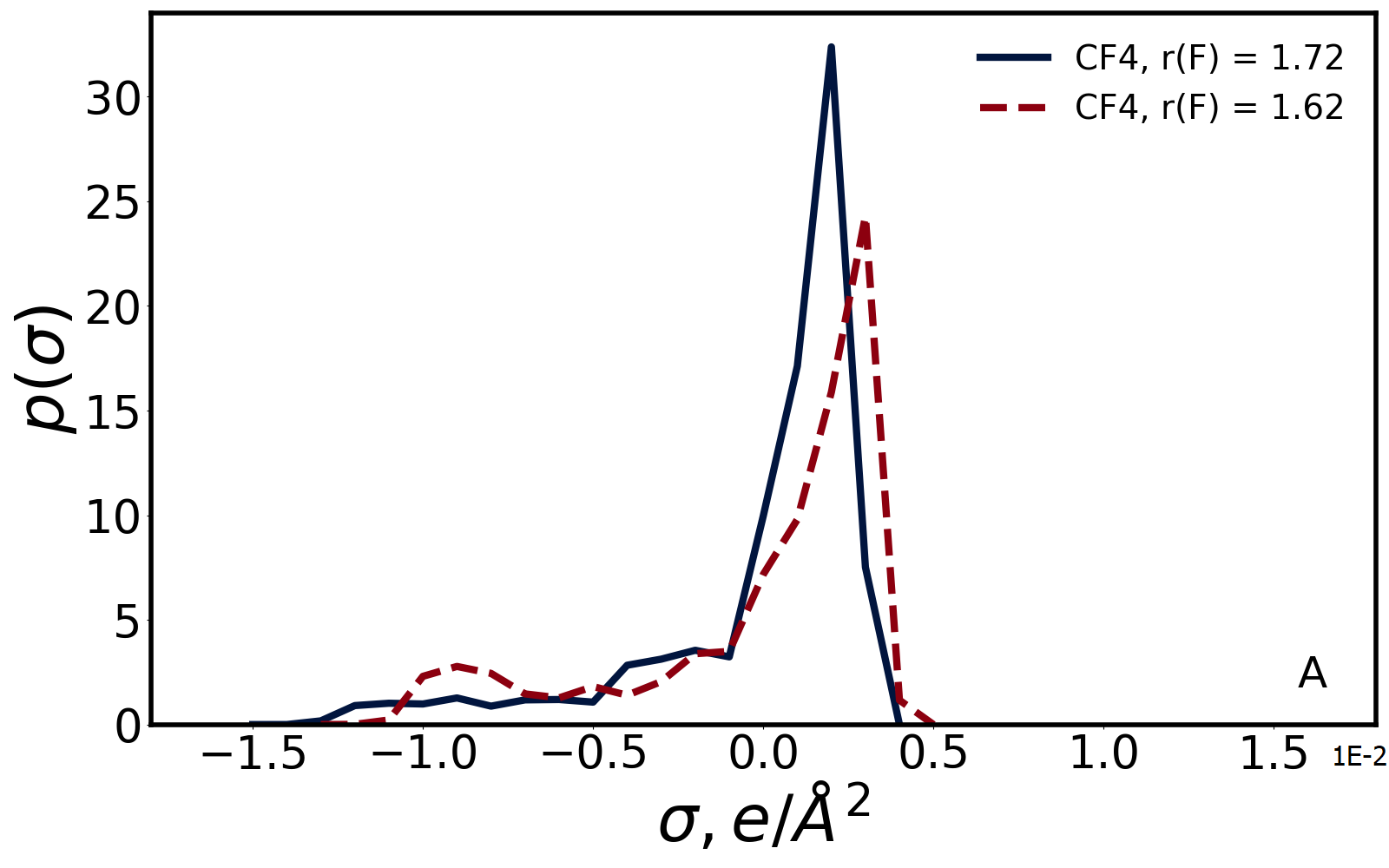}
    \includegraphics[width=0.45\linewidth]{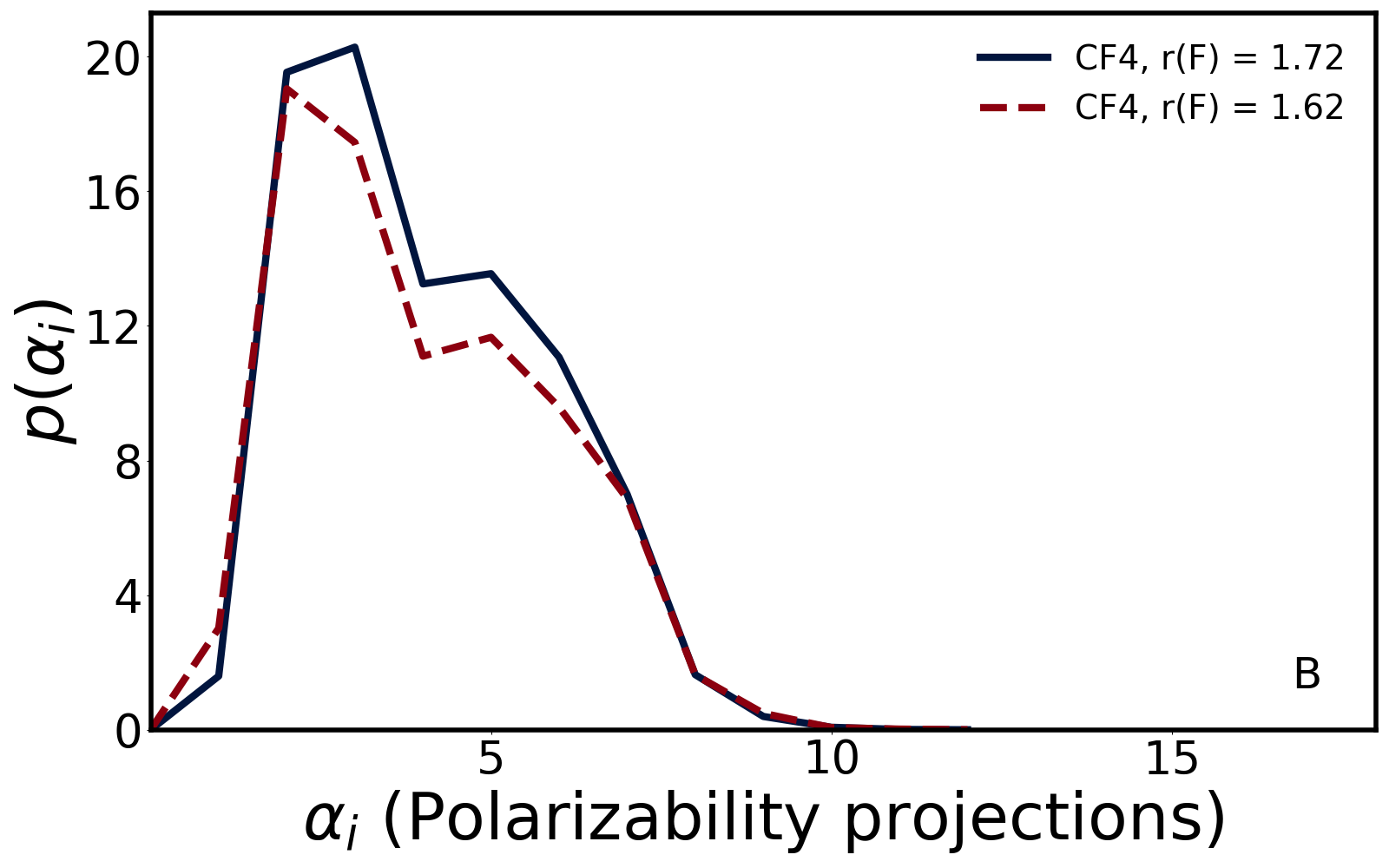}
    \caption{Comparison of the \(\sigma-\) profile (A) and \(\alpha-\)profile (B) of CF\(_4\), calculated with both the original and reduced radii of the F-atom.}
    \label{fig:CF4_sigma_profiles_diff_radius}
\end{figure}

\begin{table}[htbp]
    \centering
    \caption{Overview of parametrization approaches based on C-PCM calculations with a reduced F-atom radius of 1.62. The table includes AADs (Total, IDAC, LLE, VLE) and corresponding parameter values (\(E_{\textrm{corr}}^{\textrm{F}}\), \(m_{\textrm{vdW}}\), and \(m_{\textrm{vdW}}^{\textrm{F}}\)). Local polarizability projections with different scalings (\(w_{i}\)) on coarse \(\alpha_{i}\) grids are used in the dispersion term.}
    \begin{adjustbox}{width=\textwidth} % Adjust table width to text width
    \begin{tabular}{lccccccccc}
        \toprule
        \multirow{2}{*}{\centering{Parameterizations}} & \multirow{2}{*}{\centering{\(w_{i}\)}} & \multicolumn{3}{c}{\centering{Parameters}} & \multicolumn{4}{c}{AAD} \\
        \cmidrule(lr){3-5}
        \cmidrule(lr){6-9}
        &  & \(E_{\textrm{corr}}^{\textrm{F}}\) & \(m_{\textrm{vdW}}\) & \(m_{\textrm{vdW}}^{\textrm{F}}\) & TOT & IDAC & LLE & VLE \\
        \midrule
        no dispersion term & & 354.625 & & & 0.090 & 0.319 & 0.144 & 0.034 \\
        no misfit correction & & & 317.394 & 242.785 & 0.090 & 0.319 & 0.145 & 0.034 \\ 
        \midrule
        \multirow{4}{*}{\(E_{\textrm{vdW}} = m_{\textrm{vdW}}\sqrt{\alpha_{i}\alpha_{j}}\)} & None & \multirow{4}{*}{354.625} & 3.165 & & 0.075 & 0.255 & 0.140 & 0.029 \\        
        & \(w_{1}\) & & 816.664 & & 0.073 & 0.238 & 0.129 & 0.031 \\
        & \(w_{3}\) & & 753.313 & & 0.074 & 0.247 & 0.133 & 0.029 \\
        & \(w_{5}\) & & 3.116 & & 0.074 & 0.241 & 0.144 & 0.029 \\
        \bottomrule
    \end{tabular}
    \end{adjustbox}
    \label{tab:param_diff_F_radius}
\end{table}

Currently, the radius of the F-atom used for C-PCM calculations is set to the same value as that of the O-atom, as originally proposed by \citet{Klamt2000COSMO-RS:Liquids}. However, from a physical perspective, the radius of the F-atom should be smaller. To investigate whether this discrepancy could be contributing to modeling issues in fluorinated mixtures, we performed the same C-PCM calculations for fluorinated molecules, but with the radius of the F-atom reduced from 1.72 to 1.62. The same reduction of 6\% is observed in the C-PCM radius of the O-atom relative to the N-atom. As shown in the \(\sigma\)-profile of CF\(_4\) illustrated in Figure \ref{fig:CF4_sigma_profiles_diff_radius}A, the reduced radius of the F-atom leads to a slight increase in the molecule's polarity, and as expected, to a decrease in its total area. On the other hand, the \(\alpha\)-profile in Figure \ref{fig:CF4_sigma_profiles_diff_radius}B do not indicate any significant change in the distribution of polarizability. 

For the updated C-PCM files, tuning the \(E_{\textrm{corr}}^{\textrm{F}}\) parameter resulted in its increase from 327.76 to 354.625, compared to the results with the old radius. Additionally, the total AAD decreased by approximately 0.5\%. Subsequently, the scaling factors for the previously selected scaling approaches were regressed along with the updated \(E_{\textrm{corr}}^{\textrm{F}}\) parameter (Table \ref{tab:param_diff_F_radius}). Although there is some improvement for unscaled polarizability projections and those scaled with \(w_5\) (Eq. \ref{eq:w5}), the other two models show slightly inferior results. Generally speaking, despite the physically justified reduction in the F-atom radius, no clear advantage is observed over Klamt's conventional value. 

\subsubsection{Modification of the Square Root Combining Rule for Polarizability Projections}
\label{sec:adjusted power}

Dispersive interactions do not always align perfectly with the square root CR, therefore, alternative approaches have been proposed to address this limitation. One such approach, suggested in the molecular simulation study by \citet{Gould2016} of dispersive interactions between individual atoms, involves optimizing the power of the root used to combine the dispersion coefficients \(C_6\). We have adopted this technique for polarizability projections:

\begin{equation}
 E_{\textrm{vdW}} = m_{\textrm{vdW}}(\alpha_i, \alpha_j)^{\beta}, \label{eq:adjusted_power}
\end{equation}
where we selected the scaling factors for \(\alpha_{i}\) that showed the best performance in previous evaluations.

\begin{table}[htbp]
    \centering
    \caption{Comparison of parametrization approaches with adjusted \(\beta\) from Eq. \ref{eq:adjusted_power}, including their AADs (Total, IDAC, LLE, VLE) and corresponding parameter values (\(m_{\textrm{vdW}}, \beta\)). Local polarizability projections with various scalings (\(w_{i}\)) on coarse \(\alpha_{i}\) grids are used in the dispersion term. The value of \(E_{\textrm{corr}}^{\textrm{F}}\) is set to 327.76.}
    \begin{adjustbox}{width=\textwidth} % Adjust table width to text width
    \begin{tabular}{lccccccc}
        \toprule
        \(w_{i}\) & \(m_{\textrm{vdW}}\) & \(\beta\) & AAD\textsubscript{TOT} & AAD\textsubscript{IDAC} & AAD\textsubscript{LLE} & AAD\textsubscript{VLE}\\
        \midrule
        None & 0.845 & 0.6918 & 0.0755 & 0.255 & 0.136 & 0.0294 \\
        1 (Eq. \ref{eq:w1}) & 415.881 & 0.2594 & 0.0757 & 0.232 & 0.130 & 0.0353 \\
        3 (Eq. \ref{eq:w3}) & 339.335 & 0.1418 & 0.0729 & 0.237 & 0.132 & 0.0301 \\
        5 (Eq. \ref{eq:w5}) & 1.928 & 0.5707 & 0.0741 & 0.242 & 0.143 & 0.0289 \\  
        \bottomrule
    \end{tabular}
    \end{adjustbox}
    \label{tab:adjusted_power}
\end{table}

Interestingly, the parameter \(\beta\) regressed using unscaled \(\alpha_{i}\), shown in Table \ref{tab:adjusted_power}, is very similar to the power of 0.73 obtained in \citet{Gould2016}. Although this approach appears somewhat empirical, it may confirm the fact that combining quantities related to the dispersion energy requires more complexity than the conventional square root CR. As demonstrated in Table \ref{tab:adjusted_power}, a minor improvement is observed across all scalings, except for \(w_{1}\). This is likely because both the scaling and the adjusted power of the root aim to account for the deviations from the square root CR, and using both could potentially lead to error cancellation or accumulation. 

\subsubsection{Further Refinement Based the Atomic Ionization Potentials}
\label{sec:ionization potentials}

\begin{table}[htbp]
    \centering
    \caption{
        Comparison of parametrization approaches incorporating atomic ionization potentials into the dispersion term. The table includes AADs (Total, IDAC, LLE, VLE) and corresponding parameter values (\(m_{\textrm{vdW}}\)) for local polarizability projections using different scalings (\(w_{i}\)) on coarse \(\alpha_{i}\) grids. The value of \(E_{\textrm{corr}}^{\textrm{F}}\) is fixed at 327.76. The equations represent various methods of combining ionization potentials (\(I_{i}\), \(I_{j}\)) with polarizabilities (\(\alpha_{i}\), \(\alpha_{j}\)) to refine the dispersion term.
    }
    \begin{adjustbox}{width=\textwidth}
    \begin{tabular}{lccccccc}
        \toprule
        Dispersion Term & \(w_{i}\) & \(m_{\textrm{vdW}}\) & AAD\textsubscript{TOT} & AAD\textsubscript{IDAC} & AAD\textsubscript{LLE} & AAD\textsubscript{VLE} \\
        \midrule
        \multirow{2}{*}{
        \vbox{
            \begin{equation}            E_{\textrm{vdW}}=m_{\textrm{vdW}}\sqrt{\alpha_{i} \alpha_{j}} \frac{2\sqrt{I_{i} I_{j}}}{I_{i} + I_{j}}
            \label{eq:IP_Hudson} 
            \end{equation}}}
        & None & 3.126 & 0.076 & 0.256 & 0.140 & 0.029 \\    
        & \(w_{5}\) & 3.082 & 0.075 & 0.242 & 0.146 & 0.029 \\ 
        \midrule
        \multirow{2}{*}{\vbox{
        \begin{equation}
            E_{\textrm{vdW}}=m_{\textrm{vdW}}\sqrt{\frac{\alpha_{i} \alpha_{j}}{I_{i} I_{j}}} 
            \label{eq:IP_wi}
        \end{equation}}}
        & None & 38.652 & 0.074 & 0.248 & 0.130 & 0.029 \\    
        & \(w_{5}\) & 37.709 & 0.072 & 0.234 & 0.135 & 0.029 \\ 
        \midrule
        \multirow{2}{*}{\vbox{
        \begin{equation}
            E_{\textrm{vdW}}=m_{\textrm{vdW}}\frac{\sqrt{\alpha_{i} \alpha_{j}}}{I_{i} + I_{j}}
            \label{eq:IP_sum}
        \end{equation}}} 
        & None & 77.091 & 0.074 & 0.250 & 0.131 & 0.029 \\    
        & \(w_{5}\) & 75.135 & 0.072 & 0.234 & 0.136 & 0.029 \\ 
        \bottomrule
    \end{tabular}
    \end{adjustbox}
    \label{tab:param_IPs}
\end{table}

Dispersive interactions between two spherical atoms are influenced by their ionization potentials (IPs) \citep{London1937TheForces, Prausnitz1999MolecularEquilibria}, which quantify the energy required to remove an electron from an atom. This relationship has been employed in the development of combining rules for intermolecular interactions. For instance, IPs—often significantly different between species such as hydrocarbons and fluorocarbons—have been incorporated into the CRs proposed by \citet{Hudson1960IntermolecularRules}. This approach has also been extended to predictive frameworks, such as the Group-Contribution Statistical Association Fluid Theory model, where adjusted pseudo-ionization potentials of molecular groups are used \citep{Nguyen1, Nguyen2}. 

Incorporating IPs into the dispersion term of openCOSMO-RS could enhance the accuracy of predictions, particularly when the IPs of interacting species differ significantly. For halocarbons, the most pronounced variation is between the IP of the F-atom (17.423) and the IPs of other atoms, which range from 10 to 13.5 \citep{NCBI2024IonizationEnergy}. Although this difference is not exceedingly large, we evaluated several equations incorporating atomic IPs into segment-segment dispersion energy, as shown in Table \ref{tab:param_IPs}. Only unscaled \(\alpha_{i}\) or those scaled with \(w_{5}\) are reported, since other scalings yielded inferior AADs, potentially due to physically inconsistent overlaps of multiple corrections to the square root CR, as has also been observed in previous evaluations. Equation \ref{eq:IP_Hudson} is based on the predictive approach for estimating the \(k_{ij}\) parameter proposed in \citep{Hudson1960IntermolecularRules}, but its effect on the AADs is minimal, as the correction factor tends to one. However, this method might be more applicable to systems with greater differences in IPs. Additionally, we did not account for local IPs, which differ more profoundly within a molecule \citep{Jin2004} and therefore may be more sensitive to this approach. In Eq. \ref{eq:IP_wi}, IPs serve as scalings to \(\alpha_{i}\), while Eq. \ref{eq:IP_sum} considers only the denominator of Hudson's \(k_{ij}\). These two equations produced the best results when \(w_{5}\) was applied to \(\alpha_{i}\). While the mathematical formulation remains somewhat empirical—much like certain aspects of the model itself—it retains a physical basis. Hence, we observe the improvement in model's predictions without introducing additional adjustable parameters, but solely by integrating atomic IPs.

\subsubsection{Concluding Remarks on Incorporating Polarizabilities for Modeling Halocarbon Thermodynamic Properties}
\label{sec:conclusions_halocarbons}

\begin{table}[htbp]
    \centering
    \caption{Comparison of the new parametrization approaches, incorporating \(\alpha_{i}\) with simultaneously adjusted \(E_{\textrm{corr}}^{\textrm{F}}\) and \(m_{\textrm{vdW}}\) with the model based on atomic dispersion parameters (FH\_6). The table includes AADs (Total, IDAC, LLE, VLE) and corresponding parameter values.}
    \begin{adjustbox}{width=\textwidth} % Adjust table width to text width
    \begin{threeparttable}
    \begin{tabular}{lccccccc}
        \toprule
        Parametrization &  \(E_{\textrm{corr}}^{\textrm{F}}\) &  \(m_{\textrm{vdW}}\) & AAD\textsubscript{TOT} & AAD\textsubscript{IDAC} & AAD\textsubscript{LLE} & AAD\textsubscript{VLE}\\
        \midrule
        \(w_{1}\) & 304.92 & 842.20 & 0.069 & 0.238 & 0.102 & 0.029 \\
        \(w_{5}\) & 310.64 & 3.141 & 0.072 & 0.242 & 0.125 & 0.029 \\
        \(w_{5}\) + (Eq. \ref{eq:IP_wi}) & 313.85 & 38.044 & 0.070 & 0.234 & 0.118 & 0.029 \\
        \(w_{5}\) + (Eq. \ref{eq:IP_sum}) & 312.62 & 76.165 & 0.070 & 0.234 & 0.118 & 0.029 \\
        FH\_6\tnote{1} &   &   & 0.077 & 0.250 & 0.133 & 0.033 \\
        \bottomrule
    \end{tabular}
    \begin{tablenotes}
        \item[1] Atomic dispersion parameters [J\(^{0.5}\)/Å]: \( \tau^{\textrm{vdW}}_{\textrm{H}} =  6.745;\:\tau^{\textrm{vdW}}_{\textrm{C}} = 14.719;\:\tau^{\textrm{vdW}}_{\textrm{F}} = 2.181;\: \tau^{\textrm{vdW}}_{\textrm{Cl}} = 12.654;\:\tau^{\textrm{vdW}}_{\textrm{Br}} = 16.167;\:\tau^{\textrm{vdW}}_{\textrm{I}} = 21.211\).
    \end{tablenotes}
    \end{threeparttable}
    \end{adjustbox}
    \label{tab:concluding_remarks_halocarbons}
\end{table}

\begin{figure}
    \centering
    \includegraphics[width=0.45\linewidth]{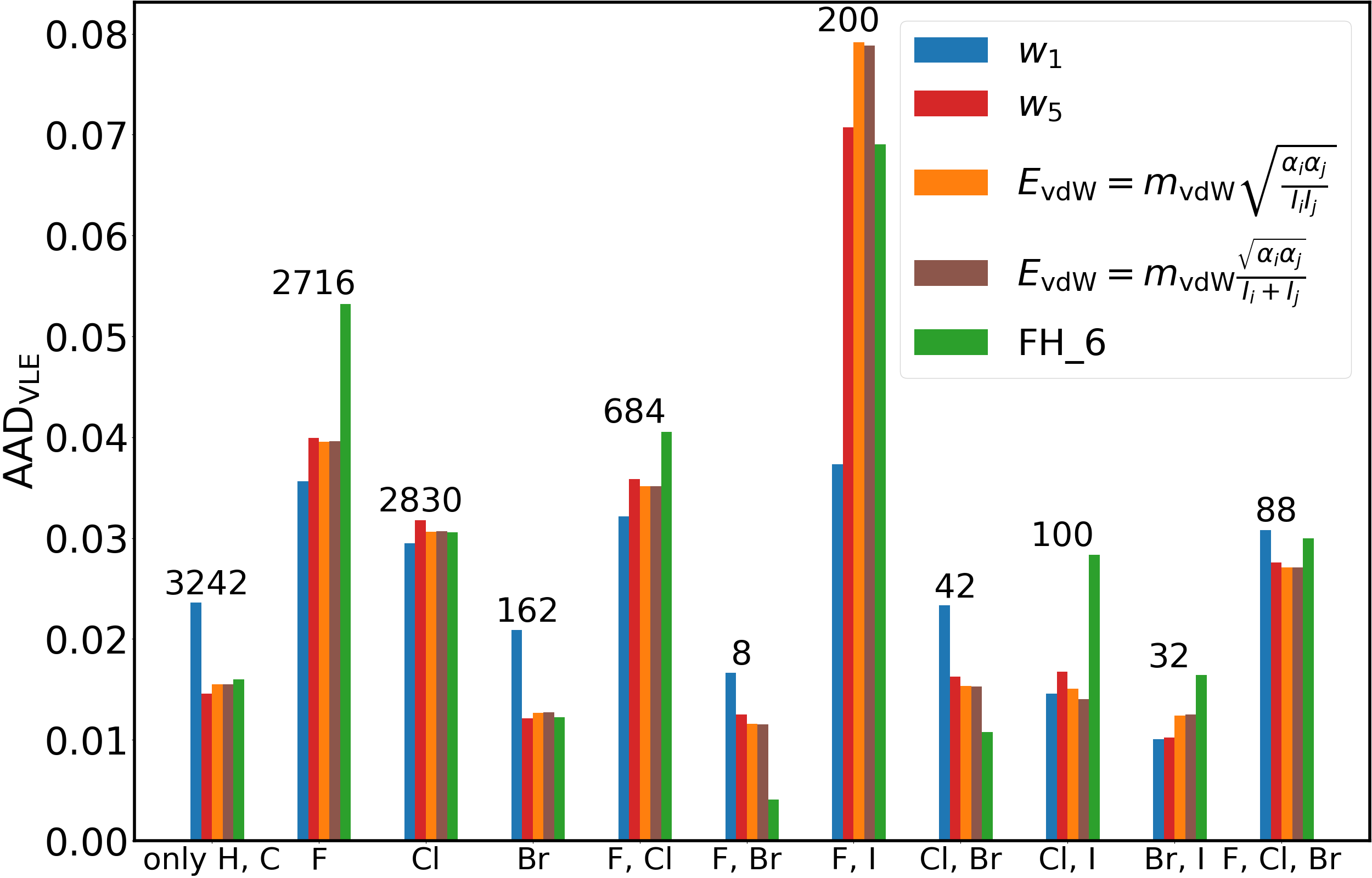}
    \includegraphics[width=0.45\linewidth]{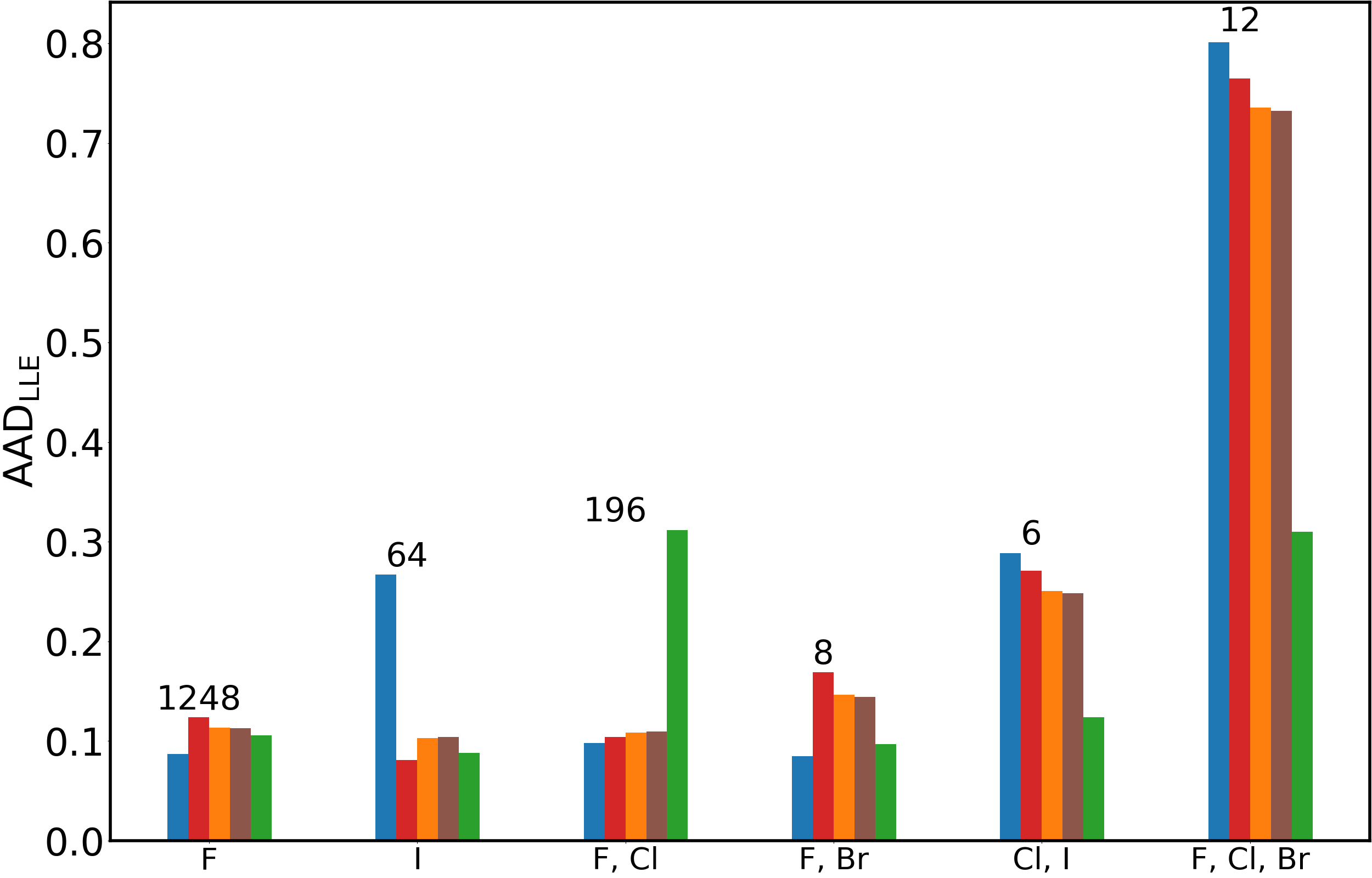}
    \includegraphics[width=0.45\linewidth]{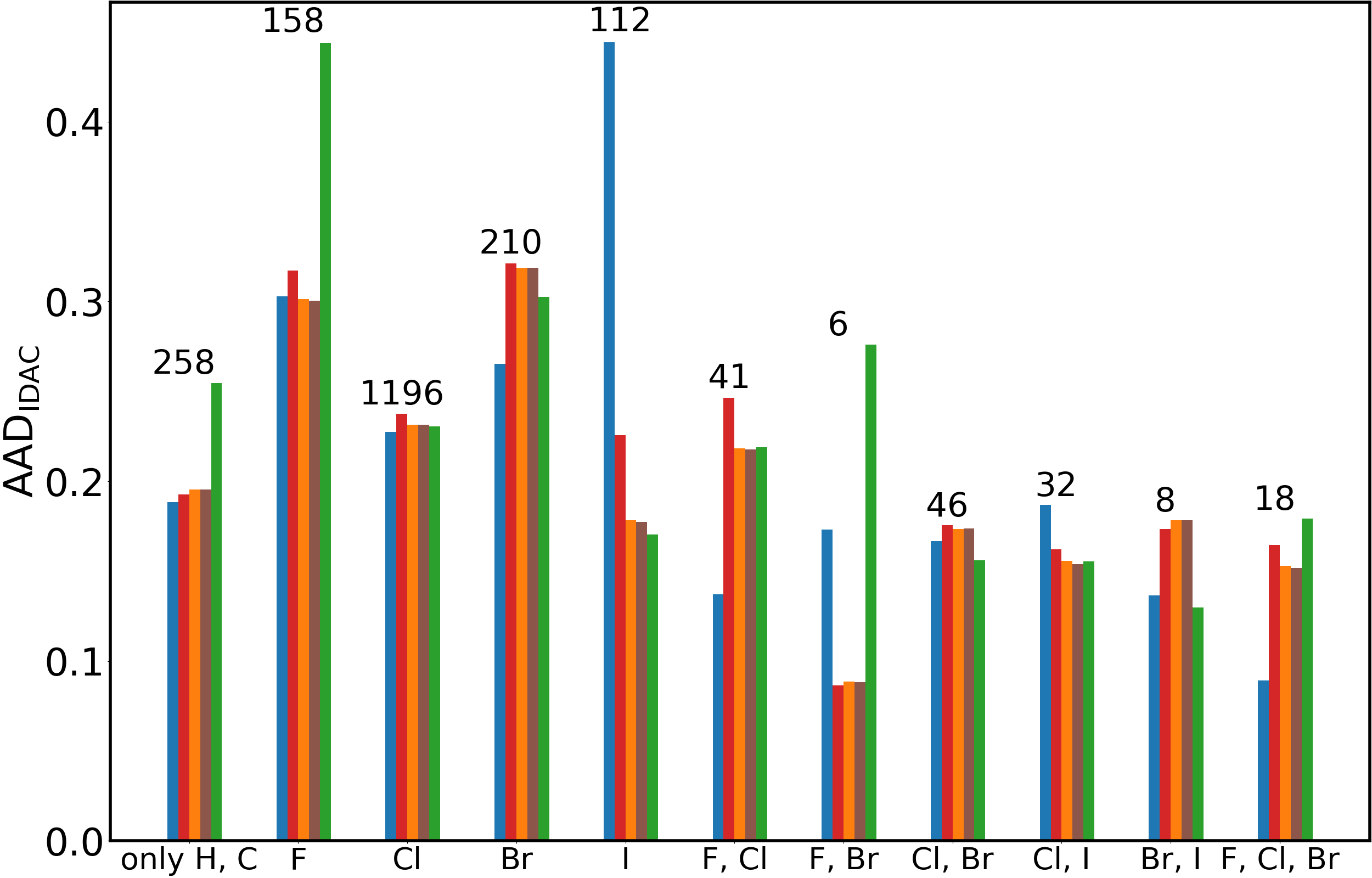}    
    \caption{Bar plots comparing the AAD for VLE (left), IDAC (middle), and LLE (right) data across five models: the models incorporating the dispersion term with different scalings and corrections for \(\alpha_{i}\) along with the \(E_{\textrm{corr}}^{\textrm{F}}\) in the misfit term, and the model based on atomic dispersion parameters (FH\_6). The evaluated data is categorized based on the types of halogen atoms present in the molecules. Numbers of data points are indicated above columns for each category.}
    \label{fig:AAD_concl_halocarbons}
\end{figure}

In the preceding discussion, we evaluated several approaches to incorporate polarizability as a first-principles molecular descriptor. Among the various combining rules, the square root emerged as the optimal choice over conventional alternatives. However, it requires corrections such as scalings, ionization potentials, or adjusting the root power. Notably, combining multiple corrections can sometimes yield inferior results. Based on prior assessments, the scaling factor \(w_{1}\) and the scaling factor \(w_{5}\), coupled with IPs using Eqs. \ref{eq:IP_wi} or \ref{eq:IP_sum}, demonstrated the best performance. 

To further refine the methodology, we simultaneously regressed two key parameters: the misfit correction \(E_{\textrm{corr}}^{\textrm{F}}\) for interactions involving the F-atom, and the \(m_{\textrm{vdW}}\) parameter, scaling the dispersion term. Additionally, we compared this new model with the FH\_6 approach developed in the first part of this study. To ensure a fair comparison, we re-optimized the dispersion atomic parameters using the updated dataset and the ORCA 6.0 C-PCM files. As in previous analyses, optimization involved the differential evolution algorithm \citep{2020SciPy-NMeth}, with physically reasonable boundaries. Table \ref{tab:concluding_remarks_halocarbons} summarizes the comparison. The new approach, despite containing three times fewer adjustable parameters, demonstrated superior performance. We further assessed the models' accuracy across different data groups. Figure \ref{fig:AAD_concl_halocarbons} reveals that the FH\_6 model underperforms for fluorinated groups, highlighting the importance of the misfit correction in addition to the dispersion term. For other non-fluorinated, halogenated groups, the dispersion term incorporating \(\alpha_{i}\) generally performs similarly to one with individual atomic parameters. An interesting observation pertains to the \(w_{1}\) model, which indicates poor predictions for purely iodinated compounds but excels in predicting datasets containing both F and I. This behavior stems from the nature of the scaling itself. Since \(w_{1}\) scales \(\alpha_{i}\) by the number of grids assigned to each atom, the large size of I - atom results in a low weight for iodinated alkanes. This likely causes an underestimation of dispersive interactions within this group. Conversely, for datasets with both F and I, containing many highly fluorinated compounds, this scaling proves beneficial, while other models overestimate attractive interactions and, consequently, activity coefficients in these mixtures.

In conclusion, the proposed models achieve significant improvements in prediction accuracy compared to both the original model and our prior FH\_6 model for halocarbons. In the following sections, we will extend these models to the complete dataset for broader validation.

\subsection{The Extended Database}
\label{sec:Parameterizations_all_data_discussion}
%\label{sec:database}

In the present study, we extended the halocarbon database comprising 11948 data points (VLE, LLE and IDAC) from the first part of this work by incorporating predominantly oxygen-containing organic compounds, as well as those containing nitrogen, sulfur, and phosphorus. The updated dataset encompasses a wide range of organic compound classes, and its chemical diversity provides a solid foundation for the development of a general model.
Additionally, as summarized in Table \ref{tab:database}, the dataset includes not only the predominant VLE, LLE, and IDAC data, but also partitioning coefficients and solvation free energies. These data types, commonly used for the parameterization and validation of thermodynamic models, capture distinct aspects of intermolecular interactions and offer valuable insights into the properties of liquid phases.
The majority of the data was obtained from various open-source databases. Specifically, most of the VLE data was extracted from the KDB database \citep{CHERIC2024}, rigorously tested for thermodynamic consistency, and used to calculate activity coefficients when it met the consistency criteria defined by the Van Ness test. Details regarding the sources and processing of the partitioning coefficients, IDAC, and solvation free energy data are available in \citep{Gerlach2022AnDescriptors, MULLER2025114250}. To extend our LLE data collection, we utilized the open-source NIST database \citep{Riccardi2021}, extracting the required information from the JSON/XML files provided. 

\begin{table}[htbp]
    \centering 
    \caption{An overview of the complete collection of experimental data.}
    \begin{adjustbox}{width=\textwidth} % Adjust table width to text width
    \begin{tabular}{lccccc}
    \toprule 
    Type & Data points & Datasets & Temperature range, K& Sources \\
    \hline
    VLE & 37730 & 1240 & 183.01 - 573.15 & \citet{grigorash2024comprehensiveapproachincorporatingintermolecular, CHERIC2024} \\
    LLE & 6882 & 541  & 87.58 - 593& \citet{grigorash2024comprehensiveapproachincorporatingintermolecular, Riccardi2021}\\
    \(\Delta G_{\textrm{solv}}\) & 2126 & 2126  & 298.15& \citet{MULLER2025114250}\\
    IDAC & 6763 & 6485 & 248.15 - 473.05& \citet{Gerlach2022AnDescriptors, Gmehling2008ActivityCoefficients} \\
    \(K_i^{\textrm{org/w}}\) & 296 & 296 & 298.15& \citet{Gerlach2022AnDescriptors} \\
    \bottomrule
    \end{tabular}
    \end{adjustbox}
    \label{tab:database}
\end{table}

The general model parameters, without both the misfit correction for the F-atom and the dispersion term, were initially re-optimized using the entire dataset. These parameters (\(a_{\textrm{eff}},\:\alpha_{\textrm{mf}},\:c_{\textrm{hb}},\:\sigma_{\textrm{hb}},\:r_{\textrm{av}},\:f_{\textrm{corr}},\:c_{\textrm{hb}}^{T}\)) pertain solely to the residual term, as the FH combinatorial term does not require additional parameters.

\begin{table}[htbp]

    \centering
    \caption{Comparison of the new parametrization approaches, incorporating \(\alpha_{i}\) with simultaneously adjusted \(E_{\textrm{corr}}^{\textrm{F}}\), \(m_{\textrm{vdW}}\), and model's general parameters. The table includes AADs (Total, \(K_i^{\textrm{org/w}}\), IDAC, LLE, VLE, \(K_i^{\textrm{org/w}}\), \(\Delta G_{\textrm{solv}}\)) and corresponding parameter values.}
    \begin{adjustbox}{width=\textwidth} % Adjust table width to text width
    \begin{tabular}{lccccc}
        \toprule 
        \multirow{2}{*}{\shortstack[c]{Parameters and\\[1.5ex] Deviations (AAD)}} & \multicolumn{5}{c}{Parametrizations} \\
        \cmidrule(lr){2-6}
        & no dispersion & \(w_{1}\) & \(w_{5}\) & \(w_{5}\) + Eq. \ref{eq:IP_wi} & \(w_{5}\) + Eq. \ref{eq:IP_sum} \\
        \midrule
        \multirow{1}{*}{\(a_{\textrm{eff}}\) [Å\textsuperscript{2}]} & 4.574 & 4.943 & 4.955 & 4.90825  & 4.706  \\
         \(r_{\textrm{av}}^{*}\) [Å] & \multicolumn{5}{c}{0.5}\\
        \multirow{1}{*}{\(\alpha_{\textrm{mf}}\) [kJ/(mol$\cdot$Å\textsuperscript{2})/e\textsuperscript{2}]} & 7613& 7294 & 7672 & 7876 &7322\\
        \(f_{\textrm{corr}}^{*}\) & \multicolumn{5}{c}{2.4}\\
        \multirow{1}{*}{\( c_{\textrm{hb}}\) [kJ/(mol$\cdot$Å\textsuperscript{2})/e\textsuperscript{2}]} & 53447& 43265&43457 & 49318 &43421 \\
        \(c_{\textrm{hb}}^{T*}\) & \multicolumn{5}{c}{1.5}\\
        \multirow{1}{*}{\(\sigma_{\textrm{hb}}\) [e/Å\textsuperscript{2}]} & 0.009739 &0.009492 & 0.00960& 0.009953 & 0.009355 \\   
        \(E_{\textrm{corr}}^{\textrm{F}}\) [J/mol] &  & 349.18 &  350.37& 346.82 & 340.73 \\
        \(m_{\textrm{vdW}}\)  &  & 339.58 & 2.4097& 29.567 & 27.853 \\
        \midrule
        AAD\textsubscript{TOT} & 0.216 & 0.178 & 0.172 & 0.169 & 0.1795 \\
        AAD\textsubscript{\(K_i^{\textrm{org/w}}\)} & 0.858& 0.790 & 0.806 &  0.812 & 0.831 \\
        AAD\textsubscript{IDAC} & 0.570 & 0.464 & 0.461 & 0.452  & 0.476 \\
        AAD\textsubscript{LLE} & 0.396 & 0.258 & 0.248 & 0.244 & 0.265 \\
        AAD\textsubscript{VLE} & 0.102 & 0.093 & 0.086 & 0.084 & 0.091 \\
        AAD\textsubscript{\(\Delta G_{\textrm{solv}}\)} & 0.444& 0.435 & 0.444 & 0.443 & 0.443 \\
        \bottomrule
    \end{tabular}
    \end{adjustbox}
    \label{tab:comparison_compl_set}

\end{table}

\begin{table}[htbp]
    \centering
    \caption{Extension of Table \ref{tab:comparison_compl_set} with parameters for calculating \(\Delta G_{\textrm{solv}}\).}
    \begin{adjustbox}{width=\textwidth} % Adjust table width to text width
    \begin{tabular}{lccccc}
        \hline
        \multirow{2}{*}{\shortstack[c]{Parameters}} & \multicolumn{5}{c}{Parametrizations} \\
        \cmidrule(lr){2-6}
        & no dispersion & \(w_{1}\) & \(w_{5}\) & \(w_{5}\) + Eq. \ref{eq:IP_wi} & \(w_{5}\) + Eq. \ref{eq:IP_sum} \\
        \midrule
        \( \tau_1~[\frac{\textrm{kJ}}{\textrm{mol}\cdot\textrm{Å}^2}] \) & 0.027 & 0.028 & 0.031 & 0.030 & 0.029 \\
        \( \tau_6~[\frac{\textrm{kJ}}{\textrm{mol}\cdot\textrm{Å}^2}] \) & 0.025 & 0.016 & 0.019 & 0.019 & 0.021 \\
        \( \tau_7~[\frac{\textrm{kJ}}{\textrm{mol}\cdot\textrm{Å}^2}] \) & 0.013 & 0.003 & 0.006 & 0.009 & 0.007 \\
        \( \tau_8~[\frac{\textrm{kJ}}{\textrm{mol}\cdot\textrm{Å}^2}] \) & 0.008 & 0.003 & 0.006 & 0.010 & 0.005 \\
        \( \tau_9~[\frac{\textrm{kJ}}{\textrm{mol}\cdot\textrm{Å}^2}] \) & 0.007 & 0.018 & 0.025 & 0.019 & 0.022 \\
        \( \tau_{17}~[\frac{\textrm{kJ}}{\textrm{mol}\cdot\textrm{Å}^2}] \) & 0.033 & 0.028 & 0.028 & 0.029 & 0.031 \\
        \( \tau_{35}~[\frac{\textrm{kJ}}{\textrm{mol}\cdot\textrm{Å}^2}] \) & 0.041 & 0.035 & 0.035 & 0.035 & 0.039 \\
        \( \tau_{53}~[\frac{\textrm{kJ}}{\textrm{mol}\cdot\textrm{Å}^2}] \) & 0.731 & 0.531 & 0.321 & 1.530 & 0.384 \\
        \( \tau_{14}~[\frac{\textrm{kJ}}{\textrm{mol}\cdot\textrm{Å}^2}] \) & 0.057 & 0.016 & 0.038 & 0.005 & 0.016 \\
        \( \tau_{15}~[\frac{\textrm{kJ}}{\textrm{mol}\cdot\textrm{Å}^2}] \) & 0.020 & 0.002 & 0.004 & 0.0001 & 0.007 \\
        \( \tau_{16}~[\frac{\textrm{kJ}}{\textrm{mol}\cdot\textrm{Å}^2}] \) & 0.035 & 0.027 & 0.026 & 0.026 & 0.031 \\
        \( \eta~[\frac{\textrm{kJ}}{\textrm{mol}}] \) & -4.271 & -4.086 & -4.134 & -4.102 & -4.252 \\
        \( \omega_{\textrm{ring}}~[\frac{\textrm{kJ}}{\textrm{mol}}] \) & 0.2423 & 0.2351 & 0.2511 & 0.2495 & 0.2491 \\
        \hline
    \end{tabular}
    \end{adjustbox}
    \label{tab:Gsolv_parameters}
\end{table}

Following this, the models that demonstrated the best performance for halocarbons were extended to the complete dataset. Subsequently, the same general parameters, along with  \(E_{\textrm{corr}}^{\textrm{F}}\) and \(m_{\textrm{vdW}}\) were optimized. This simultaneous regression was performed due to the coupling between these parameters; the introduction of the dispersion term inevitably influences both the misfit and hydrogen-bonding terms. Notably, the dispersion effects, which were previously approximated within these terms, are now explicitly accounted for.
Given the computational demands of these full-data optimizations, they were executed on a high-performance computing cluster. Depending on the number of cores (2-4), each optimization required 2–5 days of computational time, utilizing conventional algorithms like Nelder-Mead from the SciPy Python package \citep{2020SciPy-NMeth}.
Table \ref{tab:comparison_compl_set} summarizes the results of these regressions. The hydrogen-bonding parameters \(c_{\textrm{hb}}\) for models incorporating dispersion were consistently lower than for the model without dispersion. This reduction indicates that some of the attractive interactions previously attributed to hydrogen bonding are now accounted for by the dispersion term. For solvation free energy, the related parameters were independently optimized for each model, with the results provided in Table~~\ref{tab:Gsolv_parameters}. 

To provide a more detailed analysis of the modified model, we compared the AAD of the overall best-performing model from Table~\ref{tab:comparison_compl_set} (\(w_{5}\) + Eq. \ref{eq:IP_wi}) with the unmodified version, labeled as ‘no dispersion’ in the same table. This comparison was conducted across various chemical groups by calculating the difference between those two models as follows:
\begin{equation}
\Delta \textrm{AAD} = \textrm{AAD}{(\textrm{no~dispersion})} -\textrm{AAD}{(\textrm{+~dispersion})}. 
\label{eq:delta_AAD}
\end{equation}

\begin{landscape}
\begin{figure}
    \centering
    \includegraphics[width=1\linewidth]{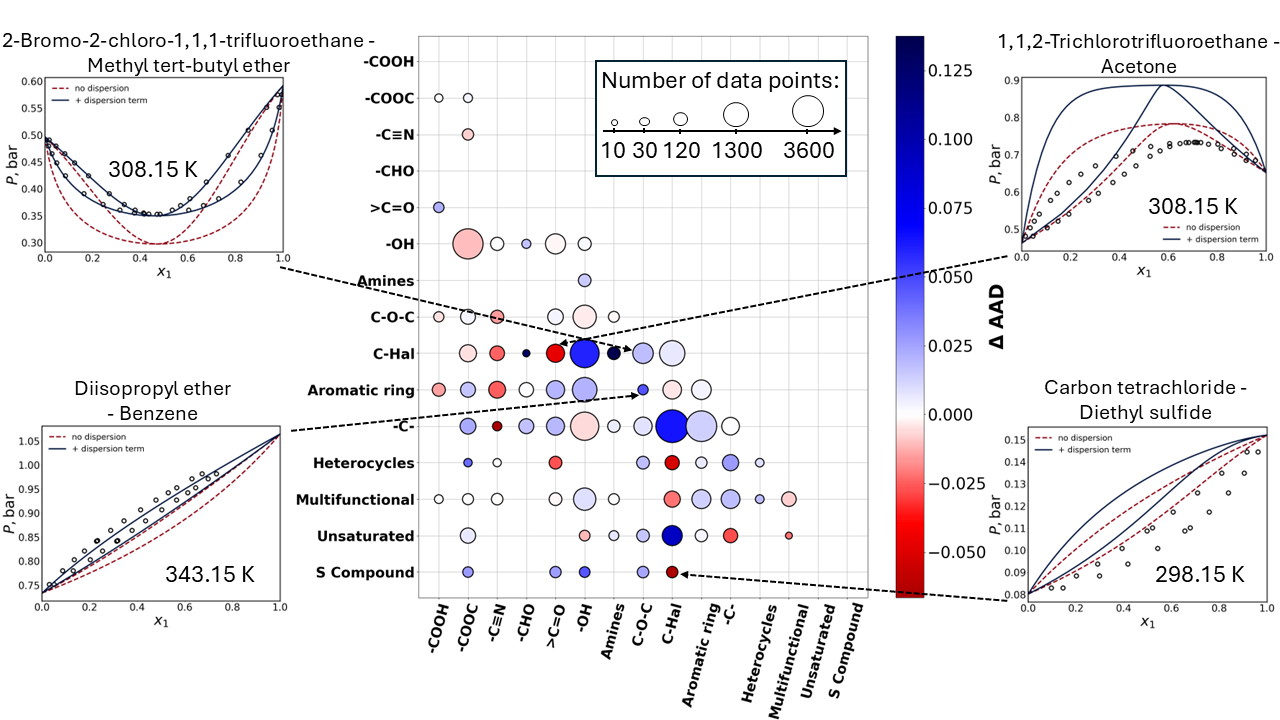}
    \caption{Visualization of \(\Delta~\textrm{AAD}\) (Eq.~\ref{eq:delta_AAD}) calculated for model \(w_{5}\) + Eq. \ref{eq:IP_wi} for binary VLE data, accompanied by \textit{P-xy} phase diagram examples. The colors in the phase diagrams indicate the order of magnitude of AAD, as shown by the color bar. Rows and columns represent the primary chemical functional groups of the components, while the circle sizes reflect the proportion of data points corresponding to molecules within each category. This plot concept is inspired by \citet{Fingerhut2017ComprehensiveEquilibria}.}
    \label{fig:circular_VLE}
\end{figure}
\end{landscape}

The results of this comparison for VLE are illustrated in Figure~\ref{fig:circular_VLE}. Positive values on the color bar indicate improved AADs, whereas negative values denote deteriorated AADs. Overall, approximately 70\% of the data groups showed improvement, with the most significant, as expected, in mixtures of halocarbons and hydrocarbons. However, this trend does not extend to all halocarbon mixtures, as only about half of the groups improved. For instance, systems involving halocarbons and ethers consistently demonstrated better accuracy, whereas mixtures of halocarbons with ketones or sulfur-containing compounds performed worse. Representative \textit{P-xy} phase diagrams illustrating these trends are shown in Figure~\ref{fig:circular_VLE}. Notably, the two \textit{P-xy} phase diagrams on the right-hand side of Figure~\ref{fig:circular_VLE} suggest a tendency to overestimate attractive interactions. This finding further highlights the potential benefits of adopting a more sophisticated combining approach. Challenges also persist in modeling mixtures of halocarbons with compounds containing unsaturated carbon atom and heteroatoms. In particular, polar oxygenated mixtures likely require additional refinement, potentially through integration with a hydrogen bonding term to ensure a more balanced representation of different contributions.

Despite these remaining challenges, we consider the modified model an improvement for VLE predictions, as the majority of data groups showed enhanced performance. Furthermore, these improvements are not limited to halogenated systems, as demonstrated by the example of the diisopropyl ether–benzene system in Figure~\ref{fig:circular_VLE}.

\begin{landscape}
\begin{figure}
    \centering
    \includegraphics[width=1\linewidth]{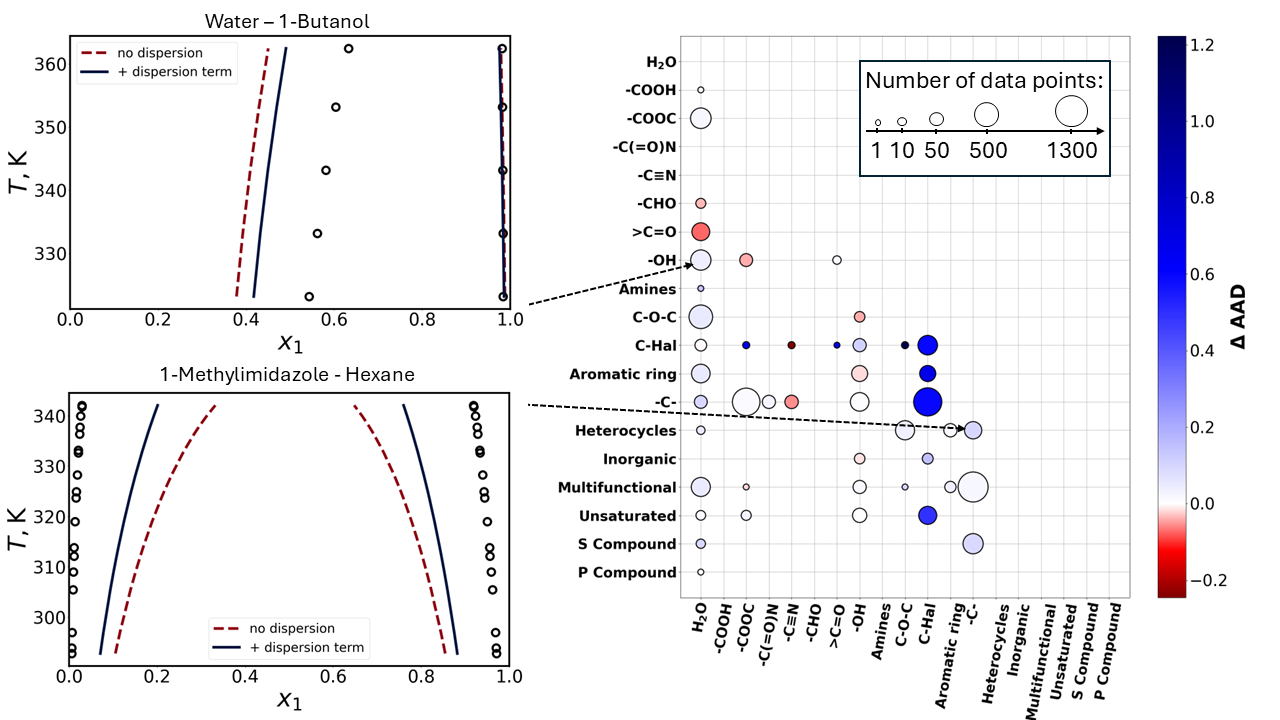}
    \caption{Visualization of \(\Delta~\textrm{AAD}\) (Eq.~\ref{eq:delta_AAD}) calculated for model \(w_{5}\) + Eq. \ref{eq:IP_wi} for binary LLE data, accompanied by LLE phase diagram examples. The colors in the phase diagrams indicate the order of magnitude of AAD, as shown by the color bar. Rows and columns represent the main chemical functional groups of the components. The size of the circles indicates the proportion of data points corresponding to molecules within each category. The concept for this plot is inspired by \citet{Fingerhut2017ComprehensiveEquilibria}.}
    \label{fig:circular_LLE}
\end{figure}
\end{landscape}

A detailed analysis for LLE systems is presented in Figure~\ref{fig:circular_LLE}, which also includes two examples of LLE phase diagrams for improved categories. Deteriorated cases are not illustrated, but often, this simply indicates the absence of a phase split. Notably, most aqueous mixtures exhibit improved performance. Furthermore, nearly all halocarbon-containing groups improved, regardless of the second compound’s chemical nature. However, challenges remain in modeling mixtures with polar oxygenated compounds. Despite these limitations, approximately 80\% of the data groups showed improved AADs. Moreover, the regression procedure applied to the LLE data, which is based on partition coefficients rather than on iterative phase split calculations, appears to yield reliable results. Combined with the proposed model modifications, this approach contributes to the overall accuracy and robustness of LLE predictions.

\begin{landscape}
\begin{figure}
    \centering
    \includegraphics[width=1\linewidth]{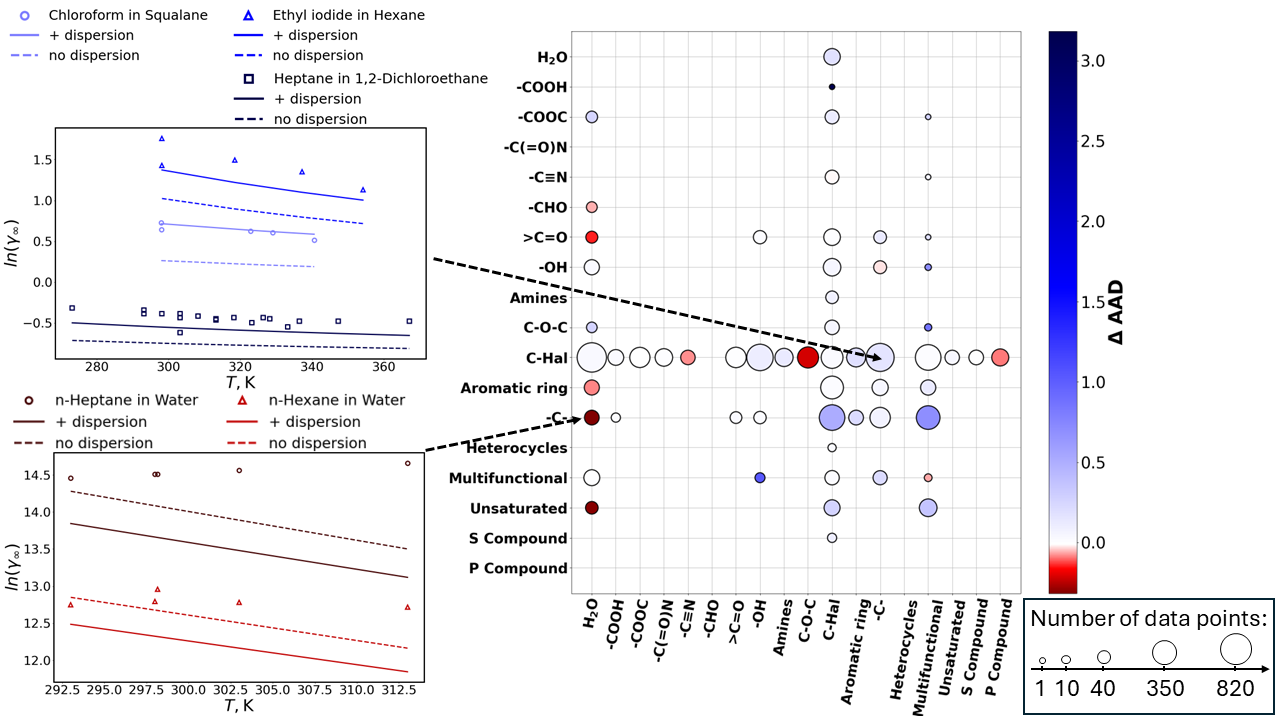}
    \caption{Visualization of \(\Delta~\textrm{AAD}\) (Eq.~\ref{eq:delta_AAD}) calculated for model \(w_{5}\) + Eq. \ref{eq:IP_wi} for binary IDAC data, accompanied by its temperature dependence examples, colored according to the order of magnitude of AAD, as shown by the color bar. Rows (solutes) and columns (solvents) represent the main chemical functional groups of the components. The size of the circles indicates the proportion of data points corresponding to molecules within each category. The concept for this plot is inspired by \citet{Fingerhut2017ComprehensiveEquilibria}.}
    \label{fig:circular_IDAC}
\end{figure}
\end{landscape}

Finally, a similar AAD analysis was conducted for the IDAC data. As illustrated in Figure~\ref{fig:circular_IDAC}, more than 80\% of the data categories showed improvement. In particular, most halocarbon systems exhibited significant decrease in AADs, both when halocarbons acted as solutes and as solvents. Since the first part of this work included relatively few halocarbon IDAC data points, we present additional examples of the temperature dependence of halocarbon–hydrocarbon systems in Figure~\ref{fig:circular_IDAC}. However, challenges remain for mixtures where water serves as the solvent. For instance, in the cases of heptane and hexane in water, the original model outperforms the revised approach. This further underscores the necessity of future work on the hydrogen bonding term in conjunction with the dispersion term.  

\begin{figure}
    \centering
    \includegraphics[width=0.45\linewidth]{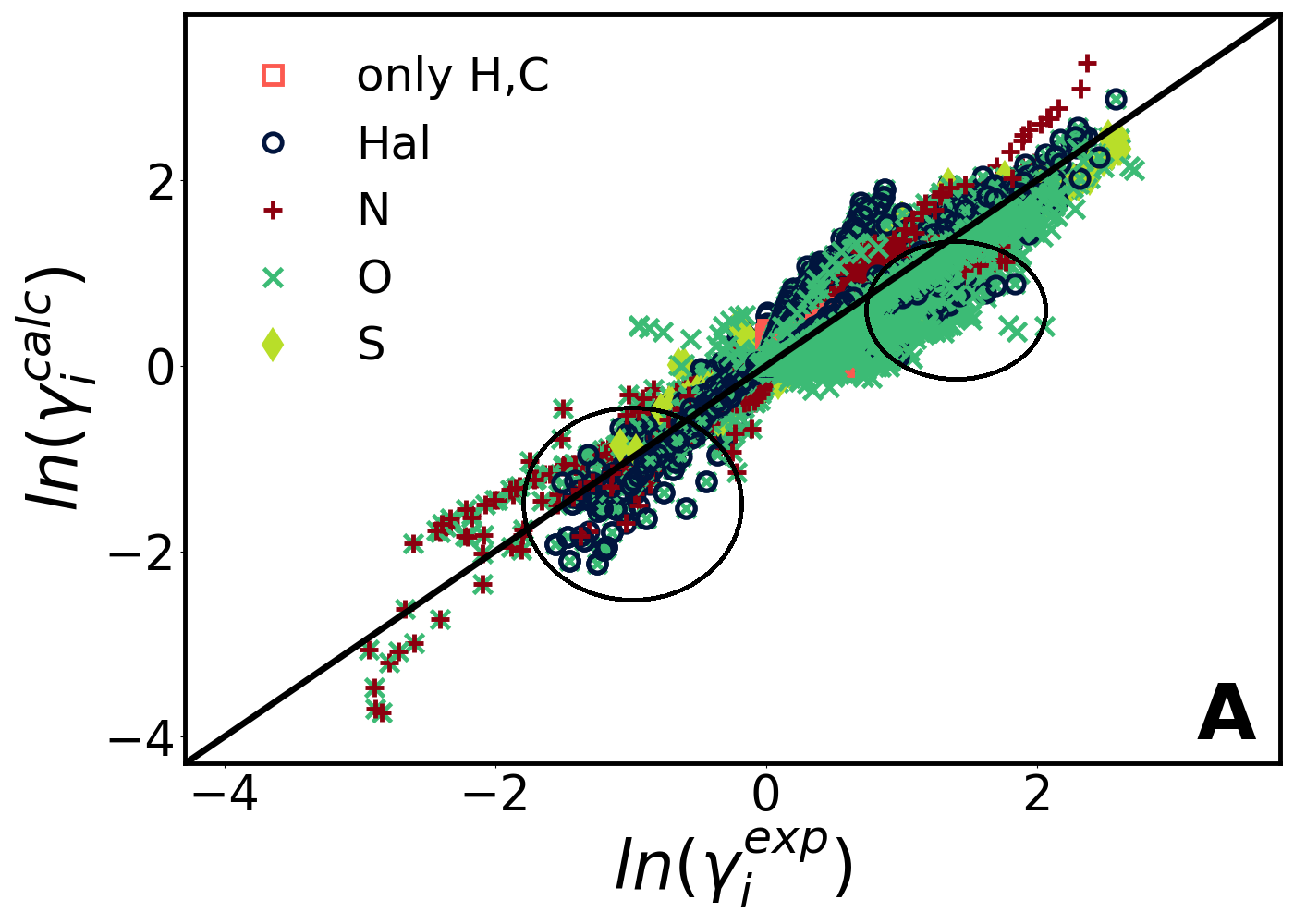}
    \includegraphics[width=0.45\linewidth]{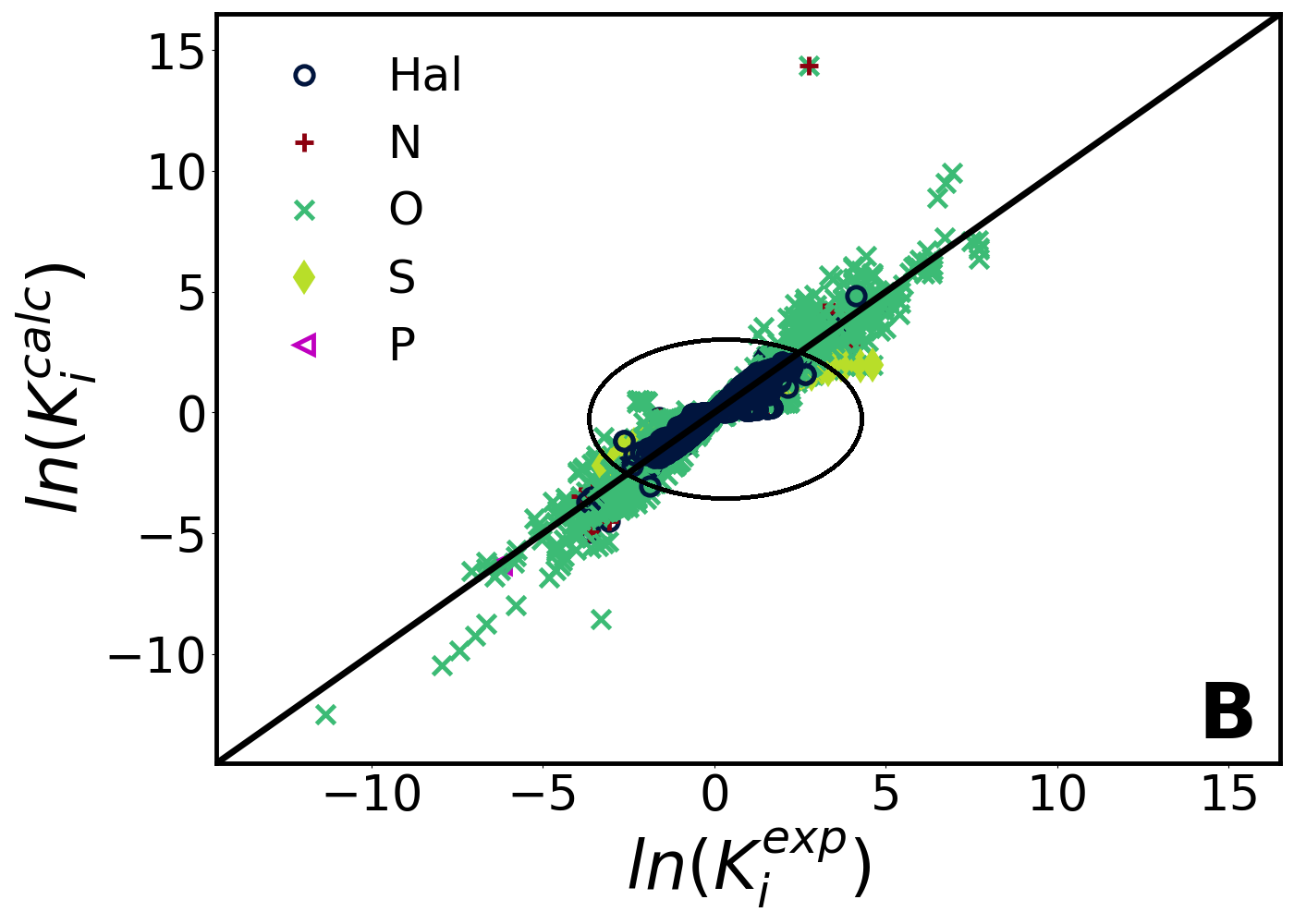}
    \includegraphics[width=0.45\linewidth]{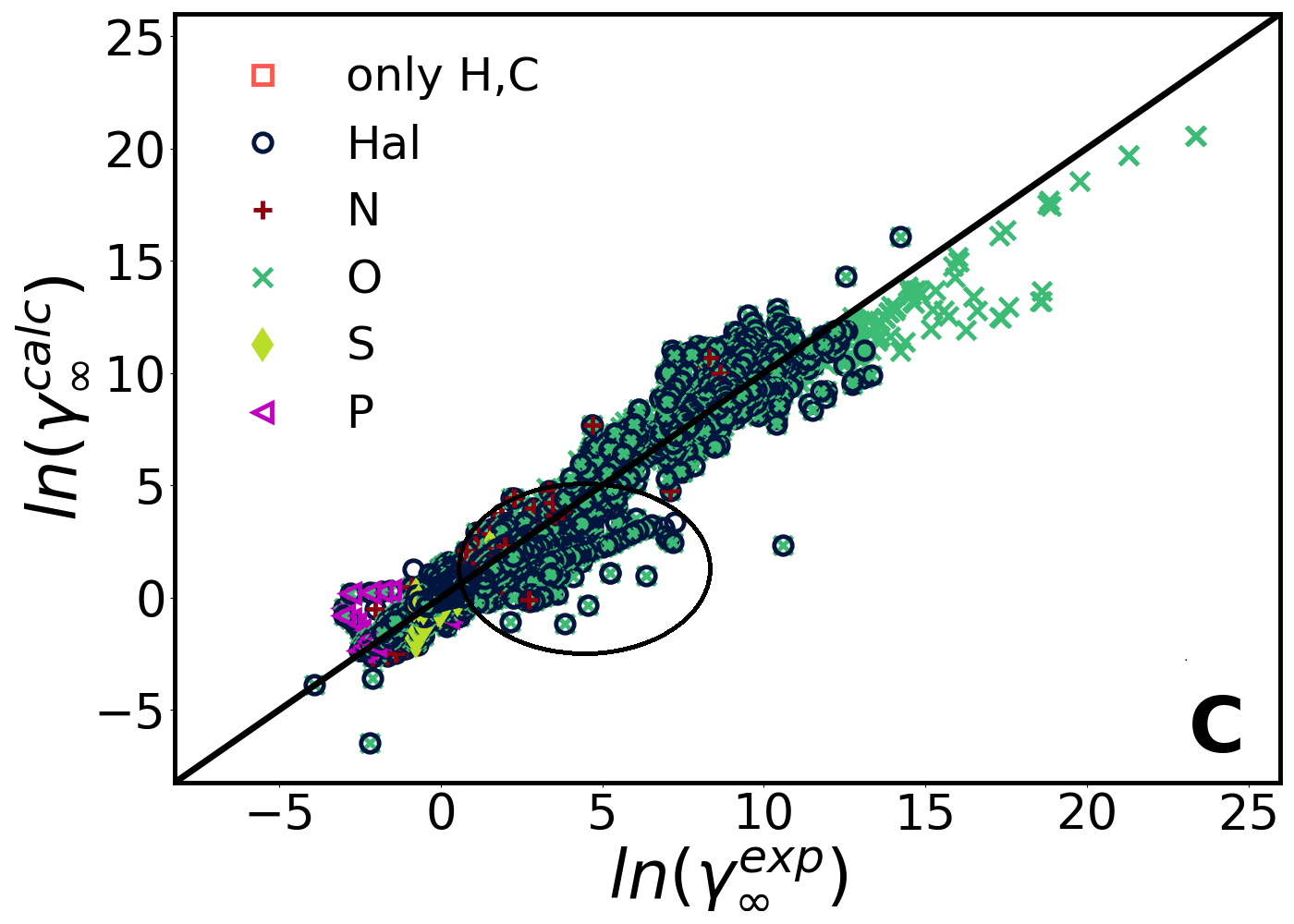}
    \includegraphics[width=0.45\linewidth]{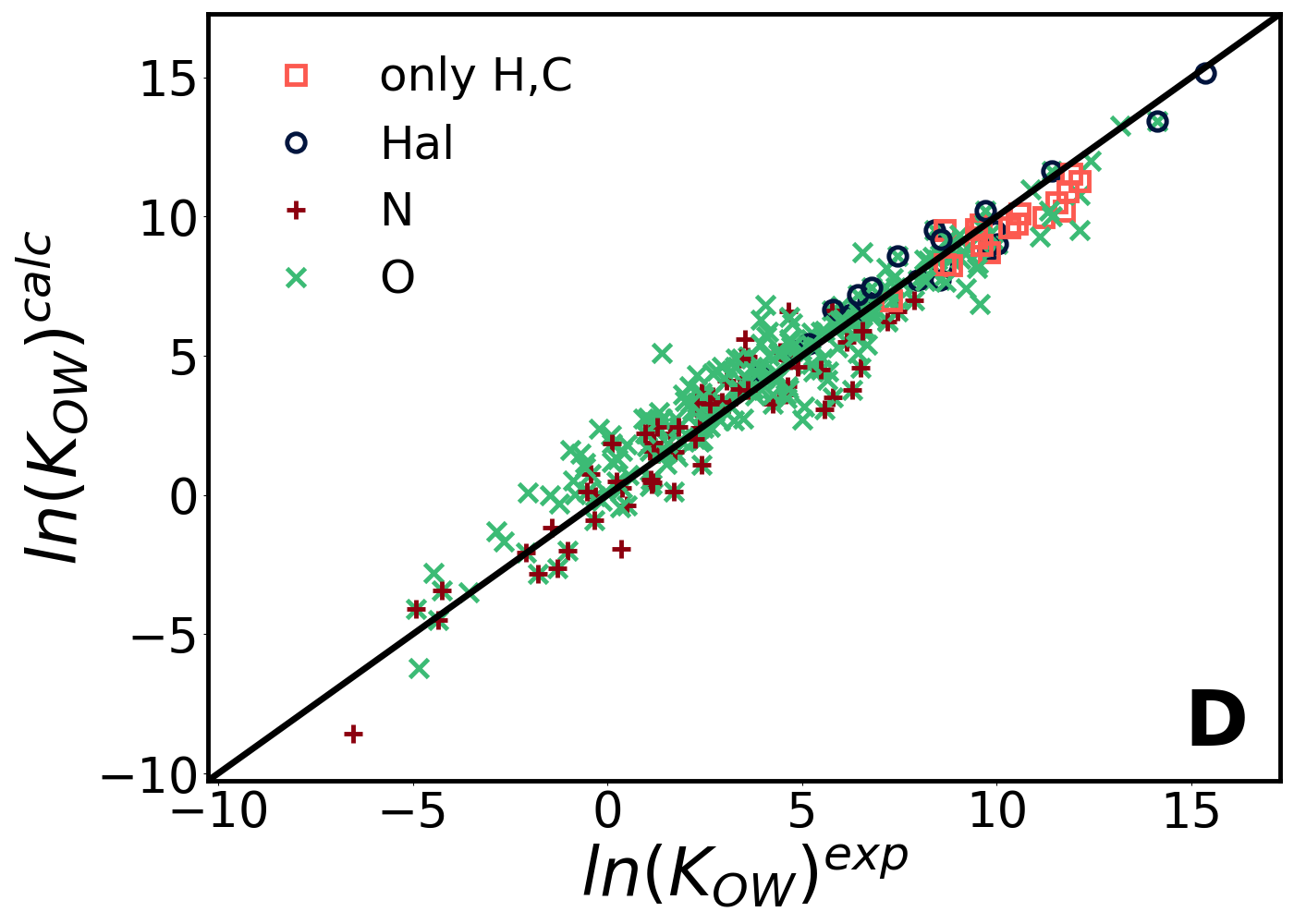} 
    \includegraphics[width=0.45\linewidth]{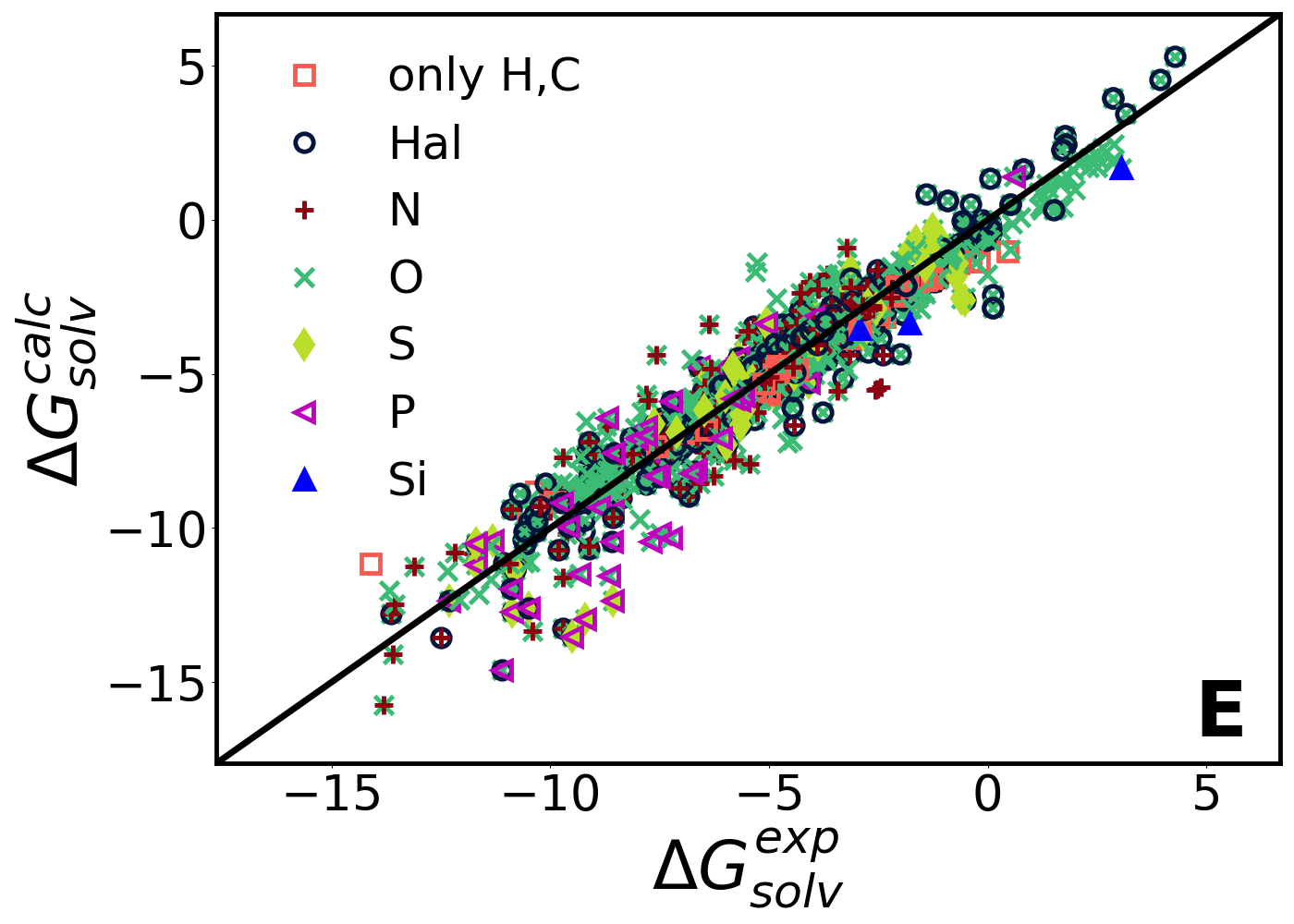} 
    \caption{Parity plots comparing predicted values (model \(w_{5}\) + Eq. \ref{eq:IP_wi}) with experimental data across various datasets: VLE activity coefficients (\textbf{A}), LLE partition coefficients (\textbf{B}), IDAC (\textbf{C}), \(K_i^{\textrm{org/w}}\) (\textbf{D}), and \(\Delta G_{\textrm{solv}}\) (\textbf{E}). The symbols are colored according to the type of atoms in the data. Regions with the most significant improvements are highlighted with circles.}
    \label{fig:all_set_parity}
\end{figure}

In regards to the solvation free energy and partitioning coefficients data, the changes in AADs are not as pronounced compared to the original model. Therefore, we do not provide a detailed group analysis, however the respective parity plots as well as ones for other data types, are illustrated in Figure \ref{fig:all_set_parity}. The circled regions in Figure \ref{fig:all_set_parity} highlight areas corresponding to data with the most notable improvements.

\clearpage

\section{Conclusions}
\label{sec:conclusions}

This study introduces an improved openCOSMO-RS framework through the development of the dispersion term based on atomic polarizability as a quantum chemical descriptor.

A significant part of this work involved developing a workflow for calculating atomic polarizabilities using the open-source quantum chemical package ORCA 6.0 and integrating these outputs into the openCOSMO-RS framework. Multiple strategies were evaluated, particularly for halocarbon systems, including polarizability projections onto molecular cavities, scaling, and combining polarizabilities to compute segment-segment dispersion energies. The most reliable results were achieved using polarizability projections, further corrected using atomic ionization potentials.

Compared to our earlier model, which relied on atomic dispersion parameters, the new approach demonstrates superior performance while requiring substantially fewer adjustable parameters. This suggests that the proposed methodology may offer a more physically grounded approach to dispersion interactions.

Lastly, applying the new model to a diverse dataset containing a wide range of chemical systems showed consistent improvements across various data types. Future work will focus on refining the openCOSMO-RS framework and exploring the additional molecular descriptors to enhance its predictive capabilities further.

\section*{Acknowledgements}

We would like to thank Edgar Ivan Sanchez Medina for generously sharing the digitalized DECHEMA database \citep{Gmehling2008ActivityCoefficients} for IDACs of halogenated compounds. 

%\bibliographystyle{elsarticle-harv} 
%\bibliography{references,cas-refs,manuscript_datarefs_LLE, manuscript_datarefs_VLE}

\bibliographystyle{elsarticle-harv} 
\bibliography{CES_ref_template}




% \end{thebibliography}
\end{document}

\begin{frontmatter}



\title{Supplementary Data for: A comprehensive approach to incorporating intermolecular dispersion into the openCOSMO-RS model. Part 1: Halocarbons}


\author[inst1,inst11]{Daria Grigorash}
\affiliation[inst1]{organization={Department of Chemistry, Technical University of Denmark},%Department and Organization
            city={Kgs. Lyngby},
            postcode={2800}, 
            country={Denmark}}
\affiliation[inst11]{organization={Center for Energy Resources Engineering, Technical University of Denmark},%Department and Organization
            city={Kgs. Lyngby},
            postcode={2800}, 
            country={Denmark}}
            
\author[inst2]{Simon Müller}
\affiliation[inst2]{organization={Institute of Thermal Separation Processes, Hamburg University of Technology},%Department and Organization
            city={Hamburg},
            postcode={21073}, 
            country={Germany}}

\author[inst3]{Patrice Paricaud}
\affiliation[inst3]{organization={UCP, ENSTA Paris, Institut Polytechnique de Paris},%Department and Organization
            city={Palaiseau},
            postcode={91762}, 
            country={France}}
            
\author[inst1,inst11]{Erling H. Stenby}
\author[inst2]{Irina Smirnova}
\author[inst1,inst11]{Wei Yan\corref{corresponding author}}
\cortext[corresponding author]{Corresponding author at: Department of Chemistry, Technical University of Denmark, 2800 Kgs. Lyngby, Denmark.\ead{weya@kemi.dtu.dk}}

\end{frontmatter}


\newpage
\section{Table S1. LLE systems and its sources.}
\label{sec:LLE_table}

%%       \citet{<label>} ==> Jones et al. [21]
%%       \citep{<label>} ==> [21]
%%
\renewcommand{\arraystretch}{1.5} % adjust the value as needed
\begin{xltabular}{\linewidth}{p{10cm}p{5cm}XX}

\toprule
System & Reference \\
\midrule
\endfirsthead

\multicolumn{2}{c}{{Table S1. LLE systems and its sources -- continued from previous page}} \\
\toprule
System & Reference \\
\midrule
\endhead

\bottomrule
\multicolumn{2}{c}{{Continued on next page}} \\
\endfoot

\bottomrule
\endlastfoot

1,1-difluoroethane - octafluoropropane & \citet{LLE2} \\ 
1,1-difluoroethane - tetrafluoromethane & \citet{LLE3} \\
1,2,2-trichloro-1,1,3,3,3-pentafluoropropane - bromine & \citet{LLE4} \\
1,3-diiodopropane - n-heptadecane & \citet{LLE5} \\
1,4-diiodobutane - n-heptadecane & \citet{LLE5} \\
1,5-diiodopentane - pentadecane & \citet{LLE5} \\
1-octene - 1,1,1,2,2,3,4,5,5,5-decafluoropentane & \citet{LLE6} \\
2-methylpropane - 1,1,2,2-tetrafluoroethane & \citet{LLE7} \\
2-methylpropane - tetrafluoromethane & \citet{LLE3} \\
chlorobenzene - perfluoro(methylcyclohexane) & \citet{LLE8} \\
chlorodifluoromethane - tetrafluoromethane & \citet{LLE3} \\
cyclohexane - diiodomethane & \citet{LLE9} \\
cyclohexane - hexadecafluoro-n-heptane & \citet{LoNostro2005PhasePerfluoroalkanes} \\
cyclohexane - octadecafluorooctane & \citet{LoNostro2005PhasePerfluoroalkanes} \\
cyclohexane - tetradecafluorohexane & \citet{LLE11}, \citet{LoNostro2005PhasePerfluoroalkanes} \\
decane - 1,1,1,2,2,3,4,5,5,5-decafluoropentane & \citet{LLE6} \\
decane - hexadecafluoro-n-heptane & \citet{LoNostro2005PhasePerfluoroalkanes} \\
decane - octadecafluorooctane & \citet{LoNostro2005PhasePerfluoroalkanes} \\
decane - tetradecafluorohexane & \citet{LoNostro2005PhasePerfluoroalkanes}, \citet{LLE44} \\
dichlorofluoromethane - tetrafluoromethane & \citet{LLE3} \\
ethane - tetrafluoromethane & \citet{LLE3} \\
ethane - trifluoromethane & \citet{LLE14}, \citet{LLE15} \\
fluoromethane - tetrafluoromethane & \citet{LLE16} \\
heptane - 1,3-diiodopropane & \citet{LLE5} \\
heptane - 1,4-diiodobutane & \citet{LLE5} \\
heptane - hexadecafluoro-n-heptane & \citet{LoNostro2005PhasePerfluoroalkanes}, \citet{LLE17}, \citet{LLE18} \\
heptane - hexafluoroethane & \citet{LLE34} \\
heptane - octadecafluorodecahydronaphthalene & \citet{LLE20}, \citet{LLE21} \\
heptane - octadecafluorooctane & \citet{LLE22}, \citet{LoNostro2005PhasePerfluoroalkanes} \\
heptane - tetradecafluorohexane & \citet{LLE23}, \citet{LLE24}, \citet{LLE44}, \citet{LLE48} \\
heptane - trifluoromethane & \citet{LLE34} \\
heptane - perfluoro(methylcyclohexane) & \citet{LLE26} \\
heptylbenzene - 1,1,1,2-tetrafluoroethane & \citet{LLE27} \\
hexadecafluoro-n-heptane - 2,2,4-trimethylpentane & \citet{LoNostro2005PhasePerfluoroalkanes}, \citet{LLE18} \\
hexadecafluoro-n-heptane - benzene & \citet{LoNostro2005PhasePerfluoroalkanes}, \citet{LLE18} \\
hexadecafluoro-n-heptane - dichloromethane & \citet{LoNostro2005PhasePerfluoroalkanes} \\
hexadecafluoro-n-heptane - tetrachloromethane & \citet{LoNostro2005PhasePerfluoroalkanes}, \citet{LLE17}, \citet{LLE18}, \citet{LLE28} \\
hexadecafluoro-n-heptane - trichloromethane & \citet{LoNostro2005PhasePerfluoroalkanes}, \citet{LLE18} \\
hexane - 1,1,1,2,2,3,4,5,5,5-decafluoropentane & \citet{LLE6} \\
hexane - octadecafluorodecahydronaphthalene & \citet{LLE20}, \citet{LLE21} \\
hexane - octadecafluorooctane & \citet{LLE22}, \citet{LoNostro2005PhasePerfluoroalkanes} \\
hexane - tetradecafluorohexane & \citet{LLE23}, \citet{LLE29}, \citet{LLE30}, \citet{LLE31}, \citet{LLE32}, \citet{LLE48}, \citet{LLE33} \\
hexane - trifluoromethane & \citet{LLE34}, \citet{LLE35} \\
hexane - perfluoro(methylcyclohexane) & \citet{LLE26} \\
methane - tetrafluoromethane & \citet{LLE36}, \citet{LLE37} \\
methylbenzene - 1,1,2,3,3,3-hexafluoro-1-propene & \citet{LLE38} \\
methylbenzene - tetradecafluorohexane & \citet{LLE39} \\
methylbenzene - perfluoro(methylcyclohexane) & \citet{LLE8} \\
n-butane - 1,1,1,2-tetrafluoroethane & \citet{LLE40} \\
n-butane - difluoromethane & \citet{LLE41} \\
n-butane - pentafluoroethane & \citet{LLE42} \\
nonane - octadecafluorodecahydronaphthalene & \citet{LLE43} \\
nonane - octadecafluorooctane & \citet{LLE22} \\
nonane - tetradecafluorohexane & \citet{LLE44} \\
nonane - perfluoro(methylcyclohexane) & \citet{LLE26} \\
octadecafluorodecahydronaphthalene - 1-heptene & \citet{LLE43} \\
octadecafluorodecahydronaphthalene - 1-hexene & \citet{LLE20}, \citet{LLE21} \\
octadecafluorooctane - 2,2,4-trimethylpentane & \citet{LoNostro2005PhasePerfluoroalkanes} \\
octadecafluorooctane - dichloromethane & \citet{LoNostro2005PhasePerfluoroalkanes} \\
octadecafluorooctane - tetrachloromethane & \citet{LoNostro2005PhasePerfluoroalkanes} \\
octadecafluorooctane - trichloromethane & \citet{LoNostro2005PhasePerfluoroalkanes} \\
octane - 1H-perfluoro-n-octane & \citet{LLE45} \\
octane - 1H,8H-perfluoro-n-octane & \citet{LLE45} \\
octane - 1Br-perfluoro-n-octane & \citet{LLE45} \\
octane - hexadecafluoro-n-heptane & \citet{LoNostro2005PhasePerfluoroalkanes} \\
octane - octadecafluorodecahydronaphthalene & \citet{LLE43} \\
octane - octadecafluorooctane & \citet{LLE46}, \citet{LLE22}, \citet{LLE32}, \citet{LoNostro2005PhasePerfluoroalkanes} \\
octane - tetradecafluorohexane & \citet{LLE23}, \citet{LLE32}, \citet{LoNostro2005PhasePerfluoroalkanes} \\
octane - perfluoro(methylcyclohexane) & \citet{LLE48}, \citet{LLE26} \\
pentane - dodecafluoropentane & \citet{LLE50} \\
propane - difluoromethane & \citet{LLE51} \\
propane - tetrafluoromethane & \citet{LLE3} \\
tetrachloromethane - iodine & \citet{LLE54} \\
tetradecafluorohexane - benzene & \citet{LoNostro2005PhasePerfluoroalkanes}, \citet{LLE56} \\
tetradecafluorohexane - dichloromethane & \citet{LoNostro2005PhasePerfluoroalkanes} \\
tetradecafluorohexane - tetrachloromethane & \citet{LoNostro2005PhasePerfluoroalkanes} \\
tetradecafluorohexane - tetradecane & \citet{LLE56} \\
tetradecafluorohexane - trichloromethane & \citet{LoNostro2005PhasePerfluoroalkanes} \\
trifluoromethane - tetrafluoromethane & \citet{LLE57} \\
perfluoro(methylcyclohexane) - 1-heptene & \citet{LLE26} \\
perfluoro(methylcyclohexane) - 1-hexene & \citet{LLE26} \\
perfluoro(methylcyclohexane) - benzene & \citet{LLE8}, \citet{LLE58} \\
perfluoro(methylcyclohexane) - tetrachloromethane & \citet{LLE48}, \citet{LLE59} \\
perfluoro(methylcyclohexane) - trichloromethane & \citet{LLE8}, \citet{LLE58} \\
undecane - 1,4-diiodobutane & \citet{LLE5} \\
\end{xltabular}
\newpage
\section{Table S2. VLE systems and its sources.}
\label{sec:VLE_table}
\begin{xltabular}{\linewidth}{lp{5cm}XX}
\toprule
System & Reference \\
\midrule
\endfirsthead

\multicolumn{2}{c}{{Table S2. VLE systems and its sources -- continued from previous page}} \\
\toprule
System & Reference \\
\midrule
\endhead

\bottomrule
\multicolumn{2}{c}{{Continued on next page}} \\
\endfoot

\bottomrule
\endlastfoot
n-hexane - 1,2-dichloroethane & \citet{CHERIC2024}, \citet{Luo2021} \\
n-heptane - 1,2-dichloroethane & \citet{CHERIC2024} \\
n-octane - 1,2-dichloroethane & \citet{CHERIC2024} \\
n-hexane - cyclohexane & \citet{Jaubert2020BenchmarkAccuracy}, \citet{CHERIC2024} \\
benzene - cyclohexane & \citet{CHERIC2024}, \citet{Jaubert2020BenchmarkAccuracy} \\
benzene - n-heptane & \citet{CHERIC2024}, \citet{Jaubert2020BenchmarkAccuracy} \\
perfluoromethylcyclohexane - n-hexane & \citet{CHERIC2024} \\
perfluoromethylcyclohexane - 1-hexene & \citet{CHERIC2024} \\
cyclohexane - monochlorobenzene & \citet{Jaubert2020BenchmarkAccuracy} \\
propylene - dichlorodifluoromethane & \citet{CHERIC2024} \\
propylene - chlorodifluoromethane & \citet{CHERIC2024}, \citet{Jaubert2020BenchmarkAccuracy} \\
propylene - 1,1-difluoroethane & \citet{CHERIC2024} \\
1,1,1,2-tetrafluoroethane - dichlorodifluoromethane & \citet{CHERIC2024} \\
propylene - 1,1,1,2-tetrafluoroethane & \citet{CHERIC2024} \\
propylene - chloropentafluoroethane & \citet{CHERIC2024} \\
1,1,1,2-tetrafluoroethane - propane & \citet{CHERIC2024}, \citet{Dong2011a}, \citet{Lim2006} \\
isoprene - n-pentane & \citet{CHERIC2024} \\
benzene - n-hexane & \citet{CHERIC2024} \\
benzene - n-octane & \citet{CHERIC2024} \\
benzene - n-decane & \citet{CHERIC2024} \\
toluene - n-hexane & \citet{CHERIC2024} \\
toluene - n-heptane & \citet{CHERIC2024} \\
toluene - n-octane & \citet{CHERIC2024} \\
toluene - n-decane & \citet{CHERIC2024} \\
p-xylene - n-hexane & \citet{CHERIC2024} \\
p-xylene - n-heptane & \citet{CHERIC2024} \\
dichloromethane - carbon tetrachloride & \citet{Jaubert2020BenchmarkAccuracy}, \citet{CHERIC2024} \\
dichloromethane - n-pentane & \citet{Jaubert2020BenchmarkAccuracy}, \citet{CHERIC2024} \\
p-xylene - n-octane & \citet{CHERIC2024} \\
p-xylene - n-decane & \citet{CHERIC2024} \\
ethylbenzene - n-hexane & \citet{CHERIC2024} \\
ethylbenzene - n-heptane & \citet{CHERIC2024} \\
ethylbenzene - n-octane & \citet{CHERIC2024} \\
ethylbenzene - n-decane & \citet{CHERIC2024} \\
chloroform - n-hexane & \citet{Jaubert2020BenchmarkAccuracy}, \citet{CHERIC2024} \\
chloroform - carbon tetrachloride & \citet{Jaubert2020BenchmarkAccuracy} \\
chloroform - 1,2-dichloroethane & \citet{Jaubert2020BenchmarkAccuracy} \\
chloroform - n-heptane & \citet{Jaubert2020BenchmarkAccuracy}, \citet{CHERIC2024} \\
chloroform - benzene & \citet{Jaubert2020BenchmarkAccuracy}, \citet{CHERIC2024} \\
1,2-dichloroethane - carbon tetrachloride & \citet{Jaubert2020BenchmarkAccuracy}, \citet{CHERIC2024} \\
1,2-dichloroethane - toluene & \citet{Jaubert2020BenchmarkAccuracy} \\
chlorodifluoromethane - dichlorodifluoromethane & \citet{Jaubert2020BenchmarkAccuracy} \\
difluoromethane - propane & \citet{Jaubert2020BenchmarkAccuracy}, \citet{Kim2003} \\
trifluoromethane - ethane & \citet{Jaubert2020BenchmarkAccuracy}, \citet{Zhang2006} \\
decafluorobutane - 1,1,1,3,3-pentafluorobutane & \citet{Jaubert2020BenchmarkAccuracy} \\
pentafluoroethane - propane & \citet{Jaubert2020BenchmarkAccuracy}, \citet{Kim2003} \\
cyclohexane - toluene & \citet{CHERIC2024} \\
ethylene - propane & \citet{CHERIC2024} \\
1-chlorobutane - n-heptane & \citet{CHERIC2024} \\
2-chlorobutane - n-heptane & \citet{CHERIC2024} \\
2-chlorobutane - methylcyclohexane & \citet{CHERIC2024} \\
tert-butyl chloride - methylcyclohexane & \citet{CHERIC2024} \\
ethylbenzene - cyclooctane & \citet{CHERIC2024} \\
n-heptane - monochlorobenzene & \citet{Jaubert2020BenchmarkAccuracy} \\
n-hexane - 2-methylpentane & \citet{CHERIC2024} \\
n-hexane - 3-methylpentane & \citet{CHERIC2024} \\
p-xylene - carbon tetrachloride & \citet{CHERIC2024} \\
chlorobenzene - p-xylene & \citet{CHERIC2024} \\
chlorobenzene - methylcyclohexane & \citet{CHERIC2024} \\
fluorobenzene - methylcyclohexane & \citet{CHERIC2024} \\
n-heptane - o-xylene & \citet{CHERIC2024} \\
chloroform - toluene & \citet{CHERIC2024} \\
2,2,4-trimethylpentane - toluene & \citet{CHERIC2024} \\
2,4-dimethylpentane - cyclohexane & \citet{CHERIC2024} \\
carbon tetrachloride - benzene & \citet{CHERIC2024} \\
carbon tetrachloride - n-hexane & \citet{CHERIC2024} \\
carbon tetrachloride - cyclohexene & \citet{CHERIC2024} \\
carbon tetrachloride - 1-hexene & \citet{CHERIC2024} \\
carbon tetrachloride - 2-methylpentane & \citet{CHERIC2024} \\
carbon tetrachloride - 2,3-dimethylbutane & \citet{CHERIC2024} \\
difluorodichloromethane - chlorodifluoromethane & \citet{CHERIC2024} \\
carbon tetrachloride - methylcyclopentane & \citet{CHERIC2024} \\
carbon tetrachloride - 2,4-dimethylpentane & \citet{CHERIC2024} \\
difluorodichloromethane - hexafluoropropylene & \citet{CHERIC2024} \\
difluorodichloromethane - octafluorocyclobutane & \citet{CHERIC2024} \\
2,3-dimethylbutane - chloroform & \citet{CHERIC2024} \\
cyclopentane - tetrachloroethylene & \citet{CHERIC2024} \\
1-chloro-1,1-difluoroethane - 1,1,1-trichloroethane & \citet{CHERIC2024} \\
cyclohexane - n-heptane & \citet{CHERIC2024} \\
1,1,2-trichlorotrifluoroethane - n-hexane & \citet{CHERIC2024} \\
ethylbenzene - styrene & \citet{CHERIC2024} \\
ethylcyclohexane - ethylbenzene & \citet{CHERIC2024} \\
benzene - 2,3-dimethylpentane & \citet{CHERIC2024} \\
n-octane - ethylcyclohexane & \citet{CHERIC2024} \\
2,2,4-trimethylpentane - 1,2-dichloroethane & \citet{CHERIC2024} \\
2,2,4-trimethylpentane - 1,1,1-trichloroethane & \citet{CHERIC2024} \\
2,2,4-trimethylpentane - 1,1,2,2-tetrachloroethane & \citet{CHERIC2024} \\
2,2,4-trimethylpentane - trichloroethylene & \citet{CHERIC2024} \\
2,2,4-trimethylpentane - tetrachloroethylene & \citet{CHERIC2024} \\
n-hexane - 1-chlorobutane & \citet{CHERIC2024} \\
n-hexane - dichloromethane & \citet{CHERIC2024} \\
n-hexane - 1,3-dichloropropane & \citet{CHERIC2024} \\
carbon tetrachloride - n-heptane & \citet{CHERIC2024} \\
toluene - 1-chlorohexane & \citet{CHERIC2024} \\
cyclohexene - 1,2-dichloroethane & \citet{CHERIC2024} \\
cyclohexane - 1,2-dichloroethane & \citet{CHERIC2024} \\
chlorodifluoromethane - octafluoropropane & \citet{CHERIC2024} \\
1-methylnaphthalene - naphthalene & \citet{CHERIC2024} \\
2,2,4-trimethylpentane - chlorobenzene & \citet{CHERIC2024} \\
benzene - benzotrifluoride & \citet{CHERIC2024} \\
benzotrifluoride - toluene & \citet{CHERIC2024} \\
benzotrifluoride - chlorobenzene & \citet{CHERIC2024} \\
2,4-dimethylpentane - benzene & \citet{CHERIC2024} \\
1,2-dichloroethane - trichloroethylene & \citet{CHERIC2024} \\
1,2-dichloroethane - tetrachloroethylene & \citet{CHERIC2024} \\
trans-1,2-dichloroethylene - cis-1,2-dichloroethylene & \citet{CHERIC2024} \\
trifluoromethane - chlorotrifluoromethane & \citet{CHERIC2024} \\
1-octene - ethylbenzene & \citet{CHERIC2024} \\
n-pentane - toluene & \citet{CHERIC2024} \\
hexafluorobenzene - methylcyclohexane & \citet{CHERIC2024} \\
n-hexane - hexafluorobenzene & \citet{CHERIC2024} \\
1,3-butadiene - n-butane & \citet{CHERIC2024} \\
1-chlorobutane - carbon tetrachloride & \citet{CHERIC2024} \\
1,2-dichloroethane - 1-chlorobutane & \citet{CHERIC2024} \\
cyclohexane - 1-octene & \citet{CHERIC2024} \\
hexafluorobenzene - isopropylcyclohexane & \citet{CHERIC2024} \\
2,2,3-trimethylbutane - benzene & \citet{CHERIC2024} \\
benzene - cyclohexene & \citet{CHERIC2024} \\
n-pentane - benzene & \citet{CHERIC2024} \\
n-heptane - chlorobenzene & \citet{CHERIC2024} \\
trifluoroiodomethane - 1,1-difluoroethane & \citet{Gong2015} \\
trifluoroiodomethane - trans-1,3,3,3-tetrafluoropropene & \citet{Guo2012} \\
trifluoroiodomethane - propane & \citet{Gong2014} \\
trifluoroiodomethane - isobutane & \citet{Guo2013} \\
m-xylene - ethylbenzene & \citet{Li2023} \\
ethyl iodide - 1-bromopropane & \citet{Artal} \\
bromotrifluoromethane - chlorodifluoromethane & \citet{Kuznetsov} \\
carbon tetrachloride - methyl iodide & \citet{Moelwyn-Hughes} \\
1-bromobenzene - fluorobenzene & \citet{Al-Hayan} \\
carbon tetrachloride - ethyl iodide & \citet{Al-Hayan} \\
methyl iodide - chloroform & \citet{Moelwyn-Hughes} \\
methyl iodide - dichloromethane & \citet{Moelwyn-Hughes} \\
ethylene dibromide - chlorobenzene & \citet{Lacher} \\
1-bromobutane - heptane & \citet{Smyth} \\
1-bromobenzene - chlorobenzene & \citet{Bourrelly} \\
1-bromopropane - cyclohexane & \citet{Wisniak} \\
heptane - 1-bromobenzene & \citet{Al-Hayan} \\
1-bromopropane - heptane & \citet{Wisniak} \\
2,3,3,3-tetrafluoropropene - isobutane & \citet{Hu2014} \\
difluoromethane - n-butane & \citet{Fedele2005b} \\
isobutane - 1,1,1,2-tetrafluoroethane & \citet{Bobbo1998},\citet{Lim1999} \\
1,1,1,2-tetrafluoroethane - 2,3,3,3-tetrafluoropropene & \citet{Abbadi2022} \\
propane - 1,1,1-trifluoroethane & \citet{Stryjek1998},\citet{Im2006} \\
difluoromethane - 1,1,1-trifluoroethane & \citet{Kim2000b} \\
propane - 1,1-difluoroethane & \citet{Dong2010} \\
trans-1,3,3,3-tetrafluoropropene - propane & \citet{Dong2011b} \\
isopentane - 1,1,1,3,3-pentafluoropropane & \citet{Ahmar2012} \\
isopentane - 1,1,1,3,3-pentafluorobutane & \citet{Ahmar2012} \\
isobutane - 1,1-difluoroethane & \citet{Lim1999} \\
2,3,3,3-tetrafluoropropene - butane & \citet{Hu2018} \\
2,3,3,3-tetrafluoropropene - propane & \citet{Zhong2017} \\
isobutane - 1,1,1,3,3-pentafluoropropane & \citet{Bobbo2001} \\
isobutane - trans-1,3,3,3-tetrafluoropropene & \citet{Dong2012} \\
propane - 3,3,3-trifluoropropene & \citet{Ding2020} \\
isobutane - 3,3,3-trifluoropropene & \citet{Deng2020} \\
3,3,3-trifluoropropene - trans-1,3,3,3-tetrafluoropropene & \citet{Yang2020} \\
difluoromethane - propylene & \citet{Ho2005} \\
propane - propylene & \citet{Harmens1985} \\

\end{xltabular}

\newpage

\section{Figure S1. Comparison of AAD values with and without the dispersion term across all data sets classified by the type of halogen atom}

\begin{figure}
    \centering
    \includegraphics[width=0.75\linewidth]{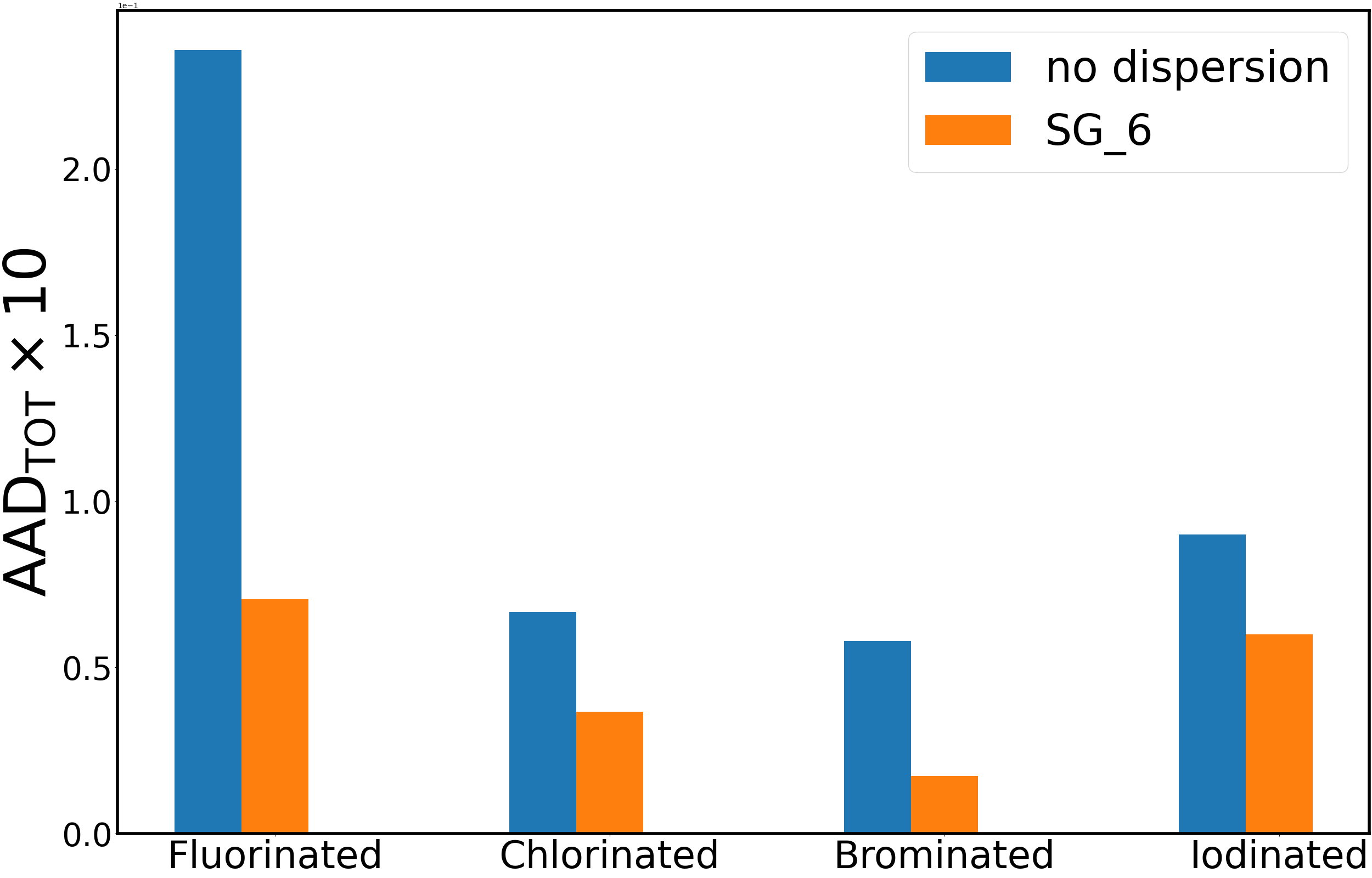}
    \label{fig:enter-label}
\end{figure}
\newpage
\section{Figure S2. Experimental IDAC (\citet{doi:10.1021/ie9503555}) of toluene and benzene in hexadecane at different temperatures.}

\begin{figure}
    \centering
    \includegraphics[width=0.75\linewidth]{plots/tol_benz_lngamma_suppl.png}
    \label{fig:enter-label}
\end{figure}

The following table is for CHBT=0
\begin{table}[htbp]
    \centering
    \caption{vdW Parameters for different atoms. For \( \tau^{\text{vdW}}_{\text{H-C}} \), \( \tau^{\text{vdW}}_{\text{H-N}} \), and \( \tau^{\text{vdW}}_{\text{H-O}} \), the second symbol indicates the atom to which hydrogen is bonded (C, N, and O, respectively).}
    \begin{adjustbox}{width=0.5\textwidth} % Adjust table width to 75% of the text width
    \begin{scriptsize} % Further reduce font size for scientific paper formatting
    \begin{tabular}{@{}lc@{}}
        \toprule
        Dispersion parameter & Value [J\textsuperscript{0.5}/Å] \\
        \midrule
        \( \tau^{\text{vdW}}_{\text{H-C}} \) & 1.130 \\
        \( \tau^{\text{vdW}}_{\text{H-N}} \) & 26.760 \\
        \( \tau^{\text{vdW}}_{\text{H-O}} \) & 3.115 \\
        \( \tau^{\text{vdW}}_{\text{C}} \) & 1.293 \\
        \( \tau^{\text{vdW}}_{\text{N}} \) & 68.673 \\
        \( \tau^{\text{vdW}}_{\text{O}} \) & 1.238 \\
        \( \tau^{\text{vdW}}_{\text{F}} \) & 18.059 \\
        \( \tau^{\text{vdW}}_{\text{Si}} \) & 16.296 \\
        \( \tau^{\text{vdW}}_{\text{P}} \) & 41.701 \\
        \( \tau^{\text{vdW}}_{\text{S}} \) & 15.477 \\
        \( \tau^{\text{vdW}}_{\text{Cl}} \) & 4.176 \\
        \( \tau^{\text{vdW}}_{\text{Br}} \) & 20.082 \\
        \( \tau^{\text{vdW}}_{\text{I}} \) & 18.289 \\
        \bottomrule
    \end{tabular}
    \end{scriptsize}
    \end{adjustbox}
    \label{tab:vdw_params}
\end{table}

the following table is for CHBT = 1.5

\begin{table}[htbp]
    \centering
    \caption{vdW Parameters for different atoms. For \( \tau^{\text{vdW}}_{\text{H-C}} \), \( \tau^{\text{vdW}}_{\text{H-N}} \), and \( \tau^{\text{vdW}}_{\text{H-O}} \), the second symbol indicates the atom to which hydrogen is bonded (C, N, and O, respectively).}
    \begin{adjustbox}{width=0.5\textwidth} % Adjust table width to 75% of the text width
    \begin{scriptsize} % Reduce font size for scientific paper formatting
    \begin{tabular}{@{}lc@{}}
        \toprule
        Dispersion parameter & Value [J\textsuperscript{0.5}/Å] \\
        \midrule
        \( \tau^{\text{vdW}}_{\text{H-C}} \) & 1.067 \\
        \( \tau^{\text{vdW}}_{\text{H-N}} \) & 12.191 \\
        \( \tau^{\text{vdW}}_{\text{H-O}} \) & 5.049 \\
        \( \tau^{\text{vdW}}_{\text{C}} \) & 1.022 \\
        \( \tau^{\text{vdW}}_{\text{N}} \) & 68.169 \\
        \( \tau^{\text{vdW}}_{\text{O}} \) & 1.213 \\
        \( \tau^{\text{vdW}}_{\text{F}} \) & 17.941 \\
        \( \tau^{\text{vdW}}_{\text{Si}} \) & 34.334 \\
        \( \tau^{\text{vdW}}_{\text{P}} \) & 55.170 \\
        \( \tau^{\text{vdW}}_{\text{S}} \) & 18.650 \\
        \( \tau^{\text{vdW}}_{\text{Cl}} \) & 5.189 \\
        \( \tau^{\text{vdW}}_{\text{Br}} \) & 19.055 \\
        \( \tau^{\text{vdW}}_{\text{I}} \) & 13.538 \\
        \bottomrule
    \end{tabular}
    \end{scriptsize}
    \end{adjustbox}
    \label{tab:vdw_params_2}
\end{table}

\bibliographystyle{elsarticle-harv} 
\bibliography{CES_ref_template}





% \end{thebibliography}